\documentclass[letterpaper,12pt]{article}
\usepackage{tabularx} 
\usepackage{amsmath}  
\usepackage{graphicx} 
\usepackage[margin=1in,letterpaper]{geometry} 
\usepackage{cite} 
\usepackage[final]{hyperref} 
\hypersetup{
	colorlinks=true,       
	linkcolor=blue,        
	citecolor=blue,        
	filecolor=magenta,     
	urlcolor=blue         
}

\usepackage{mathrsfs}
\usepackage{float}

\usepackage{revsymb}

\usepackage{amssymb}

\usepackage{exscale}

\usepackage{relsize}

\usepackage{gensymb}

\usepackage{changes}

\begin{document}



%
\title{Steady-State Micro-Bunching based on Transverse-Longitudinal Coupling}
\author{Xiujie~Deng\thanks{dengxiujie@mail.tsinghua.edu.cn},\ Alexander~Wu~Chao,\\ Institute for Advanced Study,\\ Tsinghua University, Beijing, China\\~\\ Wenhui~Huang,\ Zizheng~Li,\ Zhilong~Pan,\ Chuanxiang~Tang,\\ Department of Engineering Physics,\\ Tsinghua University, Beijing, China}

\author{Xiujie~Deng\thanks{dengxiujie@mail.tsinghua.edu.cn},\ Alexander~Wu~Chao,\ Wenhui~Huang,\\ Zizheng~Li,\ Zhilong~Pan,\ Chuanxiang~Tang,\\ Tsinghua University, Beijing, China}

\date{\today}
\maketitle

%
\begin{abstract}

	In this paper, three specific scenarios of a novel accelerator light source mechanism called steady-state micro-bunching (SSMB) have been studied, i.e., longitudinal weak focusing, longitudinal strong focusing and generalized longitudinal strong focusing (GLSF). At present, GLSF is the most promising among them in realizing high-power short-wavelength coherent radiation with a mild requirement on the modulation laser power. Its essence is to exploit the ultrasmall natural vertical emittance of an electron beam in a planar storage ring for efficient microbunching formation, like a partial transverse-longitudinal emittance exchange at the optical laser wavelength range. Based on indepth investigation of related beam physics, a solution of a GLSF SSMB storage ring which can deliver 1 kW-average-power EUV light is presented. The work in this paper, such as the generalized Courant-Snyder formalism, the analysis of theoretical minimum emittances, transverse-longitudinal coupling dynamics, and the derivation of bunching factor and modulation strengths for laser-induced microbunching schemes, is expected to be useful not only for the development of SSMB but also for future accelerator light sources in general that demand increasingly precise electron beam phase space manipulations.

%
	
\end{abstract}

\thispagestyle{empty}


\clearpage

\tableofcontents

\thispagestyle{empty}

\clearpage

\setcounter{page}{1}

\begin{figure}[tb]
	\centering
	\includegraphics[width=1\linewidth]{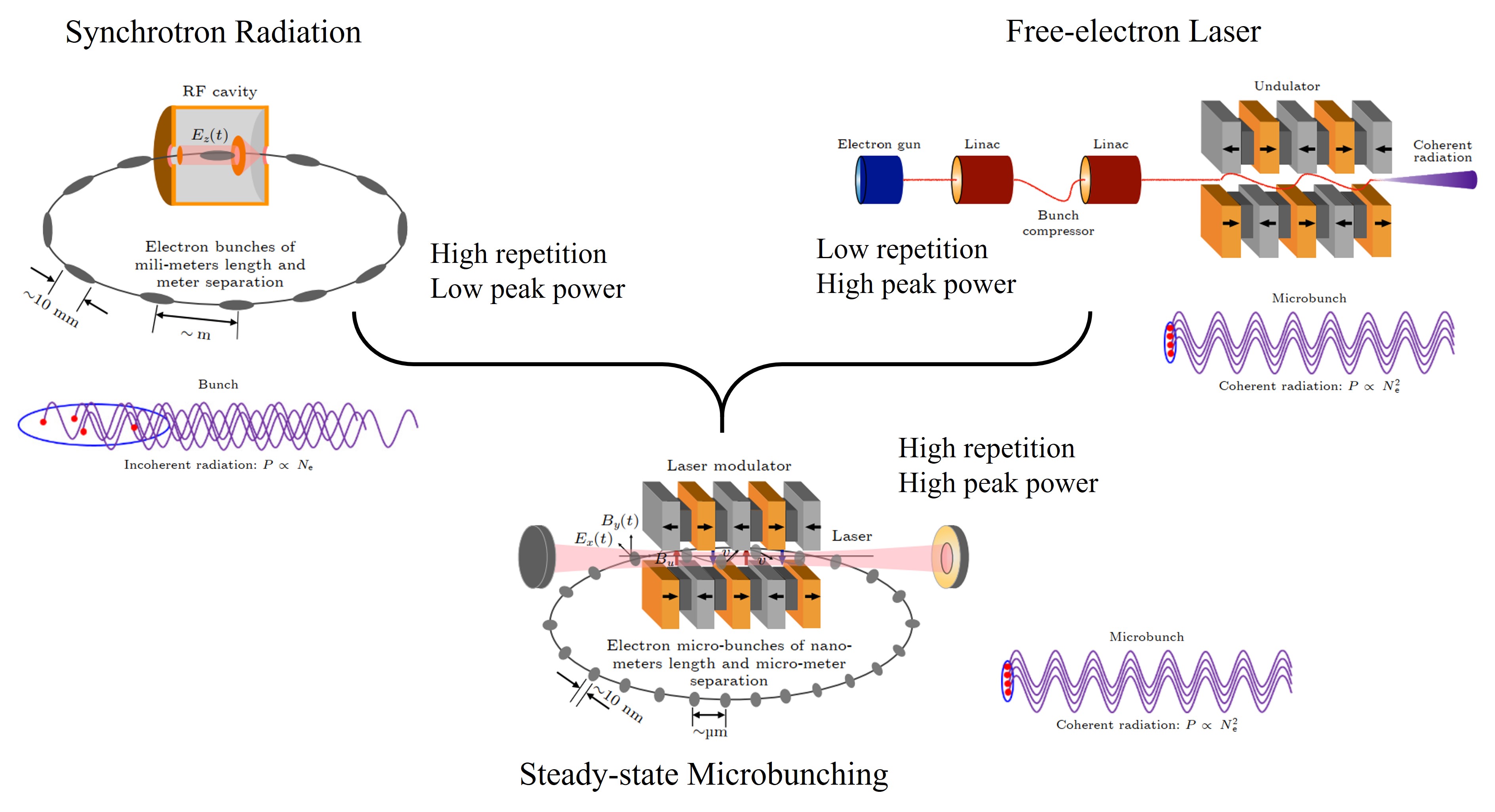}
	\caption{A schematic layout of an SSMB storage ring, and comparison to a synchrotron radiation source and free-electron laser. (Figure adapted from Ref.~\cite{Tang2022})}
	\label{fig:ssmb}
\end{figure}

\section{Introduction}

Accelerator as light source is arguably the most active driving force for accelerator development at the moment. There are presently two types of workhorses for these sources, {namely storage ring-based synchrotron radiation sources and linear accelerator (linac)-based free-electron lasers (FELs). They deliver light with high repetition rate and high peak power or brilliance, respectively.}  What we are trying to develop is a new storage ring-based light source mechanism called steady-state micro-bunching (SSMB)~\cite{Ratner2010SSMB,Jiao2021PRSTAB,Chao2016,Tang2018FLS,Deng2020Single,Deng2020Widen,Zhang2021,deng2021courant,li2023GLSF,TsaiPRAB2021,TsaiPRAB2022,TsaiNIMA2022,Zhao2024SpaceCharge,deng2021experimental,kruschinski2024confirming,Pan2020Thesis,deng2024theoretical,Zhang2022Thesis,Tang2022}, which hopefully can combine the advantages of such two kind of sources and promise both high-repetition and high-power radiation, {realizing an accelerator based fully coherent light source. The schematic layout of an SSMB storage ring and its comparison to the present synchrotron radiation source and FEL is shown in Fig.~\ref{fig:ssmb}. In a conventional storage ring, the electron bunches are longitudinally focused by one or multiple radio-frequency (RF) cavities, while in SSMB such bunching system is replaced by one or several optical laser modulation systems. The wavelength of laser is six orders of magnitude smaller than that of an RF. The bunch length or structure created by laser is very short, thus the term microbunching. When beam becomes microbunched, it can radiate coherently and strongly. Roughly speaking, if a linac-based FEL can be viewed as a free-electron based low-repetition pulsed laser, then SSMB can be viewed as a free electron-based high-repetition or continuous wave laser. But note that in SSMB, there is no exponential growth of the radiation power as that in a high-gain FEL~\cite{Bonifacio1984} or conventional quantum laser. The term laser in this context mainly reflects that the radiation is coherent, both transversely and longitudinally. To ensure the electron beam property can preserve turn by turn, the SSMB radiator length is comparatively short, typically at meter level, and the peak current of electron beam in SSMB is also lower than that in a high-gain FEL. The radiation back reaction on electron beam is therefore not violent and can be balanced by radiation damping in the ring. }

Once realized, such an SSMB ring can produce EUV radiation with greatly enhanced power and flux, allowing sub-meV energy resolution in angle-resolved photoemission spectroscopy (ARPES) and providing new opportunities for fundamental physics research, like revealing key electronic structures in topological materials. A kilowatt (kW)-level EUV source based on such a scheme is also promising to EUV lithography for high-volume chip manufacturing. The reward of such an SSMB ring is therefore tremendous. But one can imagine there are problems to be investigated and solved on the road of every new concept into a reality. To generate coherent EUV radiation in a storage ring, the electron bunch length should reach nm level, which is not at all a trivial task if one keeps in mind that the typical bunch length in present electron storage rings is at mm level. This paper is about our efforts in accomplishing this challenging goal. 


The work in this paper is organized as follows. {In Sec.~\ref{sec:formalism}, to build the foundation for the following analysis, we first introduce the generalized Courant-Snyder formalism which applies to a 3D general coupled lattice and present its application in electron storage ring physics.} In Sec.~\ref{sec:TME}, based on the formalism we derive the theoretical minimum longitudinal emittance in an electron storage ring to provide basis for later investigation since SSMB is about obtaining short bunch length and small longitudinal emittance. Following this, in Sec.~\ref{sec:SSMB}, we conduct some key analysis of three specific SSMB scenarios along the thinking of realizing nm bunch length and high-average-power EUV radiation, i.e., longitudinal weak focusing (LWF), longitudinal strong focusing (LSF) and generalized longitudinal strong focusing (GLSF). A short summary of these three schemes is: a LWF SSMB ring can be used to generate bunches with a bunch length of a  couple of 10~nm, thus can be used to generate coherent visible and infrared radiation. If we want to push the bunch length to an even shorter value, the required phase slippage factor of the LWF ring will be too small from an engineering viewpoint. As a comparison, a LSF SSMB ring can create bunches with a bunch length of nm level, thus to generate coherent EUV radiation. However, the required modulation laser power is at gigawatt (GW) level, and makes the laser modulator, which typically consists of an optical enhancement cavity with incident laser and an undulator and is used to longitudinally focus the electron beam at the laser wavelength scale, can only work at a low duty cycle pulsed mode, thus limiting the average output EUV radiation power. At present, a GLSF SSMB ring is the most promising among these three to obtain nm bunch length with a mild modulation laser power, thus allowing high-average-power radiation output. The basic idea of GLSF is to exploit the ultrasmall vertical emittance in a planar ring and apply partial transverse-longitudinal emittance for bunch compression with a shallow energy modulation strength, thus a small modulation laser power. The backbone of such an GLSF ring is the transverse-longitudinal coupling (TLC) dynamics, which is analyzed in depth in this paper. Following this analysis, before going into the concrete examples,  we prove three theorems in Sec.~\ref{sec:TLCTheorems} about TLC-based bunch compression or harmonic generation schemes. After that, we then go into the details of various TLC schemes, with Sec.~\ref{sec:TLCE} devoted to energy modulation-based schemes and Sec.~\ref{sec:TLCA} dedicated to angular modulation-based schemes. We have derived the bunching factors and the required modulation laser powers for them. The conclusion from the analysis is that the energy modulation-based coupling is favored for our application in GLSF SSMB. Based on the investigations and other critical physical considerations, an example parameters set of a 1~kW-average-power EUV light source is finally presented in Sec.~\ref{sec:application}.  A short summary is given in Sec.~\ref{sec:summary}.  

\section{{Generalized Courant-Snyder Formalism}}\label{sec:formalism}

In this section, to provide the basis for the following discussions, we introduce a generalized Courant-Snyder formalism for storage ring physics study. Particle state vector ${\bf X}=\left(x\ x'\ y\ y'\ z\ \delta\right)^{T}$ is used throughout this paper, with its components meaning the horizontal position, horizontal angle, vertical position, vertical angle, longitudinal position, and $\delta=\Delta E/E_{0}$ the relative energy deviation of a particle with respect to the reference particle, respectively. $E_{0}$ is the energy of the reference particle. The superscript $^{T}$ means the transpose of a vector or matrix. 

\subsection{Generalized Beta Functions in a General Coupled  Lattice}
Following Chao's solution by linear matrix (SLIM) formalism~\cite{chao1979evaluation}, we introduce the definition of the generalized beta functions in a 3D general coupled storage ring lattice as
\begin{equation}
	\begin{aligned}
		\beta_{ij}^{k}&=2\text{Re}\left({\bf E}_{ki}{\bf E}_{kj}^{*}\right),\\
		k&=\pm I,II,III,\\
		i,j&=1,2,3,4,5,6, 
	\end{aligned}
\end{equation}
where $^{*}$ means complex conjugate, the sub or superscript $k$ is the eigenmode index, Re() means the real component of a complex number or matrix, ${\bf E}_{ki}$ is the $i$-th component of the vector  ${\bf E}_{k}$ which is the eigenvector of the $6\times6$ symplectic one-turn map ${\bf M}$ of storage ring with eigenvalues $e^{i2\pi\nu_{k}}$
\begin{equation}\label{eq:eigentuens}
	{\bf M}{\bf E}_{k}=e^{i2\pi\nu_{k}}{\bf E}_{k},
\end{equation}
satisfying the following normalization condition
\begin{equation}\label{eq:norm}
	{\bf E}_{k}^{\dagger}{\bf S}{\bf E}_{k}=\begin{cases}
		&i,\ k=I,II,III,\\
		&-i,\ k=-I,-II,-III,
	\end{cases}
\end{equation}
and 
\begin{equation}
	{\bf E}_{k}^{\dagger}{\bf S}{\bf E}_{j}=0\ \text{for}\ k\neq j,
\end{equation}
where $i$ here means the imaginary unit, $^{\dagger}$~means complex conjugate transpose, and
\begin{equation}
	{\bf S}=\left(
	\begin{matrix}
		0 & 1 & 0 & 0 & 0 & 0\\
		-1 & 0 & 0 & 0 & 0 & 0\\
		0 & 0 & 0 & 1 & 0 & 0\\
		0 & 0 & -1 & 0 & 0 & 0\\
		0 & 0 & 0 & 0 & 0 & 1\\
		0 & 0 & 0 & 0 & -1 & 0\\
	\end{matrix}
	\right).
\end{equation} 
This normalization condition can be preserved around the ring due to the symplecticity of the transfer matrix. 
Since the one-turn map is a real symplectic matrix, for a stable motion, we have
\begin{equation}
	\nu_{-k}=-\nu_{k},\ {\bf E}_{-k}={\bf E}_{k}^{*},
\end{equation}
where $\nu_{k}$ are the eigen tunes.

Similar to the real generalized beta function, here we define the imaginary generalized beta functions as
\begin{equation}
	\hat\beta_{ij}^{k}=2\text{Im}\left({\bf E}_{ki}{\bf E}_{kj}^{*}\right),
\end{equation}
where Im() means the imaginary component of a complex number or matrix. Note that $\hat\beta_{ij}^{k}$ as a value is actually real, just like $\beta_{ij}^{k}$. The name `real' and `imaginary' here reflects that their definitions take the real and imaginary part of ${\bf E}_{ki}{\bf E}_{kj}^{*}$, respectively. We use the symbol $\hat{}$ on the top of these functions to indicate `imaginary'. Further we can define the real and imaginary generalized $6\times6$ Twiss matrices of a storage ring lattice corresponding to each eigen mode as
\begin{equation}
	\left({\bf T}_{k}\right)_{ij}=\beta_{ij}^{k},\ \left(\hat{\bf T}_{k}\right)_{ij}=\hat\beta_{ij}^{k}.
\end{equation}
From its definition we have
\begin{equation}
	\begin{aligned}
		{\bf T}_{-k}&={\bf T}_{k},\ {\bf T}^{*}_{k}={\bf T}_{k},\ {\bf T}^{T}_{k}={\bf T}_{k},\\
		\hat{\bf T}_{-k}&=-\hat{\bf T}_{k},\ \hat{\bf T}^{*}_{k}=\hat{\bf T}_{k},\ \hat{\bf T}^{T}_{k}=-\hat{\bf T}_{k}.
	\end{aligned}
\end{equation}
So we know that ${\bf T}_{k}$ is a real symmetric matrix, while $\hat{\bf T}_{k}$ is a real anti-symmetric matrix. Further, we can prove that
\begin{equation}\label{eq:ImTkSum}
	\sum_{k=I,II,III}\hat{\bf T}_{k}=-{\bf S}.
\end{equation}
The generalized Twiss matrices at different places around the ring are related to each other according to
\begin{equation}\label{eq:TkTrans}
	\begin{aligned}
		{\bf T}_{k}(s_{2})&={\bf R}(s_{2},s_{1}){\bf T}_{k}(s_{1}){\bf R}^{T}(s_{2},s_{1}),\\ \hat{\bf T}_{k}(s_{2})&={\bf R}(s_{2},s_{1})\hat{\bf T}_{k}(s_{1}){\bf R}^{T}(s_{2},s_{1}).
	\end{aligned}
\end{equation}
where ${\bf R}(s_{2},s_{1})$ is the symplectic transfer matrix of state vector ${\bf X}$ from $s_{1}$ to $s_{2}$
\begin{equation}\label{eq:XTrans}
	{\bf X}(s_{2})={\bf R}(s_{2},s_{1}){\bf X}(s_{1}),
\end{equation}
and we have 
\begin{equation}
	{\bf T}_{k}(s+C_{0})={\bf T}_{k}(s),\ \hat{\bf T}_{k}(s+C_{0})=\hat{\bf T}_{k}(s)
\end{equation}
where $C_{0}$ is the circumference of the ring.

With the help of the generalized Twiss matrices and eigen tunes, the one-turn map ${\bf M}$ can be parametrized as
\begin{equation}
	\begin{aligned}
		{\bf M}&=e^{{\bf S}\left(\sum_{k=I,II,III}{\bf G}_{k}\Phi_{k}\right)}\\
		&=\left(\sum_{k=I,II,III}\left[{\bf T}_{k}\sin\Phi_{k}+{\bf \hat{T}}_{k}\cos\Phi_{k}\right]\right){\bf S},
	\end{aligned}
\end{equation} 
where 
\begin{equation}
	{\bf G}_{k}\equiv{\bf S}^{T}{\bf T}_{k}{\bf S},
\end{equation}
and $\Phi_{k}=2\pi\nu_{k}$ the phase advance of the corresponding mode in one turn.
For ${\bf M}^{n}$ we only need to replace the above $\Phi_{k}$ with $n\Phi_{k}$.
We can write the matrix terms of the one-turn map more explicitly as
\begin{equation}
	\begin{aligned}
		{\bf M}_{ij}
		&=(-1)^{j}\sum_{k{=I,II,III}}\left[\beta_{i(j-(-1)^{j})}^{k}\sin\Phi_{k}\right.\\
		&\left.\ \ \ \ +\hat\beta_{i(j-(-1)^{j})}^{k}\cos\Phi_{k}\right].
	\end{aligned}
\end{equation}
Applying $\beta_{ij}^{k}=\beta_{ji}^{k}$, $\hat\beta_{ij}^{k}=-\hat\beta_{ji}^{k}$, we have
\begin{equation}
	\begin{aligned}
		{\bf M}_{12}&=\sum_{k{=I,II,III}}\beta_{11}^{k}\sin\Phi_{k},\\
		{\bf M}_{21}&=-\sum_{k{=I,II,III}}\beta_{22}^{k}\sin\Phi_{k},\\
		{\bf M}_{34}&=\sum_{k{=I,II,III}}\beta_{33}^{k}\sin\Phi_{k},\\
		{\bf M}_{43}&=-\sum_{k{=I,II,III}}\beta_{44}^{k}\sin\Phi_{k},\\
		{\bf M}_{56}&=\sum_{k{=I,II,III}}\beta_{55}^{k}\sin\Phi_{k},\\
		{\bf M}_{65}&=-\sum_{k{=I,II,III}}\beta_{66}^{k}\sin\Phi_{k}.\\
	\end{aligned}
\end{equation}

Using the generalized Twiss matrices, the actions or generalized Courant-Snyder invariants of a particle are defined according to
\begin{equation}
	J_{k}\equiv\frac{{\bf X}^{T}{\bf G}_{k}{\bf X}}{2}.
\end{equation}
It is easy to prove that $J_{k}$ are invariants of a particle when it travels around the ring, from Eqs.~(\ref{eq:TkTrans}), (\ref{eq:XTrans}), and the symplecticity of the transfer matrix
\begin{equation}
	{\bf R}^{T}{\bf S}{\bf R}={\bf S}.
\end{equation}
For a beam with $N_{p}$ particles, the three beam invariants can then be defined according to
\begin{equation}\label{eq:eigenemittance}
	\epsilon_{k}\equiv\langle J_{k}\rangle=\frac{\sum_{i=1}^{N_{p}}J_{k,i}}{N_{p}},\ k=I,II,III,
\end{equation}
where $J_{k,i}$ means the $k$-th mode invariant of the $i$-th particle. These beam invariants can be viewed as the eigen emittances in the ring. These invariants are the generalized root-mean-square (RMS) emittances of beam in the ring.


The beam emittances defined above are based on the eigenmode motion of particles in the storage ring. Another definition of emittance is based directly on the second moments matrix of a particle beam
\begin{equation}
	{\bf \Sigma}=\langle{\bf X}{\bf X}^{T}\rangle,
\end{equation}
where $\langle\rangle$ here means particle ensemble average. The beam second moment matrix at different places are related according to
\begin{equation}
	{\bf \Sigma}(s_{2})={\bf R}(s_{2},s_{1}){\bf \Sigma}(s_{1}){\bf R}^{T}(s_{2},s_{1}).
\end{equation}
From the symplecticity of ${\bf R}$, we can prove that the eigenvalues of ${\bf \Sigma}(-i{\bf S})$ are unchanged with the beam transport in a linear symplectic lattice. The beam eigen emittances can thus be defined as the positive eigen values of ${\bf \Sigma}(-i{\bf S})$.

When the particle beam matches the storage ring lattice, which means the beam distribution at a given location repeats itself turn by turn, we have
\begin{equation}
	{\bf \Sigma}(s+C_{0})={\bf M}(s){\bf \Sigma}(s){\bf M}^{T}(s)={\bf \Sigma}(s).
\end{equation}
It can be proven that for a matched beam the RMS emittances defined in Eq.~(\ref{eq:eigenemittance}) are the eigenvalues of ${\bf \Sigma}(-i{\bf S})$ with the eigenvector ${\bf E}_{k}$, i.e.,
\begin{equation}\label{eq:eigenemittance2}
	{\bf \Sigma}(-i{\bf S}){\bf E}_{k}=\text{sgn}(k)\epsilon_{k}{\bf E}_{k},
\end{equation}
where we have used $\epsilon_{-k}\equiv\epsilon_{k}$ and
\begin{equation}\label{eq:sgnk}
	\text{sgn}(k)=\begin{cases}
		&1,\ k=I,II,III,\\
		&-1,\ k=-I,-II,-III.
	\end{cases}
\end{equation}
Using the generalized Twiss and second moments matrices, the eigen emittances for a matched beam can be calculated as
\begin{equation}\label{eq:emtanceTr}
	\epsilon_{k}=\text{Tr}\left({\bf T}_{k}{\bf S}{\bf \Sigma}{\bf S}^{T}\right)=\text{Tr}\left({\bf G}_{k}{\bf \Sigma}\right).
\end{equation}
where  Tr() means the trace of a matrix.

To make sure that the eigenvectors ${\bf E}_{k}$ are uniquely defined all around the ring once they are determined at a given location, we will let
\begin{equation}\label{eq:EVPhase}
	{\bf E}_{k}(s_{2})=e^{-i\frac{s_{2}-s_{1}}{C_{0}}\Phi_{k}}{\bf R}(s_{2},s_{1}){\bf E}_{k}(s_{1}).
\end{equation}
Following this definition, we have
\begin{equation}
	{\bf E}_{k}(s+C_{0})=e^{-i\Phi_{k}}{\bf M}(s){\bf E}_{k}(s)={\bf E}_{k}(s).
\end{equation}
Using the generalized beta function, we can write the eigenvector component in an amplitude-phase form
\begin{equation}
	{\bf E}_{kj}=\sqrt{\frac{\beta_{jj}^{k}}{2}}e^{i\phi_{j}^{k}}.
\end{equation}
And according to definition we have
\begin{equation}
	\beta_{ij}^{k}=\sqrt{\beta_{ii}^{k}\beta_{jj}^{k}}\cos(\phi_{i}^{k}-\phi_{j}^{k}).
\end{equation}
Using the generalized Courant-Snyder invariants and beta functions, we can express the phase space coordinate of a particle at $s$ as
\begin{equation}
	{\bf X}_{i}(s)=\sum_{k=I,II,III}\sqrt{2J_{k}\beta_{ii}^{k}(s)}\cos\left[\psi_{i}^{k}(s)\right],
\end{equation} 
with $\psi_{i}^{k}$ determined by the initial condition of particle state. The phase term $\psi_{i}^{k}$ at different locations are related according to
\begin{equation}
	\psi_{i}^{k}(s_{2})=\psi_{i}^{k}(s_{1})+\frac{s_{2}-s_{1}}{C_{0}}\Phi_{k}+\phi_{i}^{k}(s_{2})-\phi_{i}^{k}(s_{1}).
\end{equation}
In particular, after $n$ revolutions in the ring, we have
\begin{equation}
	{\bf X}_{i}(s+nC_{0})=\sum_{k=I,II,III}\sqrt{2J_{k}\beta_{ii}^{k}(s)}\cos\left[\psi_{i}^{k}(s)+n\Phi_{k}\right].
\end{equation} 


\subsection{Perturbations}
After considering the parametrization of a general coupled lattice and the prescribed particle motion in it, let us now add perturbations, that from the lattice and also that from the beam. Assume there is a perturbation ${\bf K}$ to the one-turn map ${\bf M}$, i.e.,
\begin{equation}
	{\bf M}_{\text{per}}=({\bf I}+{\bf K}){\bf M}_{\text{unp}},
\end{equation}
where ${\bf I}$ is the identity matrix, the subscripts `per' and `unp' mean `perturbed' and `unperturbed', respectively. When the perturbation is small, from cannonical perturbation theory,  the tune shift of the $k$-th eigen mode can be calculated as
\begin{equation}
	\Delta\nu_{k}=-\frac{1}{4\pi}\text{Tr}\left[\left(\text{sgn}(k){\bf T}_{k}+i\hat{\bf T}_{k}\right){\bf S}{\bf K}\right].
\end{equation}
This tune shift formula can be used to calculate the real and imaginary tune shifts due to symplectic (for example lattice error) and non-symplectic (for example radiation damping) perturbations.  
For example, given the radiation damping matrix ${\bf D}$ around the ring,  the damping rate of each eigen mode per turn is
\begin{equation}\label{eq:dampC}
	\alpha_{k}=-\frac{1}{2}\oint\text{Tr}\left(\hat{\bf T}_{k}{\bf S}{\bf D}\right)ds, k=I,II,III,
\end{equation}
where $\oint$ means integration around the ring.
Note that the damping rates here are that for the corresponding eigenvectors. The damping rates for particle action or beam eigen emittance is a factor of two larger, since they are quadratic with respect to the phase space coordinates. 

Apart from radiation damping, there are also various beam diffusion effects in the ring, like quantum excitation and intra-beam scattering. Using Eq.~(\ref{eq:emtanceTr}), once we know the diffusion and damping matrix ${\bf N}$ around the ring, the emittance growth per turn due to diffusion is 
\begin{equation}
	\begin{aligned}
		\Delta \epsilon_{k}&=\frac{1}{2}\oint\text{Tr}\left({\bf G}_{k}{\bf N}\right)ds=-\frac{1}{2}\oint\text{Tr}\left({\bf T}_{k}{\bf S}{\bf N}{\bf S}\right)ds,\\
	\end{aligned}
\end{equation}
The equilibrium eigen emittance between a balance of diffusion and damping can be calculated as
\begin{equation}
	\begin{aligned}
		\epsilon_{k}
		&=\frac{\Delta\epsilon_{k}}{2\alpha_{k}}
		=\frac{\frac{1}{2}\oint\text{Tr}\left({\bf G}_{k}{\bf N}\right)ds}{-\oint\text{Tr}\left(\hat{\bf T}_{k}{\bf S}{\bf D}\right)ds}\\
		&=\frac{-\frac{1}{2}\sum_{i,j}\oint\beta^{k}_{ij}\left({\bf S}{\bf N}{\bf S}\right)_{ij}ds}{\sum_{i,j}\oint\hat{\beta}^{k}_{ij}\left({\bf S}{\bf D}\right)_{ij} ds}.
	\end{aligned}
\end{equation} 
It can be proven that the equilibrium beam distribution in 6D phase space given by such a balance in a linear lattice is Gaussian, with $\langle{\bf X}\rangle={\bf 0}$.
After getting the equilibrium eigen emittances, the second moments of beam can be written as 
\begin{equation}\label{eq:2ndMoments}
	\Sigma_{ij}=\langle {\bf X}_{i}{\bf X}_{j}\rangle=\sum_{k=I,II,III}\epsilon_{k}\beta_{ij}^{k},
\end{equation}
or in matrix form as
\begin{equation}
	{\bf \Sigma}=\sum_{k=I,II,III}\epsilon_{k}{\bf T}_{k}.
\end{equation}

\subsection{Application to Electron Storage Rings}
In an electron storage ring, the intrinsic diffusion and damping are both from the emission of photons, namely the so-called quantum excitation and radiation damping.
For quantum excitation, we have 
all the other components of the diffusion matrix ${\bf N}$ zero except that
\begin{equation}
	\begin{aligned}
		N_{66}&=\frac{\left\langle\mathcal{\dot{N}}\frac{u^2}{E_{0}^{2}}\right\rangle}{c}=\frac{2C_{L}\gamma^{5}}{c|\rho|^{3}},\\
	\end{aligned}
\end{equation}
where $\mathcal{\dot{N}}$ is the number of photons emitted per unit time, $u$ is the photon energy, $E_{0}$ is the particle energy, $C_{L}=\frac{55}{48\sqrt{3}}\frac{r_{e}\hbar}{m_{e}}$ with $r_{e}$ the classical electron radius, $\hbar$ the reduced Planck's constant, $m_{e}$ the electron mass, $\gamma$ is the Lorentz factor, $c$ is the speed of light in free space,  $\rho$ is the bending radius. We take the convention that the sign of $\rho$ is positive when the bending is inward, and negative when the bending is outward. $\langle\mathcal{\dot{N}}u^2\rangle$ in the above formula is a result of Campbell's theorem~\cite{Campbell1909,Sands1955}. 

For damping effect, we have two sources, i.e., dipole magnets and RF cavity. 
For a horizontal dipole, we have all the other components of damping matrix ${\bf D}$ zero except that 
\begin{equation}\label{eq:DD}
	\begin{aligned}
		D_{66}&=-\frac{C_{\gamma}E_{0}^{3}}{\pi}\frac{1}{\rho^{2}},\ D_{61}=-\frac{C_{\gamma}E^{3}_{0}}{2\pi}\frac{1-2n}{\rho^{3}},
	\end{aligned}
\end{equation}
where  $C_{\gamma}=\frac{4\pi}{3}\frac{r_{e}}{\left(m_{e}c^2\right)^3}=8.85\times10^{-5}\frac{\text{m}}{\text{GeV}^{2}}$, $n=-\frac{\rho}{B_{y}}\frac{\partial{B_{y}}}{\partial{x}}$ is the transverse field gradient index. The physical origin of $D_{66}$ is the fact that a higher energy particle tends to radiate more photon energy in a given magnetic field, while $D_{61}$ is due to the fact that a transverse displacement of particle will affect its path length in the dipole and when there is transverse field gradient also the magnetic field strength observed, thus the radiation energy loss.
For an RF cavity,  we have all the other damping matrix terms of ${\bf D}$ zero except that 
\begin{equation}\label{eq:RFD}
	D_{22}=D_{44}=-\frac{eV_{RF}\sin\phi_{RF}}{E_{0}}\delta(s_\text{RF}),
\end{equation}
where $e$ is the elementary charge and is positive in this paper, $V_{RF}$ and $\phi_{RF}$ are the RF voltage and phase, respectively, $\delta(s)$ means Dirac's delta function. Here we have assumed that the RF cavity is a zero-length one.  The physical origin of these damping terms is that the momentum boost of a particle in the RF cavity is along the longitudinal direction, while the transverse momentums of the particle are unchanged. Therefore, there is a damping effect on the horizontal and vertical angle of the particle. We remind the readers that if we use the horizontal and vertical particle momentum $p_{x}$ and $p_{y}$, instead of $x'$ and $y'$, as the phase space coordinates, then the damping happens only at bending magnets since the RF acceleration does not affect $p_{x,y}$.

In an electron storage ring, the RF acceleration compensates the radiation energy loss of electrons. If there is $N$ cavities in the ring, we have
\begin{equation}\label{eq:RFCom}
	\sum_{i=1}^{N} eV_{RF,i}\sin\phi_{RF,i}=U_{0},
\end{equation}
where 
\begin{equation}
	U_{0}=\frac{C_{\gamma}E_{0}^{4}}{2\pi}I_{2}
\end{equation}
is the radiation energy loss per particle per turn, with
\begin{equation}
	I_{2}=\oint\frac{1}{\rho^{2}}ds.
\end{equation}
If the ring consists of iso-bending magnets, then $U_{0}=C_{\gamma}\frac{E^{4}_{0}}{\rho}$.  From Eq.~(\ref{eq:DD}), we have 
\begin{equation}
	\oint D_{66}(s)ds=-\frac{2U_{0}}{E_{0}}.
\end{equation}
Similarly, based on Eqs.~(\ref{eq:RFD}) and (\ref{eq:RFCom}), we have
\begin{equation}
	\oint D_{22}(s)ds=\oint D_{44}(s)ds=-\frac{U_{0}}{E_{0}}.
\end{equation}
Combing with Eqs.~(\ref{eq:ImTkSum}) and (\ref{eq:dampC}), it is easy to show that for radiation damping, we have
\begin{equation}
	\begin{aligned}
		\sum_{k=I,II,III}\alpha_{k}&=-\frac{1}{2}\oint\text{Tr}\left[\left(\sum_{k=I,II,III}\hat{\bf T}_{k}\right){\bf S}{\bf D}\right]ds\\
		&=-\frac{1}{2}\oint\text{Tr}({\bf D})ds\\
		&=-\frac{1}{2}\oint \left(D_{22}+D_{44}+D_{66}\right)ds\\
		&=\frac{2U_{0}}{E_{0}},
	\end{aligned}
\end{equation}
which is the well-known Robinson's sum rule~\cite{robinson1958radiation}. 

The above formulation applies for a 3D general coupled lattice. For a ring without $x$-$y$ coupling and when the RF cavity is placed at dispersion-free location, we can express the normalized eigenvectors using the classical Courant-Snyder functions~\cite{courant1958theory} $\alpha$, $\beta$, $\gamma$ and dispersion $D$ and dispersion angle $D'$ as
{}
	\begin{equation}\label{eq:eigenvector}
		\begin{aligned}
			{\bf E}_{I}&=\frac{1}{\sqrt{2}}\left(\begin{matrix}
				\sqrt{\beta_{x}}\\
				\frac{-\alpha_{x}+i}{\sqrt{\beta_{x}}}\\
				0\\
				0\\
				\frac{-(\alpha_{x}D_{x}+\beta_{x}D_{x}')+iD_{x}}{\sqrt{\beta_{x}}}\\
				0\\
			\end{matrix}\right)e^{i\Psi_{I}},\
			{\bf E}_{II}=\frac{1}{\sqrt{2}}\left(\begin{matrix}
				0\\
				0\\
				\sqrt{\beta_{y}}\\
				\frac{-\alpha_{y}+i}{\sqrt{\beta_{y}}}\\	
				\frac{-(\alpha_{y}D_{y}+\beta_{y}D_{y}')+iD_{y}}{\sqrt{\beta_{y}}}\\
				0\\
			\end{matrix}\right)e^{i\Psi_{II}},\\
			{\bf E}_{III}&=\frac{1}{\sqrt{2}}\left(\begin{matrix}
				\frac{-\alpha_{z}+i}{\sqrt{\beta_{z}}}D_{x}\\
				\frac{-\alpha_{z}+i}{\sqrt{\beta_{z}}}D_{x}'\\
				\frac{-\alpha_{z}+i}{\sqrt{\beta_{z}}}D_{y}\\
				\frac{-\alpha_{z}+i}{\sqrt{\beta_{z}}}D_{y}'\\
				\sqrt{\beta_{z}}\\
				\frac{-\alpha_{z}+i}{\sqrt{\beta_{z}}}\\
			\end{matrix}\right)e^{i\Psi_{III}},
		\end{aligned}
	\end{equation}
	where the subscripts $x,y,z$ correspond to the horizontal, vertical and longitudinal dimensions, respectively, and $\Psi_{I,II,III}$ are the phase factors  of the eigenvectors. There is flexibility in choosing these phase factors as they does not affect the calculation of our defined Twiss matrices and physical quantities. But note that once they are set at one location, then their values all around the ring are determined according to Eq.~(\ref{eq:EVPhase}).  In this case, the real and imaginary generalized Twiss matrices can be obtained explicitly
	\begin{equation}\label{eq:Twiss}
		\begin{aligned}
			{\bf T}_{I}&=\left(
			\begin{array}{cccccc}
				\beta _x & -\alpha _x & 0 & 0 & - \alpha _x D_x-\beta _x D_x' & 0 \\
				-\alpha _x & \gamma_{x} & 0 & 0 & \gamma_{x}D_x+\alpha _x D_x' & 0 \\
				0 & 0 & 0 & 0 & 0 & 0 \\
				0 & 0 & 0 & 0 & 0 & 0 \\
				-\alpha _x D_x -\beta _x D_x' & \gamma_{x}D_x+\alpha _x D_x' &  0 & 0 & \mathcal{H}_{x} & 0 \\
				0 & 0 & 0 & 0 & 0 & 0 \\
			\end{array}
			\right),\\
			 \hat{\bf T}_{I}&=\left(
			\begin{array}{cccccc}
				0 & -1 & 0 & 0 & -D_x & 0 \\
				1 & 0 & 0 & 0 & -D_x' & 0 \\
				0 & 0 & 0 & 0 & 0 & 0 \\
				0 & 0 & 0 & 0 & 0 & 0 \\
				D_x & D_x' & 0 & 0 & 0 & 0 \\
				0 & 0 & 0 & 0 & 0 & 0 \\
			\end{array}
			\right),\\
			{\bf T}_{II}&=\left(
			\begin{array}{cccccc}
				0 & 0 & 0 & 0 & 0 & 0 \\
				0 & 0 & 0 & 0 & 0 & 0 \\
				0 & 0 & \beta _y & -\alpha _y & - \alpha _y D_y-\beta _y D_y' & 0 \\
				0 & 0 & -\alpha _y & \gamma_{y} &  \gamma_{y}D_y+\alpha _y D_y' & 0 \\
				0 & 0 & -\alpha _y D_y -\beta _y D_y' & \gamma_{y}D_y+\alpha _y D_y' &   \mathcal{H}_{y} & 0 \\
				0 & 0 & 0 & 0 & 0 & 0 \\
			\end{array}
			\right),\\
			\hat{\bf T}_{II}&=\left(
			\begin{array}{cccccc}
				0 & 0 & 0 & 0 & 0 & 0 \\
				0 & 0 & 0 & 0 & 0 & 0 \\
				0 & 0 & 0 & -1 & -D_y & 0 \\
				0 & 0 & 1 & 0 & -D_y' & 0 \\
				0 & 0 & D_y & D_y' & 0 & 0 \\
				0 & 0 & 0 & 0 & 0 & 0 \\
			\end{array}
			\right),\\
			{\bf T}_{III}&=\left(
			\begin{array}{cccccc}
				\gamma_{z} D_x^2 & \gamma _z D_{x}D_{x}' & \gamma_{z}D_{x}D_{y} & \gamma_{z}D_{x}D_{y}' & -\alpha _z D_x  & \gamma _z D_{x} \\
				\gamma _z D_{x}D_{x}' & \gamma _zD_{x}'^2 & \gamma_{z}D_{x}'D_{y} & \gamma_{z}D_{x}'D_{y}' & -\alpha _z D_x' & \gamma _z D_{x}' \\
				\gamma_{z}D_{x}D_{y} & \gamma_{z}D_{x}'D_{y} &  \gamma_{z} D_y^2 & \gamma _z D_{y}D_{y}' & -\alpha _z D_y  & \gamma _z D_{y} \\
				\gamma_{z}D_{x}D_{y}' & \gamma_{z}D_{x}'D_{y}' & \gamma _z D_{y}D_{y}' & \gamma _zD_{y}'^2 &  -\alpha _z D_y' & \gamma _z D_{y}' \\
				- \alpha _z D_x & -\alpha _z D_x' & - \alpha _z D_y & -\alpha _z D_y' & \beta _z & -\alpha _z \\
				\gamma _z D_{x}  & \gamma _z D_{x}' & \gamma _z D_{y}  & \gamma _z D_{y}' & -\alpha _z & \gamma_{z} \\
			\end{array}
			\right),\\ \hat{\bf T}_{III}&=\left(
			\begin{array}{cccccc}
				0 & 0 & 0 & 0 & D_x & 0 \\
				0 & 0 & 0 & 0 & D_x' & 0 \\
				0 & 0 & 0 & 0 & D_y & 0 \\
				0 & 0 & 0 & 0 & D_y' & 0 \\
				-D_x & -D_x' & -D_y & -D_y' & 0 & -1 \\
				0 & 0 & 0 & 0 & 1 & 0 \\
			\end{array}
			\right).
		\end{aligned}
	\end{equation}
	Correspondingly, and generalized Courant-Snyder invariants are given by 
	\begin{equation}\label{eq:invarinats}
		\begin{aligned}
			J_{I}&\equiv\frac{{\bf (SX)}^{T}{\bf T}_{I}{\bf SX}}{2}=\frac{\left(x-D_{x}\delta\right)^{2}+\left[\alpha_{x}\left(x-D_{x}\delta\right)+\beta_{x}\left(x'-D_{x}'\delta\right)\right]^{2}}{2\beta_{x}},\\
			J_{II}&\equiv\frac{{\bf (SX)}^{T}{\bf T}_{II}{\bf SX}}{2}=\frac{\left(y-D_{y}\delta\right)^{2}+\left[\alpha_{y}\left(y-D_{y}\delta\right)+\beta_{y}\left(y'-D_{y}'\delta\right)\right]^{2}}{2\beta_{y}},\\
			J_{III}&\equiv\frac{{\bf (SX)}^{T}{\bf T}_{III}{\bf SX}}{2}\\
			&=\frac{\left(z+D_{x}'x-D_{x}x'+D_{y}'y-D_{y}y'\right)^{2}+\left[\alpha_{z}\left(z+D_{x}'x-D_{x}x'+D_{y}'y-D_{y}y'\right)+\beta_{z}\delta\right]^{2}}{2\beta_{z}},\\
		\end{aligned}
	\end{equation}
{}
The equilibrium emittance determine by the balance of quantum excitation and radiation damping in an electron storage in this case reduce to the classical Sands  radiation integrals formalism found in textbooks~\cite{sands1970physics}
\begin{equation}\label{eq:Sands}
	\begin{aligned}
		\epsilon_{x}&\equiv\langle J_{I}\rangle=\frac{C_{L}\gamma^{5}}{2c\alpha_{I}}\oint \frac{\beta_{55}^{I}}{|\rho|^{3}}ds=C_{q}\frac{\gamma^{2}}{J_{x}}\frac{I_{5x}}{I_{2}},\\ 
		\epsilon_{y}&\equiv\langle J_{II}\rangle=\frac{C_{L}\gamma^{5}}{2c\alpha_{II}}\oint \frac{\beta_{55}^{II}}{|\rho|^{3}}ds=C_{q}\frac{\gamma^{2}}{J_{y}}\frac{I_{5y}}{I_{2}},\\ 
		\epsilon_{z}&\equiv\langle J_{III}\rangle=\frac{C_{L}\gamma^{5}}{2c\alpha_{III}}\oint \frac{\beta_{55}^{III}}{|\rho|^{3}}ds=C_{q}\frac{\gamma^{2}}{J_{z}}\frac{I_{5z}}{I_{2}},\\ 
	\end{aligned}
\end{equation}
with $C_{q}=\frac{55\lambdabar_{e}}{32\sqrt{3}}=3.8319\times10^{-13}$~m for electrons, ${\lambdabar}_{e}=\frac{\lambda_{e}}{2\pi}=\frac{\hbar c}{m_{e}c^{2}}=386$ fm is the reduced Compton wavelength of electron,  and the radiation integrals given by 
\begin{equation}
	\begin{aligned}
		I_{2}&=\oint\frac{1}{\rho^{2}}ds,\
		I_{4x}=\oint D_{x}\left(\frac{1-2n}{\rho^{3}}\right)ds,\\
		I_{5x}&=\oint \frac{\mathcal{H}_{x}}{|\rho|^{3}}ds,\
		I_{5y}=\oint \frac{\mathcal{H}_{y}}{|\rho|^{3}}ds,\
		I_{5z}=\oint \frac{\beta_{z}}{|\rho|^{3}}ds,
	\end{aligned}
\end{equation}
where
\begin{equation}
	\mathcal{H}_{x,y}=\beta_{55}^{I,II}=\frac{D_{x,y}^{2}+\left(\alpha_{x,y} D_{x,y}+\beta_{x,y} D_{x,y}'\right)^2}{\beta_{x,y}}
\end{equation}
are the horizontal and vertical dispersion invariant, and $\beta_{z}=\beta_{55}^{III}$ is the longitudinal beta function~\cite{deng2021courant}. The damping rate of three eigen modes are given by
\begin{equation}\label{eq:dampingrates}
	\alpha_{I}=\frac{U_{0}}{2E_{0}}J_{x},\ \alpha_{II}=\frac{U_{0}}{2E_{0}}J_{y},\ \alpha_{III}=\frac{U_{0}}{2E_{0}}J_{z},\\
\end{equation} 
where $J_{x,y,z}$ are the damping partition numbers 
\begin{equation}
	J_{x}=1-\frac{I_{4x}}{I_{2}},\ J_{y}=1,\ J_{z}=2+\frac{I_{4x}}{I_{2}}.
\end{equation} 
The horizontal, vertical, longitudinal radiation damping times are 
\begin{equation}
	\tau_{x,y,z}=\frac{T_{0}}{\alpha_{x,y,z}},
\end{equation}
where $T_{0}$ the particle revolution period time in the ring. 

After getting the equilibrium emittance, the beam second moments can be obtained using Eq.~(\ref{eq:2ndMoments}). For example, 
\begin{equation}
	\begin{aligned}
		\sigma_{x}^{2}&=\langle x^{2}\rangle=\epsilon_{x}\beta_{x}+\epsilon_{z}\gamma_{z}D_{x}^{2}=\epsilon_{x}\beta_{x}+\sigma_{\delta}^{2}D_{x}^{2},\\
		\sigma_{z}^{2}&=\langle z^{2}\rangle=\epsilon_{z}\beta_{z}+\epsilon_{x}\mathcal{H}_{x}+\epsilon_{y}\mathcal{H}_{y}.
	\end{aligned}
\end{equation}
We remind the readers that the generalized Courant-Snyder formalism presented in this section has been briefly reported before in Ref.~\cite{Deng2023FLSOSC}.

\section{Theoretical Minimum Emittances}\label{sec:TME}
After introducing the generalized Courant-Snyder formalism, we now apply it to analyze the theoretical minimum longitudinal emittance in an electron storage ring. This work serves as the basis for the following investigation of SSMB, since the longitudinal weak focusing and strong focusing SSMB to be introduced soon is about lowering the equilibrium bunch length and longitudinal emittance in an electron storage ring.  For completeness, here in this section we also present the analysis of theoretical minimum transverse emittance since they can be treated within a single framework.


\subsection{Theoretical Minimum Horizontal Emittance}
From Eq.~(\ref{eq:Sands}) we can see that $\mathcal{H}_{x}$ and $\beta_{z}$, i.e., $\beta_{55}^{I}$ and $\beta_{55}^{III}$ as defined by us, at the bending magnets are of vital importance in determining the horizontal and longitudinal emittance. Therefore, we need to know how they evolve inside a bending magnet.
The transfer matrix of particle state vector for a sector bending magnet is
{}
	\begin{equation}\label{eq:sec}
		\begin{aligned}
			&{\bf B}(\alpha)=
			\left(
			\begin{array}{cccccc}
				\cos \alpha & \rho  \sin \alpha & 0 & 0 & 0 & \rho (1-\cos \alpha)   \\
				-\frac{\sin \alpha }{\rho} & \cos \alpha & 0 & 0 & 0 & \sin \alpha \\
				0 & 0 & 1 & \rho\alpha & 0 & 0\\
				0 & 0 & 0 & 1 & 0 & 0\\
				-\sin \alpha & -\rho (1-\cos \alpha) & 0 & 0  & 1 & \rho  \left(\frac{\alpha }{\gamma ^2}-\alpha +\sin \alpha\right) \\
				0 & 0 & 0 & 0  & 0 & 1 \\
			\end{array}
			\right),
		\end{aligned}
	\end{equation}
{}
with $\rho$ and $\alpha$ the bending radius and angle of the bending magnet.
In a planar uncoupled ring, the normalized eigenvectors of the one-turn map are given by Eq.~(\ref{eq:eigenvector}). 
Assuming the Courant-Snyder functions, dispersion and dispersion angle at the dipole center where we set to be $\alpha=0$ are given by $\alpha_{x0},\beta_{x0},\alpha_{z0},\beta_{z0},D_{x0},D_{x0}'$.
From Eqs.~(\ref{eq:sec}) and (\ref{eq:eigenvector}) we have the evolution of $\mathcal{H}_{x}$ {expression} in the dipole 
{}
	\begin{equation}\label{eq:Hx}
		\begin{aligned}
			&\mathcal{H}_{x}(\alpha)\equiv\beta_{55}^{I}(\alpha)=2|{\bf E}_{I5}(\alpha)|^{2}=2|\left({\bf B}(\alpha){\bf E}_{I}(0)\right)_{5}|^{2}\\
			&=\left(\sqrt{\beta_{x0}}\left(\sin \alpha+D_{x0}'\right)+\frac{\alpha_{x0}}{\sqrt{\beta_{x0}}}\left[D_{x0}-\rho (1-\cos \alpha)\right]\right)^{2}+\left(\frac{D_{x0}-\rho (1-\cos \alpha)}{\sqrt{\beta_{x0}}}\right)^{2}.
		\end{aligned}
	\end{equation}
	Similarly, the evolution of $\beta_{z}$ in the dipole is given by
	\begin{equation}\label{eq:betaZ}
		\begin{aligned}
			\beta_{z}(\alpha)&\equiv\beta_{55}^{III}(\alpha)=2|{\bf E}_{III5}(\alpha)|^{2}=2|\left({\bf B}(\alpha){\bf E}_{III}(0)\right)_{5}|^{2}\\
			&=\left(\sin \alpha\frac{\alpha_{z0}}{\sqrt{\beta_{z0}}}D_{x0}+\rho (1-\cos \alpha)\frac{\alpha_{z0}}{\sqrt{\beta_{z0}}}D_{x0}'+\sqrt{\beta_{z0}}-\rho  \left(-\alpha +\sin \alpha\right)\frac{\alpha_{z0}}{\sqrt{\beta_{z0}}}\right)^{2}\\
			&\ \ \ \ +\left(-\sin \alpha\frac{1}{\sqrt{\beta_{z0}}}D_{x0}-\rho (1-\cos \alpha)\frac{1}{\sqrt{\beta_{z0}}}D_{x0}'+\rho  \left(-\alpha +\sin \alpha\right)\frac{1}{\sqrt{\beta_{z0}}}\right)^{2}.\\
		\end{aligned}
	\end{equation}
{}
For simplicity we have neglected the contribution of $\frac{\rho\alpha}{\gamma^{2}}$ to $R_{56}$ of dipole in the above calculation of $\beta_{z}$, since we are interested in the relativistic cases. But we remind the readers that in a quasi-isochronous ring, the contribution of $\frac{C_{0}}{\gamma^{2}}$ to the ring $R_{56}$ or phase slippage may not be negligible.

With the evolution of $\mathcal{H}_{x}$ and $\beta_{z}$ known, now we derive the theoretical minimum emittances. For simplicity, we assume the ring consists of iso-bending-magnets,  with the bending angle induced by each bending magnet $\theta$,  and the optical functions are identical in each bending magnets. 
Then from Eq.~(\ref{eq:Sands}), we have
\begin{equation}
	\begin{aligned}
		\epsilon_{x}
		&=C_{q}\frac{\gamma^{2}}{J_{x}}\frac{1}{\rho}f_{x}(\alpha_{x0},\beta_{x0},D_{x0},D_{x0}'), 
	\end{aligned}
\end{equation}
with
\begin{equation}\label{eq:fx}
	\begin{aligned}
		&f_{x}(\alpha_{x0},\beta_{x0},D_{x0},D_{x0}')=\frac{1}{\theta}\int_{-\frac{\theta}{2}}^{\frac{\theta}{2}}\mathcal{H}_{x}(\alpha)d\alpha.
	\end{aligned}
\end{equation}
$f_{x}$ can be interpreted as the average $\mathcal{H}_{x}$ in dipoles.
The lengthy explicit expression of $f_{x}(\alpha_{x0},\beta_{x0},D_{x0},D_{x0}')$ is omitted here. It can be obtained straightforwardly by inserting Eq.~(\ref{eq:Hx}) in Eq.~(\ref{eq:fx}).  The mathematical problem we are trying to solve is then to minimize $f_{x}$, by adjusting $\alpha_{x0}$, $\beta_{x0}$, $D_{x0}$, $D_{x0}'$.
From Eq.~(\ref{eq:fx}) we have
{}
	\begin{equation}
		\begin{aligned}
			\frac{\partial f_{x}}{\partial\alpha_{x0}}&=\alpha_{x0} \frac{2 D_{x0} \left(D_{x0}  -2 \rho +2 \rho  \frac{\sin \frac{\theta }{2}}{\frac{\theta }{2}}\right)+\rho ^2 \left(3  -4 \frac{\sin \frac{\theta }{2}}{\frac{\theta}{2}}+\frac{\sin \theta }{\theta}\right)}{\beta_{x0}}+D_{x0}'2 \left(D_{x0}  - \rho + \rho  \frac{\sin \frac{\theta }{2}}{\frac{\theta }{2}}\right),\\
			\frac{\partial f_{x}}{\partial D_{x0}'}&= \alpha_{x0} 2 \left(D_{x0}  - \rho + \rho  \frac{\sin \frac{\theta }{2}}{\frac{\theta }{2}}\right)+ D_{x0}'2 \beta_{x0}.  \\
		\end{aligned}
	\end{equation}
{}
We notice that the requirement of $\frac{\partial f_{x}}{\partial\alpha_{x0}}=0$ and $\frac{\partial f_{x}}{\partial D_{x0}'}=0$ leads to $\alpha_{x0}=0$ and $D_{x0}'=0$. Under the above conditions, then 
\begin{equation}
	\begin{aligned}
		\frac{\partial f_{x}}{\partial D_{x0}}&=\frac{2 \left(D_{x0}  - \rho + \rho  \frac{\sin \frac{\theta }{2}}{\frac{\theta }{2}}\right)}{\beta_{x0}}.\\
	\end{aligned}
\end{equation}
The requirement of $\frac{\partial f_{x}}{\partial D_{x0}}=0$ leads to
\begin{equation}
	\begin{aligned}
		D_{x0}&=\rho\left(1-\frac{\sin \frac{\theta }{2}}{\frac{\theta}{2}} \right)\approx\frac{\rho\theta^2}{24}.
	\end{aligned}
\end{equation} 
Under the above conditions, then 
\begin{equation}
	\begin{aligned}
		\frac{\partial f_{x}}{\partial\beta_{x0}}&= \frac{\theta  (\theta -\sin \theta )-\frac{\rho ^2 \left(\theta ^2+\theta  \sin \theta +4 \cos \theta -4\right)}{\beta _{x0}^2}}{2 \theta ^2}.
	\end{aligned}
\end{equation}
The requirement of $\frac{\partial f_{x}}{\partial \beta_{x0}}=0$ leads to
\begin{equation}
	\begin{aligned}
		\beta_{x0}&=\rho\sqrt{\frac{\left(\theta ^2+\theta  \sin \theta +4 \cos \theta -4\right)}{\theta  (\theta -\sin \theta )}}\approx\frac{\rho\theta}{2\sqrt{15}}.
	\end{aligned}
\end{equation} 
Summarizing, the extreme value of $f_{x}$ is realized when
\begin{equation}
	\begin{aligned}
		\alpha_{x0}&=0,\ \beta_{x0}\approx\frac{\rho\theta}{2\sqrt{15}},\
		D_{x0}\approx\frac{\rho\theta^2}{24},\
		D_{x0}'=0,\ 
	\end{aligned}
\end{equation} 
which means
\begin{equation}
	\begin{aligned}
		\mathcal{H}_{x0}&=\rho\frac{\left(1-\frac{\sin \frac{\theta }{2}}{\frac{\theta}{2}} \right)^{2}}{\sqrt{\frac{\left(\theta ^2+\theta  \sin \theta +4 \cos \theta -4\right)}{\theta  (\theta -\sin \theta )}}}\approx\frac{5\rho\theta ^3 }{96 \sqrt{15}},
	\end{aligned}
\end{equation}
and 
\begin{equation}
	\begin{aligned}
		f_{x,\text{min}}&=\rho\left(1-\frac{\sin \theta }{\theta}\right)\sqrt{\frac{\left(\theta ^2+\theta  \sin \theta +4 \cos \theta -4\right)}{\theta  (\theta -\sin \theta )}}\\
		&\approx\frac{\rho\theta ^3 }{12 \sqrt{15}}.
	\end{aligned}
\end{equation}
One can check that this is the minimum value of $f_{x}$.
Note that $f_{x,\text{min}}=\frac{8}{5}\mathcal{H}_{x0}$. Under these conditions we get the minimum horizontal emittance
\begin{equation}\label{eq:TMEx}
	\epsilon_{x,\text{min}}=C_{q}\frac{\gamma^{2}}{J_{x}}\frac{\theta^{3}}{12\sqrt{15}}.
\end{equation}
The above results are consistent with the classical results of Teng~\cite{Teng1984}.   For practical use, and considering nominally $J_{x}\approx1$, the above scaling can be written as
\begin{equation}
	\epsilon_{x,\text{min}}[\text{nm}]=31.6E^{2}_{0}[\text{GeV}]\theta ^3[\text{rad}].
\end{equation}
For example, if $E_{0}=6$ GeV and $\theta=\frac{2\pi}{300}$ rad, we have $\epsilon_{x,\text{min}}=10.4$ pm.

\subsection{Theoretical Minimum Longitudinal Emittance}
Now we analyze the theoretical minimum longitudinal emittance. Similar to the horizontal direction,
we have
\begin{equation}
	\begin{aligned}
		\epsilon_{z}&=C_{q}\frac{\gamma^{2}}{J_{z}}\frac{1}{\rho}f_{z}(\alpha_{z0},\beta_{z0},D_{x0},D_{x0}'),
	\end{aligned}
\end{equation}
with
\begin{equation}\label{eq:fz}
	\begin{aligned}
		&f_{z}(\alpha_{z0},\beta_{x0},D_{x0},D_{x0}')=\frac{1}{\theta}\int_{-\frac{\theta}{2}}^{\frac{\theta}{2}}\beta_{z}(\alpha)d\alpha,
	\end{aligned}
\end{equation}
which can be interpreted as the average $\beta_{z}$ in the dipole.
Then
\begin{equation}
	\begin{aligned}
		\frac{\partial f_{z}}{\partial\alpha_{z0}}&= \alpha _{{z0}} {\mathcal{G}}(\rho,\theta,D_{x0},D_{x0}',\beta_{z0})+D_{x0}' 2 \rho \left(1 - \frac{\sin \frac{\theta }{2}}{\frac{\theta }{2}}\right),\\
		\frac{\partial f_{z}}{\partial D_{x0}'}&= \alpha _{{z0}}2 \rho \left(1 - \frac{\sin \frac{\theta }{2}}{\frac{\theta }{2}}\right)\\
		&+ D_{x0}' \frac{ \left(1+\alpha _{{z0}}^2\right)\rho ^2 \left(3  -4 \frac{\sin \frac{\theta }{2}}{\frac{\theta}{2}}+\frac{\sin \theta }{\theta}\right)}{\beta _{z0}},\\
	\end{aligned}
\end{equation}
where the lengthy explicit expression of ${\mathcal{G}}(\rho,\theta,D_{x0},D_{x0}',\beta_{z0})$ is omitted here. 
Similar to the analysis of transverse minimum emittance, we notice again that the requirement of $\frac{\partial f_{z}}{\partial\alpha_{z0}}=0$ and $\frac{\partial f_{z}}{\partial D_{x0}'}=0$ leads to $\alpha_{z0}=0$ and $D_{x0}'=0$. Under the above conditions, then 
{}
	\begin{equation}
		\begin{aligned}
			\frac{\partial f_{z}}{\partial D_{x0}}&=\frac{D_{x0} \left(1 -\frac{\sin \theta }{\theta}\right)-\rho  \left(1-\frac{\sin \theta }{\theta} -2\frac{ \sin \frac{\theta }{2}}{\frac{\theta}{2}}+2  \cos \frac{\theta }{2}\right)}{\beta _{z0}}.\\
		\end{aligned}
	\end{equation}
	The requirement of $\frac{\partial f_{z}}{\partial D_{x0}}=0$ leads to 
	\begin{equation}
		\begin{aligned}
			D_{x0}&=\frac{\rho  \left(1-\frac{\sin \theta }{\theta} -2\frac{ \sin \frac{\theta }{2}}{\frac{\theta}{2}}+2  \cos \frac{\theta }{2}\right)}{1 -\frac{\sin \theta }{\theta}}\approx-\frac{\rho \theta ^2 }{40}.\\
		\end{aligned}
	\end{equation} 
	Under the above conditions, then 
	\begin{equation}
		\begin{aligned}
			\frac{\partial f_{z}}{\partial\beta_{z0}}&=1-\frac{\rho ^2  \left(\theta ^4-12 \theta ^2-\left(\theta ^2-48\right) \theta  \sin \theta -12 \left(\theta ^2-4\right) \cos \theta -48\right)}{12 \theta  (\theta -\sin \theta ) \beta _{z0}^2}.
		\end{aligned}
	\end{equation}
	The requirement of $\frac{\partial f_{z}}{\partial\beta_{z0}}=0$ leads to
	\begin{equation}
		\begin{aligned}
			\beta_{z0}&=\rho\sqrt{\frac{ \left(\theta ^4-12 \theta ^2-\left(\theta ^2-48\right) \theta  \sin \theta -12 \left(\theta ^2-4\right) \cos \theta -48\right)}{12\theta  (\theta -\sin \theta )}}\approx\frac{\rho\theta^{3}}{120\sqrt{7}},
		\end{aligned}
	\end{equation} 
	Summarizing, the extreme value of $f_{z}$ is realized when
	\begin{equation}\label{eq:TMEzCOnditions}
		\begin{aligned}
			\alpha_{z0}&=0,\
			\beta_{z0}\approx\frac{\rho\theta^{3}}{120\sqrt{7}},\
			D_{x0}\approx-\frac{\rho \theta ^2 }{40} ,\
			D_{x0}'=0,\\ 
		\end{aligned}
	\end{equation} 
	and 
	\begin{equation}
		\begin{aligned}
			f_{z,\text{min}}&=\rho\sqrt{\frac{ \left(\theta ^4-12 \theta ^2-\left(\theta ^2-48\right) \theta  \sin \theta -12 \left(\theta ^2-4\right) \cos \theta -48\right)}{3\theta  (\theta -\sin \theta )}}\approx\frac{\rho\theta ^3 }{60 \sqrt{7}}.
		\end{aligned}
	\end{equation}
{}
One can check that this is the minimum value of $f_{z}$. Note that $f_{z,\text{min}}=2\beta_{z0}$.  Under these conditions we get the minimum longitudinal emittance
\begin{equation}\label{eq:TMEz0}
	\epsilon_{z,\text{min}}=C_{q}\frac{\gamma^{2}}{J_{z}}\frac{\theta^{3}}{60\sqrt{7}}.
\end{equation} 
The above result has also been given in Ref.~\cite{Zhang2021}.
For practical use, and considering nominally $J_{z}\approx2$, the above scaling can be written as
\begin{equation}\label{eq:TMEz}
	\epsilon_{z,\text{min}}[\text{nm}]=4.62E^{2}_{0}[\text{GeV}]\theta ^3[\text{rad}].
\end{equation}
For example, if $E_{0}=0.6$ GeV and $\theta=\frac{2\pi}{50}$ rad, we have $\epsilon_{z,\text{min}}=3.3$ pm.

Here we remind the readers that  in reality it may not {be} easy to reach the optimal conditions Eq.~(\ref{eq:TMEzCOnditions}) for all the dipoles in a ring. {This is based on the observation that by realizing Eq.~(\ref{eq:TMEzCOnditions}) at the dipole center, the dipole as a whole will have a nonzero $R_{56}$. So to make the longitudinal optics identical in different dipoles, there should be RF or laser modulator kick between neighboring dipoles, otherwise the required drift space will be too long to accumulate the required opposite $R_{56}$~\cite{Zhang2021}. It may not easy to apply too many radio-frequency (RF) cavities or laser modulators in a ring to manipulate the longitudinal optics, while in the transverse dimension it is straightforward to implement many quadrupoles to manipulate the transverse optics.} Instead, we may choose a more practical strategy to realize small longitudinal emittance, which is letting each half of the bending magnet be isochronous, and the longitudinal optics for each dipole can then be identical. This can be realized by requiring 
\begin{equation}\label{eq:TMEzCOnditionsDeng}
	\begin{aligned}
		\alpha_{z0}&=0,\
		\beta_{z0}\approx\frac{\rho\theta^{3}}{12\sqrt{210}},\
		D_{x0}\approx-\frac{\rho \theta ^2 }{24} ,\
		D_{x0}'=0.
	\end{aligned}
\end{equation} 
In this case, we still have $f_{z}=2\beta_{z0}$. Under such condition, the minimum longitudinal emittance is~\cite{deng2024theoretical}
\begin{equation}\label{eq:TMEzDeng0}
	\epsilon_{z,\text{min},\text{ISO}}=C_{q}\frac{\gamma^{2}}{J_{z}}\frac{\theta^{3}}{6\sqrt{210}}.
\end{equation}
The emittance given in Eq.~(\ref{eq:TMEzDeng0}) is larger than the real theoretical minimum Eq.~(\ref{eq:TMEz}), but offers a more practical reference.
For practical use, and considering nominally $J_{z}\approx2$,  the above scaling can be written as
\begin{equation}\label{eq:TMEzDeng}
	\epsilon_{z,\text{min},\text{ISO}}[\text{nm}]=8.44E^{2}_{0}[\text{GeV}]\theta ^3[\text{rad}].
\end{equation}

Based on this emittance and $\beta_{z0}$, we can know the bunch length at the dipole center contributed from the longitudinal emittance.  We remind the readers that the bunch length can in principle be even smaller than this value, by pushing $\beta_{z0}$ to an even smaller value, although the longitudinal emittance will actually grow then since $\beta_{z}$ will diverge faster when going away from the dipole center. If the ring works in a longitudinal weak focusing regime to be introduced in next section, there is actually a lower limit of bunch length in this process, which is a factor of $\sqrt{2}$ smaller than the bunch length given by conditions of Eqs.~(\ref{eq:TMEzDeng}) and (\ref{eq:TMEzCOnditionsDeng}). The energy spread will diverge when we push the bunch length to this limit~\cite{deng2023breakdown}.  Putting in the numbers, we have this lower limit of bunch length in a longitudinal weak focusing ring~\cite{deng2024theoretical}
\begin{equation}\label{eq:BLDeng}
	\sigma_{z,\text{min},\text{ISO}}[\mu\text{m}]=4.93\rho^{\frac{1}{2}}[\text{m}]E_{0}[\text{GeV}]\theta ^3[\text{rad}].
\end{equation}
For example, if $E_{0}=0.6$ GeV, $\rho=1.5$ m  which corresponds to the bending field strength $B_{0}=1.33$ T and $\theta=\frac{2\pi}{50}$ rad, then $\sigma_{z,\text{min},\text{ISO}}=7.2$~nm.

\subsection{Application of Transverse Gradient Bend}
The previous analysis assumes that the transverse gradient of the bending field is zero. Now we consider the application of transverse gradient bending (TGB) magnets to lower the horizontal and longitudinal emittances. For simplicity, we will consider the case of a constant gradient. The transfer matrix of a sector bending magnet with a constant transverse gradient $n=-\frac{\rho}{B_{y}}\frac{\partial B_{y}}{\partial x}$ is
{}
	\begin{equation}
		\begin{split}
			{\bf B}_{\text{TGB}}(\alpha)=
			&\left(
			\begin{array}{cccccc}
				\cos \left(\sqrt{1-n}\alpha\right) & \frac{\rho}{\sqrt{1-n}}  \sin \left(\sqrt{1-n}\alpha\right) & 0 & 0 \\
				-\frac{\sqrt{1-n}}{\rho}\sin \left(\sqrt{1-n}\alpha\right)  & \cos \left(\sqrt{1-n}\alpha\right) & 0 & 0 \\
				0 & 0 & \cos\left(\sqrt{n}\alpha\right) & \frac{\rho}{\sqrt{n}}\sin\left(\sqrt{n}\alpha\right)\\
				0 & 0 & -\frac{\sqrt{n}}{\rho}\sin\left(\sqrt{n}\alpha\right) & \cos\left(\sqrt{n}\alpha\right) \\
				-\frac{1}{\sqrt{1-n}}\sin \left(\sqrt{1-n}\alpha\right) & -\frac{\rho}{1-n} \left(1-\cos \left(\sqrt{1-n}\alpha\right)\right) & 0  & 0  \\
				0 & 0 & 0  & 0 \\
			\end{array}\right.\\
			&\ \ \ \ \ \ \ \ \ \left.
			\begin{array}{cccccc}
				0 & \frac{\rho}{1-n} \left(1-\cos \left(\sqrt{1-n}\alpha\right)\right)   \\
				0 & \frac{1}{\sqrt{1-n}}\sin \left(\sqrt{1-n}\alpha\right) \\
				0 & 0\\
				0 & 0\\
				1 & \frac{\rho\alpha }{\gamma ^2}+  \frac{\rho}{\left(1-n\right)^{\frac{3}{2}}}\left(-\sqrt{1-n}\alpha +\sin \left(\sqrt{1-n}\alpha\right)\right) \\
				0 & 1 \\
			\end{array}
			\right).
		\end{split}
	\end{equation}
{}

\subsubsection{Horizontal Emittance}

Following similar steps presented in the above analysis, we find the minimum value of $f_{x}$ in Eq.~(\ref{eq:fx}) is now realized when
\begin{equation}
	\begin{aligned}
		\alpha_{x0}&=0,\ \beta_{x0}\approx\frac{\rho\theta}{2\sqrt{15}}\left[1+\frac{(1-n)\theta^{2}}{140}\right]+\mathcal{O}(\theta^{6}),\\
		D_{x0}&\approx\frac{\rho\theta^2}{24}\left[1-\frac{(1-n)\theta^{2}}{48}\right]+\mathcal{O}(\theta^{6}),\
		D_{x0}'=0,\ 
	\end{aligned}
\end{equation} 
where $\mathcal{O}(\theta^{n})$ means terms of order $\theta^{n}$ and higher,
which means
\begin{equation}
	\mathcal{H}_{x0}=\frac{D_{x0}^{2}}{\beta_{x0}}\approx\frac{5\rho\theta ^3 }{96 \sqrt{15}}\left[1-\frac{9(1-n)\theta^{2}}{280}\right]+\mathcal{O}(\theta^{7}),
\end{equation}
and 
\begin{equation}
	\begin{aligned}
		f_{x,\text{min}}&\approx\frac{\rho\theta ^3 }{12 \sqrt{15}}\left[1-\frac{3(1-n)\theta^{2}}{70}\right]+\mathcal{O}(\theta^{7}).
	\end{aligned}
\end{equation}
So we have
\begin{equation}
	\epsilon_{x,\text{min},\text{TGB}}=C_{q}\frac{\gamma^{2}}{J_{x}}\frac{\theta^{3}}{12\sqrt{15}}\left[1-\frac{3(1-n)\theta^{2}}{70}\right].
\end{equation}
Therefore, the impact of the transverse gradient on the theoretical minimum emittance is on the higher order of the bending angle of each magnet. However, we should recognize that $n$ can be a quite large value in practice. So its impact may actually {be} not small. In addition, a transverse gradient bend can also affect the damping partition, whose details we do not go into here. 


\subsubsection{Longitudinal Emittance}

Similarly, using TGB, the minimum value of $f_{z}$ in Eq.~(\ref{eq:fz}) is realized when
\begin{equation}
	\begin{aligned}
		\alpha_{z0}&=0,\ \beta_{z0}\approx\frac{\rho\theta^{3}}{120\sqrt{7}}\left[1+\frac{(1-n)\theta^{2}}{90}\right]+\mathcal{O}(\theta^{7}),\\
		D_{x0}&\approx-\frac{\rho \theta ^2 }{40}\left[1+\frac{19(1-n)\theta^{2}}{1680}\right]+\mathcal{O}(\theta^{6}),\
		D_{x0}'=0,\ 
	\end{aligned}
\end{equation} 
and
\begin{equation}
	\begin{aligned}
		f_{z,\text{min}}&\approx\frac{\rho\theta ^3 }{60 \sqrt{7}}\left[1+\frac{(1-n)\theta^{2}}{90}\right]+\mathcal{O}(\theta^{7}).
	\end{aligned}
\end{equation}
Note that $f_{z,\text{min}}=2\beta_{z0}$ still holds here. So we have
\begin{equation}
	\epsilon_{z,\text{min},\text{TGB}}=C_{q}\frac{\gamma^{2}}{J_{z}}\frac{\theta^{3}}{60\sqrt{7}}\left[1+\frac{(1-n)\theta^{2}}{90}\right].
\end{equation}

\subsection{Application of Longitudinal Gradient Bend}

We can also apply longitudinal gradient bends (LGB) to lower the transverse and longitudinal emittance. For simplicity, we will study the case of a LGB consisting of several sub-dipoles with each a constant bending radius. Further, we will assume each sub-dipole is a sector dipole. The analysis for the case of rectangular dipoles is similar, as long as the impact of edge angles on transfer matrix and damping partition have been properly handled. Now we investigate the case of sector sub-dipoles. For example, we may choose to let the LGB has a symmetric structure
\begin{equation}\label{eq:TGBStructure}
	(\rho_{2},\theta_{2}),(\rho_{1},\theta_{1}),(\rho_{0},2\theta_{0}),(\rho_{1},\theta_{1}),(\rho_{2},\theta_{2})
\end{equation}
The total bending angle of such a LGB is $\theta_{T}=2(\theta_{0}+\theta_{1}+\theta_{2})$, and the total length is $2(\rho_{0}\theta_{0}+\rho_{1}\theta_{1}+\rho_{2}\theta_{2})$. Note that $\rho_{i}\theta_{i}\geq0$. We use this structure as an example for the analysis. The presented method however applies also to a more general setup.

\subsubsection{Horizontal Emittance}
Now we calculate the theoretical minimum horizontal emittance by invoking LGBs with each the structure given in Eq.~(\ref{eq:TGBStructure}). Still we assume all the LGB setup in the ring and optical functions in each LGB are identical, then 

\begin{equation}
	\begin{aligned}
		\epsilon_{x}
		=C_{q}\frac{\gamma^{2}}{J_{x}}I_{5x}(\alpha_{x0},\beta_{x0},D_{x0},D_{x0}')\\
	\end{aligned}
\end{equation}
with
{}
	\begin{equation}
		\begin{aligned}
			I_{5x}(\alpha_{x0},\beta_{x0},D_{x0},D_{x0}')=\frac{1}{2\left(\frac{\theta_{0}}{\rho_{0}}+\frac{\theta_{1}}{\rho_{1}}+\frac{\theta_{2}}{\rho_{2}}\right)}\int_{-(\rho_{0}\theta_{0}+\rho_{1}\theta_{1}+\rho_{2}\theta_{2})}^{(\rho_{0}\theta_{0}+\rho_{1}\theta_{1}+\rho_{2}\theta_{2})}\frac{\mathcal{H}_{x}(s)}{|\rho(s)|^{3}}ds.
		\end{aligned}
	\end{equation}
{}
Note that the dimension of $I_{5x}$ here is equivalent to $\frac{f_{x}}{\rho}$ given in previous sections.
The mathematical problem is then to minimize $I_{5x}$, by adjusting $\alpha_{x0}$, $\beta_{x0}$, $D_{x0}$, $D_{x0}'$. The calculation method of $\mathcal{H}_{x}$ evolution in a LGB is the same as that given in Eq.~(\ref{eq:Hx}), but note that here for different part (sub-dipole) of the LGBs, we should apply the transfer matrix from the middle point of LGB to the corresponding location.
Following similar procedures, we find that to get the minimum emittance, we still need $\alpha_{x0}=0$ and $D_{x0}'=0$. Then we have $\mathcal{H}_{x}(-s)=\mathcal{H}_{x}(s)$, which means
{}
	\begin{equation}
		\begin{aligned}
			&I_{5x}=\frac{1}{\left(\frac{\theta_{0}}{\rho_{0}}+\frac{\theta_{1}}{\rho_{1}}+\frac{\theta_{2}}{\rho_{2}}\right)}\left(\int_{0}^{\rho_{0}\theta_{0}}\frac{\mathcal{H}_{x}(s)}{|\rho_{0}|^{3}}ds+\int_{\rho_{0}\theta_{0}}^{\rho_{0}\theta_{0}+\rho_{1}\theta_{1}}\frac{\mathcal{H}_{x}(s)}{|\rho_{1}|^{3}}ds+\int_{\rho_{0}\theta_{0}+\rho_{1}\theta_{1}}^{\rho_{0}\theta_{0}+\rho_{1}\theta_{1}+\rho_{2}\theta_{2}}\frac{\mathcal{H}_{x}(s)}{|\rho_{2}|^{3}}ds\right).
		\end{aligned}
	\end{equation}
{}
We find a general analytical discussion of the combination of $\rho_{i}$ and $\theta_{i}$ cumbersome. Here for simplicity and to get a concrete feeling, we first consider one specific case: $\rho_{i}=\rho_{0}2^{i}$. The physical consideration behind this choice is that $\mathcal{H}_{x}$ will be smaller in the central part of the bending magnet, and larger at the entrance and exit. So we make the bending field in the center stronger and smaller at the entrance and exit region, such that to minimize the quantum excitation of horizontal emittance. For example, we may choose
\begin{equation}\label{eq:LGBSet}
	\left(4\rho,\frac{\theta}{8}\right),\left(2\rho,\frac{\theta}{8}\right),\left(\rho,\frac{\theta}{4}\right),\left(\rho,\frac{\theta}{4}\right),\left(2\rho,\frac{\theta}{8}\right),\left(4\rho,\frac{\theta}{8}\right).
\end{equation}
The total length of such a LGB is $2\rho\theta$.
The minimum value of $I_{5x}$ and horizontal emittance $\epsilon_{x}$ in this case is realized when
\begin{equation}
	\begin{aligned}
		\alpha_{x0}&=0,\ \beta_{x0}\approx\frac{\sqrt{\frac{35977}{133755}}\rho\theta}{4},\
		D_{x0}\approx\frac{127\rho\theta^{2}}{7104},\
		D_{x0}'=0,\ 
	\end{aligned}
\end{equation} 
which means
\begin{equation}
	\mathcal{H}_{x0}=\frac{D_{x0}^{2}}{\beta_{x0}}\approx0.183\times\frac{5 }{96 \sqrt{15}}\rho\theta ^3, 
\end{equation}
and 
\begin{equation}
	\begin{aligned}
		I_{5x,\text{min}}&\approx0.344\times\frac{\theta ^3 }{12 \sqrt{15}}, 
	\end{aligned}
\end{equation}
which compared to Eq.~(\ref{eq:TMEx}) means the theoretical minimum horizontal emittance can now become about one third of the case with no longitudinal gradient. So application of LGB is quite effective in lowering the transverse emittance.

\begin{figure}[tb]
	\centering
	\includegraphics[width=0.6\textwidth]{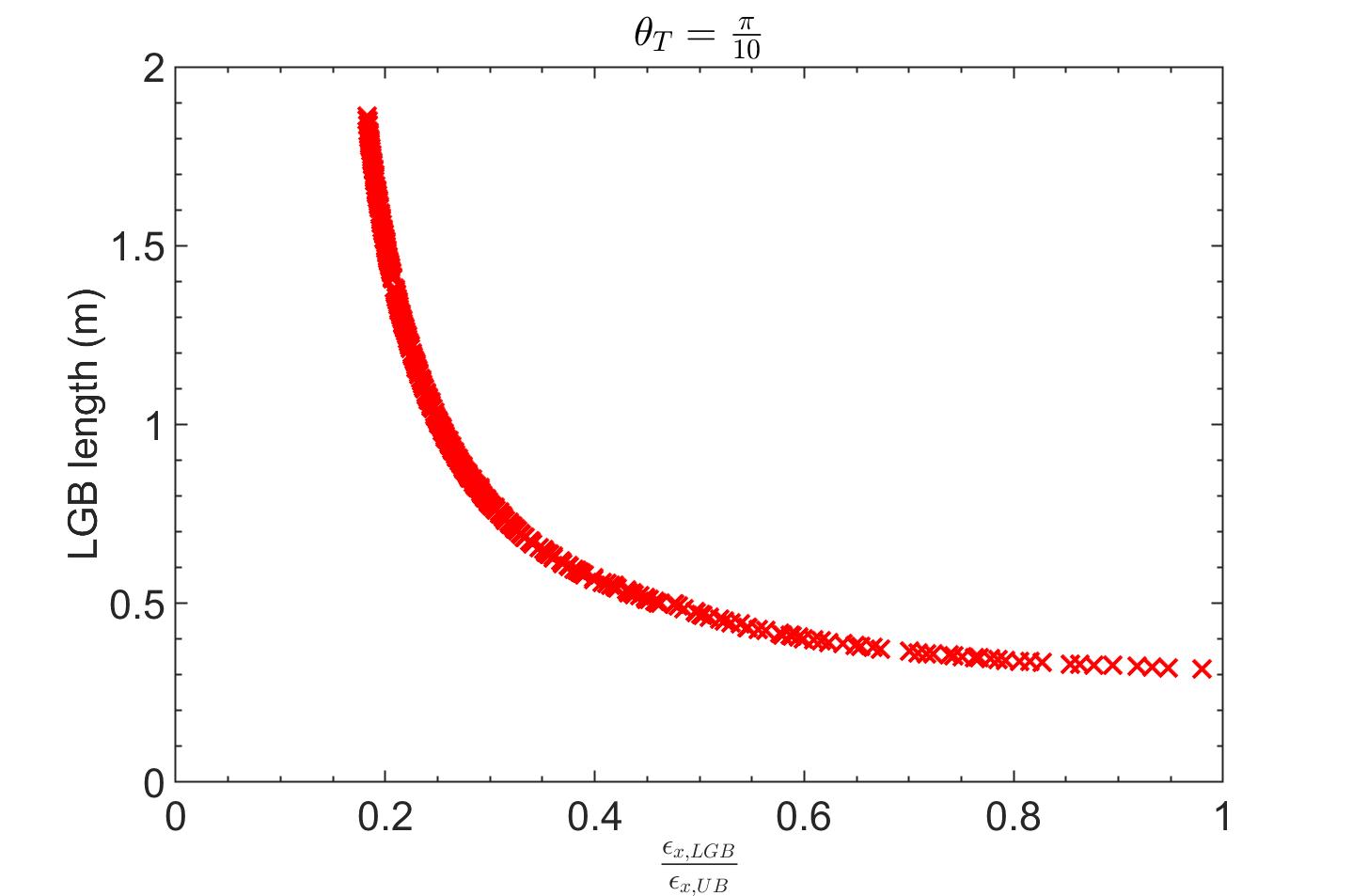}
	\includegraphics[width=0.6\textwidth]{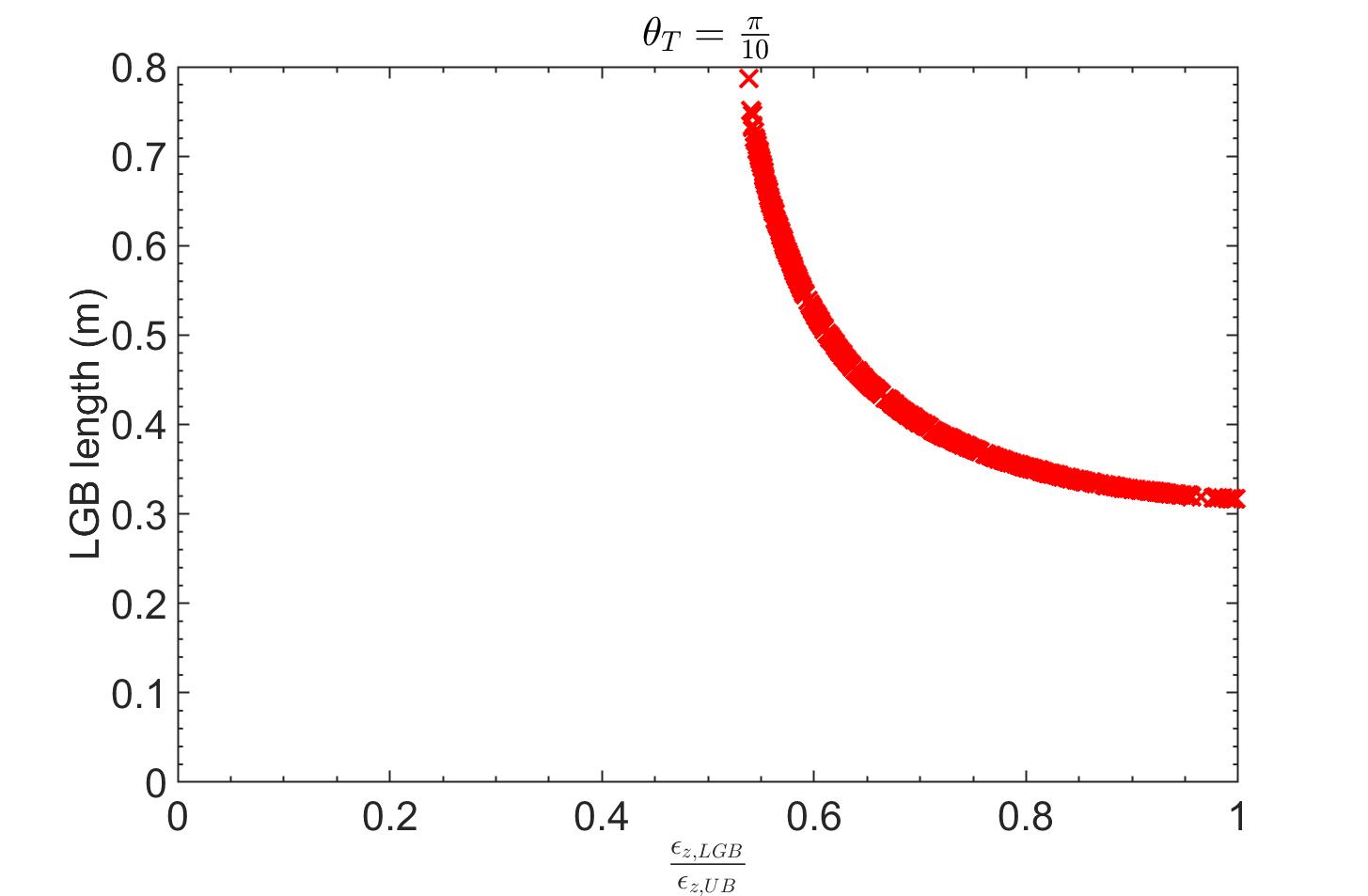}
	\caption{Application of LGB to minimize horizontal (top) and longitudinal (bottom) emittance, respectively. The subscripts ``LGB" and ``UB" represent longitudinal gradient bend and uniform bend, respectively. }
	\label{fig:emitxmoga}
\end{figure}

The next question is: what is the optimal combination of $\theta_{0,1,2}$ and $\rho_{0,1,2}$? This questions is not straightforward to answer by analytical method. Here we refer to numerical method to do the optimization directly. $\alpha_{x0}$ and $D_{x0}'$ are set to be zero in the optimization.  The variables in the numerical optimization are $(\theta_{0},\theta_{1},\theta_{2},\rho_{0},\rho_{1},\rho_{2},\beta_{x0},D_{x0})$. Two optimization goals are $\frac{\epsilon_{x,\text{LGB}}}{\epsilon_{x,\text{UB}}}$ and the length of a LGB $L_{\text{LGB}}$, where $\epsilon_{x,\text{UB}}$ is the theoretical minimum emittance of applying bending magnet without longitudinal gradient. We require $\rho_{i}\theta_{i}>0$. In the optimization, we keep the total bending angle of a LGB a constant value. 
The optimization result of one specific case where $\theta_{T}=2(\theta_{0}+\theta_{1}+\theta_{2})=\frac{\pi}{10}$ is presented in the upper part of Fig.~\ref{fig:emitxmoga}, from which we can see that in this case by applying LGBs, in principle we can lower the horizontal emittance by a factor of five with a reasonable length of the LGB.  Note that in this optimization and the one in the following section, we have assumed that $\rho_{i}>0$. But the formalism applies also the case with anti-bends.

\subsubsection{Longitudinal Emittance}
Similarly, for longitudinal emittance we have
\begin{equation}
	\begin{aligned}
		\epsilon_{z}&=C_{q}\frac{\gamma^{2}}{J_{z}}I_{5z}(\alpha_{z0},\beta_{z0},D_{x0},D_{x0}')
	\end{aligned}
\end{equation}
with
{}
	\begin{equation}
		\begin{aligned}
			&I_{5z}(\alpha_{z0},\beta_{z0},D_{x0},D_{x0}')=\frac{1}{2\left(\frac{\theta_{0}}{\rho_{0}}+\frac{\theta_{1}}{\rho_{1}}+\frac{\theta_{2}}{\rho_{2}}\right)}\oint_{-(\rho_{0}\theta_{0}+\rho_{1}\theta_{1}+\rho_{2}\theta_{2})}^{(\rho_{0}\theta_{0}+\rho_{1}\theta_{1}+\rho_{2}\theta_{2})}\frac{\beta_{z}(s)}{|\rho(s)|^{3}}ds.
		\end{aligned}
	\end{equation}
{}
For example, if we still choose the setup given in Eq.~(\ref{eq:LGBSet}).
The minimum value of $I_{5z}$ and longitudinal emittance in this case is realized when
\begin{equation}
	\begin{aligned}
		\alpha_{z0}&=0,\ \beta_{z0}\approx \frac{\sqrt{\frac{36233641}{62419}} \rho \theta ^3  }{7680 },\
		D_{x0}\approx-\frac{169 \rho \theta ^2 }{9640},\
		D_{x0}'=0,\ 
	\end{aligned}
\end{equation} 
and 
\begin{equation}
	\begin{aligned}
		F_{z,\text{min}}&\approx0.838\times\frac{\theta^{3}}{60\sqrt{7}}, 
	\end{aligned}
\end{equation}
which compared with Eq.~(\ref{eq:TMEz0}) means the theoretical minimum longitudinal emittance now can become a bit smaller than the case of applying a constant bending field. 


Similar to what presented just now about lowering transverse emittance, we also apply the numerical optimization to choose a better combination of $\rho_{i}$ and $\theta_{i}$ in lowering longitudinal emittance. Presented in the bottom part of Fig.~\ref{fig:emitxmoga} is the result of one specific case where $\theta_{T}=2(\theta_{0}+\theta_{1}+\theta_{2})=\frac{\pi}{10}$, from which we can see that in this case by applying LGBs, in principle we can lower the longitudinal emittance by a factor of two with a reasonable length of the LGB. So generally, LGB is more effective in lowering the horizontal emittance, compared to lowering the longitudinal emittance.  



\section{Steady-State Micro-Bunching Storage Rings}\label{sec:SSMB}

\begin{figure}[tb] 
	\centering 
	\includegraphics[width=0.8\columnwidth]{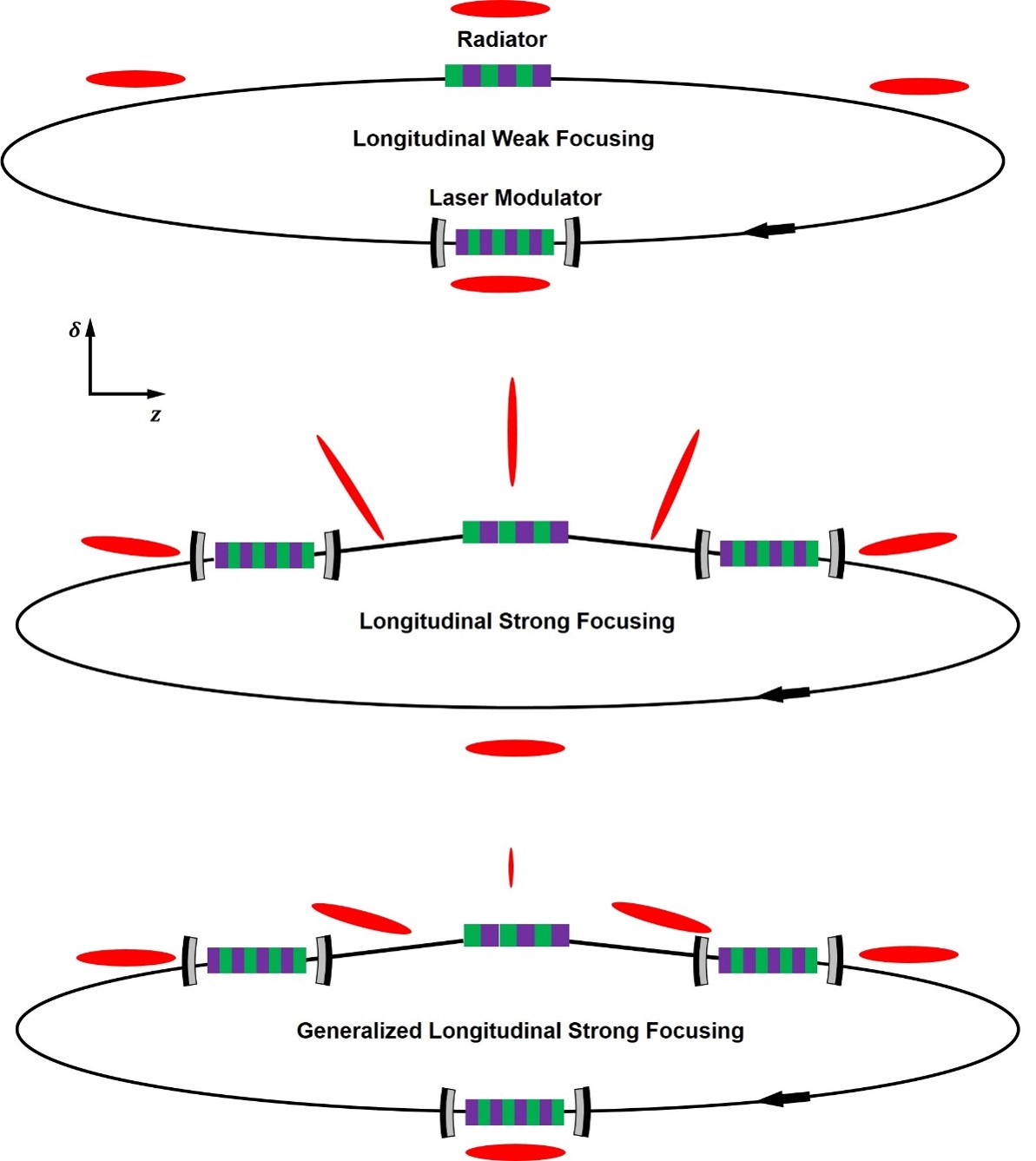}
	\caption{
		\label{fig:GLSF} 
		Schematic layout of longitudinal weak focusing (LWF), longitudinal strong focusing (LSF), generalized longitudinal strong focusing (GLSF) SSMB storage rings. The red ellipses illustrate the beam distribution in longitudinal phase space. To realize the same bunch length compression ratio, the required energy chirp strength in the GLSF scheme is much smaller than that of LSF.
	}
\end{figure}

In this section, based on the theoretical minimum emittance derived in {the} last section, we conduct some key analysis of three specific SSMB scenarios along the thinking of realizing nm bunch length and high-average-power EUV radiation, i.e., longitudinal weak focusing (LWF), longitudinal strong focusing (LSF) and generalized longitudinal strong focusing (GLSF). The analysis aims to answer the question why GLSF SSMB is our present choice in realizing high-average-power EUV radiation. Before going into the details, here first we use {Table~\ref{tab:Table1} and Fig.~\ref{fig:GLSF}} to briefly summarize the characteristics  of these three scenarios.  Note that in Fig.~\ref{fig:GLSF}, the beam distribution in longitudinal phase space are all of the microbunch, whose length is at the laser wavelength range. {We also remind the readers that in the figure, the energy chirp strength in GLSF is much smaller than that of LSF. The physical reason should be clear with the analysis in this section conducted.}

\begin{table}[b]\centering
	\caption{
		\label{tab:Table1} 
		Main characteristics of LWF, LSF and GLSF SSMB storage rings.
	}
	\begin{tabular}{|c|c|c|c|}
		\hline 
		LWF	& $\nu_{s}\ll1$ & $\frac{\beta_{z,\text{max}}}{\beta_{z,\text{min}}}\approx1$ & 2D phase space dynamics\\ 
		\hline 
		LSF	& $\nu_{s}\sim1$ &  $\frac{\beta_{z,\text{max}}}{\beta_{z,\text{min}}}\gg1$ & 2D phase space dynamics \\ 
		\hline 
		GLSF	& - &   $\frac{\sigma_{z,\text{max}}}{\sigma_{z,\text{min}}}\gg1$  & 4D or 6D phase space dynamics\\ 
		\hline 
	\end{tabular} 
\end{table}

In all the example calculations to be shown in the following part of this paper, we set the electron energy to be $E_{0}=600$~MeV, and modulation laser wavelength to be $\lambda_{L}=1064$~nm. The choice of this beam energy is because it is an appropriate energy for EUV generation using an undulator as radiator. On one hand, it is not too high, otherwise the laser modulation will become more difficult, which means more laser power is needed to imprint a given modulation strength. On the other hand, it is not too low otherwise intra-beam scattering (IBS) could become too severe. {Actually as we will see in Sec.~\ref{sec:application}, IBS is a fundamental issue in SSMB storage rings which require at least one of the three eigen emittances to be small.} The reason for choosing this laser wavelength is due to the fact that it is the common wavelength range for high-power optical enhancement cavity, which is used together with an undulator to form the laser modulator of SSMB.

\subsection{Longitudinal Weak Focusing}

Now we start the quantitative analysis. We start from the longitudinal weak focusing (LWF) SSMB ring. In a LWF ring with a single laser modulator (LM) as shown in Fig.~\ref{fig:GLSF}, the single-particle longitudinal dynamics turn by turn is modeled as
{
	\begin{equation}\label{eq:standMap}
		\begin{aligned}
			\delta_{n+1}&=\delta_{n}+\frac{h}{k_{L}}\sin(k_{L}z_{n}),\\
			z_{n+1}&=z_{n}-\eta C_{0}\delta_{n+1},
		\end{aligned}
	\end{equation}
}
{where the subscripts ${n,n+1}$ means the number of revolutions}, $h$ is the energy chirp strength around the zero-crossing phase, $k_{L}=\frac{2\pi}{\lambda_{L}}$ is the wavenumber of modulation laser and $\lambda_{L}$ is the laser wavelength, $C_{0}$ is the ring circumference, and
\begin{equation}
	\eta=\frac{\Delta T/T_{0}}{\Delta E/E_{0}}=\frac{1}{C_{0}}\oint\left(\frac{D_{x}}{\rho}-\frac{1}{\gamma^{2}}\right)ds
\end{equation}
is the phase slippage of the ring. Note that in the above model, we have assumed that the radiation energy loss in an SSMB storage ring will be compensated by other system instead of the laser modulators, so the microbunching will be formed around the laser zero-crossing phase. The laser in principle can also be used for energy compensation, but is not a cost-effective choice and will also limit the output radiation power. We can use induction linacs or RF cavities to supply the radiation energy loss. Linear stability of Eq.~(\ref{eq:standMap}) around zero-crossing phase requires that $0<h\eta C_{0}<4$. To avoid strong chaotic dynamics which may destroy the regular longitudinal phase space structure, an empirical criterion is
\begin{equation}\label{eq:LWFChaos}
	0<h\eta C_{0}\lesssim 0.1.
\end{equation} 
In a LWF ring, when the synchrotron tune $|\nu_{s}|\approx{\sqrt{h\eta C_{0}}}/{2\pi}\ll1$, the longitudinal beta function at the center of laser modulator is~\cite{deng2024theoretical}
\begin{equation}\label{eq:LWFbetaZ}
	\beta_{z\text{M}}\approx\sqrt{\frac{\eta C_{0}}{h}}.
\end{equation}
Note that in this paper, we will use subscript `M' to represent modulator, and `R' to represent radiator.  From Eq.~(\ref{eq:LWFChaos}) we then require
\begin{equation}\label{eq:LWFbetaZ2}
	|\eta C_{0}|\lesssim \frac{\beta_{z\text{M}}}{\sqrt{10}}.
\end{equation}

Now we can use the previous analysis of theoretical minimum longitudinal emittance, more specifically Eq.~(\ref{eq:BLDeng}), and the above result to do some evaluation for a LWF ring.  If $E_{0}=600$ MeV, the bending radius of dipoles in the ring $\rho_{\text{ring}}=1.5$ m which corresponds to a bending field strength $B_{\text{ring}}=1.33$ T, and if our desired bunch length is $\sigma_{z}=50$~nm ($\sigma_{z}\lesssim{\lambda_{L}}/{20}$ for a microbunches to be safely stored in the optical microbuckets for $\lambda_{L}=1064\ $nm), then we may need $\sigma_{z,\text{min},\text{ISO}}\leq{50}/{\sqrt{2}}$ nm to avoid significant energy widening when we reach the desired bunch length. From Eq.~(\ref{eq:BLDeng}) we then need $\theta\leq\frac{2\pi}{30}$ rad. Therefore, we need at least 30 bending magnets in the ring. Assuming the length of each isochronous cell containing a bending magnet is 3 m, then the arc section of such a storage ring has a length of about 90 m. Considering the straight section for beam injection/extraction, radiation energy loss compensation, and insertion device for radiation generation, the circumference of such a ring is 100~m to 120~m.

To reach the desired bunch length, according to Eq.~(\ref{eq:TMEzCOnditionsDeng}) we need $\beta_{z\text{M}}=\beta_{z0}={\rho\theta^{3}}/{12\sqrt{210}}$. In our present example case, if $\rho=1.5$ m, $\theta=\frac{2\pi}{30}$, then this value we need is $\beta_{z\text{M}}=79.2\ \mu$m. 
Then from Eq.~(\ref{eq:LWFbetaZ2}) we have
\begin{equation}
	|\eta C_{0}|\lesssim25\ \mu\text{m}.
\end{equation}
If $C_{0}=100$~m, then it means we need a phase slippage factor $|\eta|\lesssim2.5\times10^{-7}$, which is a quite small value. If we want a bunch length even smaller than 50 nm at a beam energy of 600 MeV, then the required phase slippage will be too demanding to  be realized using present technology. More details on the lattice design of an LWF SSMB ring which can store microbunches with a couple of 10 nm bunch length can be found in Ref.~\cite{Pan2020Thesis}.


\subsection{Longitudinal Strong Focusing}

\begin{figure}[tb] 
	\centering 
	\includegraphics[width=0.6\columnwidth]{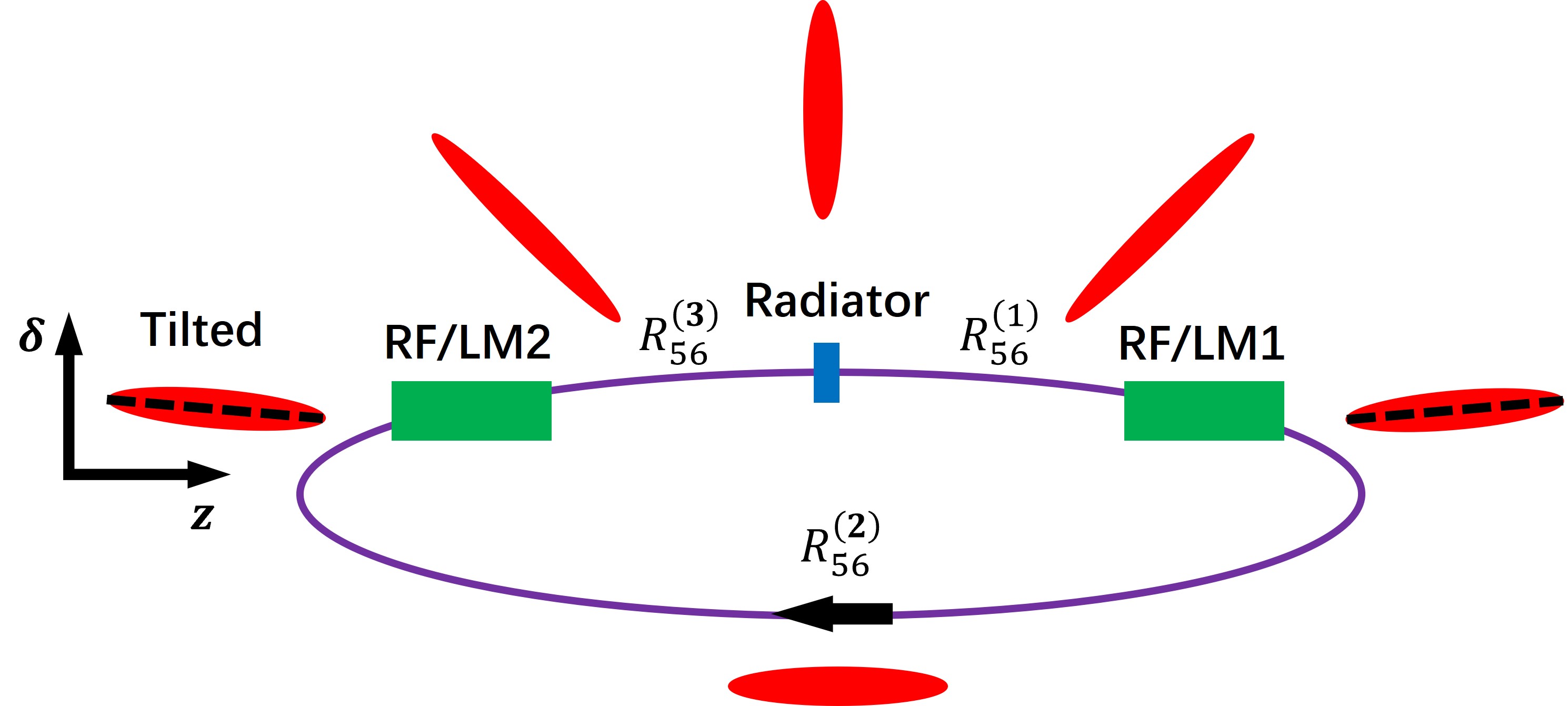}
	\caption{
		\label{fig:Chap2-TwoRFs} 
		A schematic layout of a storage ring using two RF systems for longitudinal strong focusing and an example beam distribution evolution in longitudinal phase space. (Figure from Ref.~\cite{deng2021courant})
	}
\end{figure}

After discussing the LWF SSMB ring, we now start the analysis of LSF. First we observe that the above analysis of LWF SSMB considers the case with only a single LM. When there are multiple LMs, for the longitudinal dynamics, it is similar to implementing multiple quadrupoles in the transverse dimension, and the beam dynamics can have more possibilities. Longitudinal strong focusing scheme for example can be invoked, not unlike its transverse counterpart which is the foundation of modern high-energy accelerators.  Here we use a setup with two LMs for SSMB as an example to show the scheme of manipulating $\beta_{z}$ around the ring using strong focusing regime. The schematic layout of the ring is shown in Fig.~\ref{fig:Chap2-TwoRFs}. The treatment of cases with more LMs is similar.

We divide the ring into five sections from the transfer matrix viewpoint, i.e., three longitudinal drifts ($R_{56}$) and two LM kicks ($h$), with the linear transfer matrices of state vector in longitudinal $(z,\delta)^{T}$ given by
\begin{equation}
	\begin{aligned}
		&{\bf T}_{\text{D1}}=\left(\begin{matrix}
			1&R_{56}^{(1)}\\
			0&1
		\end{matrix}\right),\ {\bf T}_{\text{LM1}}=\left(\begin{matrix}
			1&0\\
			h_{1}&1
		\end{matrix}\right),\\
		&{\bf T}_{\text{D2}}=\left(\begin{matrix}
			1&R_{56}^{(2)}\\
			0&1
		\end{matrix}\right),\
		{\bf T}_{\text{LM2}}=\left(\begin{matrix}
			1&0\\
			h_{2}&1
		\end{matrix}\right),\\
		&{\bf T}_{\text{D3}}=\left(\begin{matrix}
			1&R_{56}^{(3)}\\
			0&1
		\end{matrix}\right).
	\end{aligned}
\end{equation}
Then the one-turn map at the radiator center is
\begin{equation}
	\begin{aligned}
		{\bf M}_{\text{R}}&={\bf T}_{\text{D3}}{\bf T}_{\text{LM2}}{\bf T}_{\text{D2}}{\bf T}_{\text{LM1}}{\bf T}_{\text{D1}}.\\
	\end{aligned}
\end{equation}
For the generation of coherent radiation, we usually want the bunch length to reach its minimum at the radiator, then we need $\alpha_{z}=0$ at the radiator.


With the primary goal of presenting the principle, instead of a detailed design, here for simplicity we focus on one special case: $R_{56}^{(1)}=R_{56}^{(3)}$, $h_{1}=h_{2}=h$.  Denote 
\begin{equation}
	\zeta_{1}\equiv1+R_{56}^{(1)}h,\ \zeta_{2}\equiv1+\frac{R_{56}^{(2)}}{2}h,
\end{equation}
we then have 
\begin{equation}
	\begin{aligned}
		{\bf M}_{\text{R}}=\left(\begin{matrix}
			2\zeta_{1}\zeta_{2}-1&2\frac{\zeta_{1}^{2}\zeta_{2}-\zeta_{1}}{h}\\
			2h\zeta_{2}&2\zeta_{1}\zeta_{2}-1
		\end{matrix}\right).
	\end{aligned}
\end{equation}
The linear stability requires 
$
|2\zeta_{1}\zeta_{2}-1|<1
$ which means $0<\zeta_{1}\zeta_{2}<1$.

We remind the readers that the analysis above in this subsection about LSF has been presented before in Ref.~\cite{deng2024theoretical} and is presented here again for completeness. Now we try to go further to gain more insight.
The longitudinal beta function at the radiator is 
\begin{equation}\label{eq:betaZR}
	\beta_{z\text{R}}=\frac{2\frac{\zeta_{1}^{2}\zeta_{2}-\zeta_{1}}{h}}{\sin\Phi_{z}}=\frac{1}{|h|}\sqrt{\frac{\zeta_{1}\left(1-\zeta_{1}\zeta_{2}\right)}{\zeta_{2}}},
\end{equation}
where $\Phi_{z}=2\pi\nu_{s}$ is the synchrotron phase advance per turn. The longitudinal beta function at the opposite of the radiator, from where to LM2 has an $R_{56}=\frac{R_{56}^{(2)}}{2}$,  is
\begin{equation}
	\beta_{z\text{RO}}=\frac{1}{|h|}\sqrt{\frac{\zeta_{2}\left(1-\zeta_{1}\zeta_{2}\right)}{\zeta_{1}}},
\end{equation}
which can be obtained by switching $\zeta_{1}$ and $\zeta_{2}$ in the expression of $\beta_{z\text{R}}$ in Eq.~(\ref{eq:betaZR}).
So we have
\begin{equation}
	\frac{\beta_{z\text{RO}}}{\beta_{z\text{R}}}=\frac{\zeta_{2}}{\zeta_{1}}.
\end{equation}
The longitudinal beta function at the LM1 and LM2 (here we name them as modulators) is
\begin{equation}\label{eq:betazM}
	\beta_{z\text{M}}=\beta_{z\text{R}}+\frac{{R_{56}^{(1)}}^{2}}{\beta_{z\text{R}}}.
\end{equation}
For efficient bunch compression from the place of modulator to the place of radiator, we have $|\zeta_{1}|\ll1$. 
Then
\begin{equation}
	\frac{\beta_{z\text{M}}}{\beta_{z\text{R}}}=\frac{\zeta_{1}-2\zeta_{1}\zeta_{2}+\zeta_{2}}{\zeta_{1}\left(1-\zeta_{1}\zeta_{2}\right)}\approx\frac{\zeta_{2}}{\zeta_{1}\left(1-\zeta_{1}\zeta_{2}\right)}=\frac{1}{h^{2}\beta_{z\text{R}}^{2}},
\end{equation}
{ from which we have the required energy chirp strength}
\begin{equation}\label{eq:LSFh}
	|h|\approx\frac{1}{\sqrt{\beta_{z\text{R}}\beta_{z\text{M}}}}.
\end{equation}
%
This relation is similar to the theorems to be presented in Sec.~\ref{sec:TLCTheorems} about transverse-longitudinal coupling-based bunch compression schemes which is the backbone of the GLSF scheme. The relation can also be casted as
\begin{equation}\label{eq:LSFh2}
	|h|\approx\frac{\sqrt{\epsilon_{z}/\beta_{z\text{M}}}}{\sqrt{\epsilon_{z}\beta_{z\text{R}}}}=\frac{\sqrt{\epsilon_{z}/\beta_{z\text{M}}}}{\sigma_{z\text{R}}},
\end{equation}
with $\sigma_{z\text{R}}=\sqrt{\epsilon_{z}\beta_{z\text{R}}}$ the bunch length at the radiator.

We now investigate the required energy chirp strength and modulation laser power based on the analysis. We assume that in a LSF ring, the longitudinal emittance is dominantly from the quantum excitation in ring dipoles. This assumption will be {justified} later to see if it is really the case. Further, we assume that the average longitudinal beta function at the dipoles equals that at the modulator, i.e., $\langle\beta_{z}\rangle=\beta_{z\text{M}}$. Then the equilibrium longitudinal emittance from the balance of quantum excitation and radiation damping in a ring consisting of iso-bending-magnets has the scaling
\begin{equation}\label{eq:emitZLSF}
	\epsilon_{z,\text{LSF}}=C_{q}\frac{\gamma^{2}}{J_{z}}\frac{I_{5z}}{I_{2}}=C_{q}\frac{\gamma^{2}}{J_{z}}\frac{\frac{\langle\beta_{z}\rangle}{\rho_{\text{ring}}^{3}}2\pi\rho_{\text{ring}}}{\frac{1}{\rho_{\text{ring}}^{2}}2\pi\rho_{\text{ring}}}\propto\gamma^{2}\frac{\beta_{z\text{M}}}{\rho_{\text{ring}}},
\end{equation}
which combining with Eq.~(\ref{eq:LSFh2}) gives
\begin{equation}
	|h|\propto\frac{\gamma}{\sigma_{z\text{R}}\sqrt{\rho_{\text{ring}}}}.
\end{equation}
{Here}, $\sigma_{z\text{R}}$ is determined by our desired radiation wavelength. Therefore, given the beam energy and desired bunch length, to lower the required energy chirp strength in LSF, we should use as large $\rho_{\text{ring}}$ as possible which means as weak as bending magnet as possible.  But note that the total length of the bending magnets should be within a reasonable range. Also note that when the bending magnets in the ring are very weak, the assumption that the longitudinal emittance in a LSF ring is dominantly from them will fail, {since the quantum excitation in the bending magnets will become weaker, while there are other contributions like quantum excitation at the laser modulators.}


{We have mentioned in the introduction section that a short bunch can generate coherent radiation. The parameter used to quantify the capability of beam for coherent radiation generation is called bunching factor, and in the 1D case is defined as 
	\begin{equation}
		b(\omega)=\int_{-\infty}^{\infty}\psi(z)e^{-i\frac{\omega}{c}z}dz,
	\end{equation}
	with $\omega$ the radiation frequency, $\psi(z)$ the longitudinal charge density distribution satisfying the normalization $\int_{-\infty}^{\infty}\psi(z)dz=1$. We will present more indepth discussion of bunching factor in Sec.~\ref{sec:BF}. The coherent radiation power of a beam with $N_{p}$ particles at frequency $\omega$ is related to the radiation of a single particle according to
	\begin{equation}
		P_{\text{beam}}(\omega)=N_{e}^{2}b^{2}(\omega)P_{\text{single}}(\omega).
	\end{equation}
	For a Gaussian bunch with RMS bunch length of $\sigma_{z}$, we have $b(\omega)=\text{exp}\left[-{\left(\frac{\omega}{c}\sigma_{z}\right)^{2}}/{2}\right].$
	For significant 13.5~nm-wavelength coherent EUV radiation generation, we may need $\sigma_{z\text{R}}\lesssim4$~nm which corresponds to $b_{\text{13.5 nm}}\gtrsim0.18$.} To increase the radiation power, we may need the radiator, which is assumed to be an undulator, has a large period number $N_{u}$ (for example $N_{u}\sim300$). To avoid significant bunch lengthening from the energy spread and the undulator $R_{56}=2N_{u}\lambda_{R}$, we then need  $N_{u}\lambda_{R}\sigma_{\delta}\lesssim\sigma_{z\text{R}}$, which then requires
the energy spread at the radiator $\sigma_{\delta\text{R}}\lesssim1\times10^{-3}$.  So we need $\epsilon_{z}=\sigma_{z\text{R}}\sigma_{\delta\text{R}}\lesssim4$~pm. Since in a LSF ring, as explained we cannot make the optimal conditions Eq.~(\ref{eq:TMEzCOnditions}) be satisfied in all the bending magnets, a reasonable argument is that the real longitudinal emittance should be at least a factor of two larger than the true theoretical minimum, which then requires
\begin{equation}
	\epsilon_{z,\text{min}}\lesssim2\ \text{pm}.
\end{equation}
Then according to Eq.~(\ref{eq:TMEz}), if $E_{0}=600$ MeV, we need $\theta\lesssim\frac{2\pi}{59}$, which means 59 bending magnets are needed. Assuming the length of each isochronous cell containing a bending magnet is 3 m, then the arc section of such a ring has a length of about 177 m. Considering the straight section for beam injection/extraction, radiation energy loss compensation, and longitudinal strong focusing section, the circumference of such a ring is about 200 m.  

If $\epsilon_{z}=4$ pm, to get $\sigma_{z\text{R}}\lesssim4$ nm, we then need
\begin{equation}\label{eq:betaLSFZR}
	\beta_{z\text{R}}=\frac{\sigma_{z\text{R}}^{2}}{\epsilon_{z}}\lesssim4\ \mu\text{m}.
\end{equation}
From Eq.~(\ref{eq:LSFh}), given $\beta_{z\text{R}}$, we should apply as large $\beta_{z\text{M}}$ as possible to decrease the energy chirp strength $h$. Since $\beta_{z0}\propto\rho$ with $\theta$ given, we will choose a reasonable large bending radius for the dipoles in the ring. If $\rho=10$~m which corresponds to $B_{0}=0.2$~T and total  bending magnet length 62.8 m, then the longitudinal beta function at the dipole center required to reach the practical theoretical minimum longitudinal emittance (see Eqs.~(\ref{eq:TMEzDeng}) and~(\ref{eq:TMEzCOnditionsDeng})) is $\beta_{z0}={\rho\theta^{3}}/{12\sqrt{210}}=69.5\ \mu$m. Then we may let 
\begin{equation}\label{eq:betaLSFZM}
	\beta_{z\text{M}}\approx2\beta_{z0}\approx139\ \mu\text{m}.
\end{equation}
Then from Eqs.~(\ref{eq:LSFh}), (\ref{eq:betaLSFZR}) and (\ref{eq:betaLSFZM}), the energy chirp strength required in such a LSF SSMB ring is
\begin{equation}
	|h|\approx\frac{1}{\sqrt{\beta_{z\text{R}}\beta_{z\text{M}}}}\gtrsim4.24\times10^{4}\ \text{m}^{-1}.
\end{equation}
According to Eq.~(\ref{eq:TEM00h}) to be presented later about the laser modulator induced energy {chirp} strength, if $E_{0}=600$ MeV, and for a modulator undulator with period $\lambda_{u\text{M}}=8$~cm ($B_{0\text{M}}=1.13$~T), length $L_{u\text{M}}=1.6$ m, and for a laser with wavelength $\lambda_{L}=1064$~m, to {introduce} the required energy chirp strength, we need a laser power $P_{L}\approx1$ GW. This is a large value and is three orders of magnitude higher than the average stored laser power reachable in an optical enhancement cavity at the moment~\cite{Lu2024OEC}, which is at the level of one megawatt (MW). This makes the optical enhancement cavity only work in a low duty cycle pulsed mode, thus limiting the filling factor of microbunched electron beam in the ring, and thus limiting the average output EUV power.  

We remind the readers that there is a subtle point in a LSF SSMB ring if we take the nonlinear sinusoidal modulation waveform into account, since the dynamic system is then strongly chaotic and requires careful {analysis} to ensure a large enough stable region for particle motion in the longitudinal phase space. More details in this respect can be found in Ref.~\cite{Zhang2022Thesis}.

Now for completeness of discussion, let us evaluate the contribution of modulator undulators to longitudinal emittance in our above example case, since there {is} also quantum excitation at the modulators. 
The quantum excitation contributions of two modulators to $\epsilon_{z}$ in a LSF ring are
{
	\begin{equation}\label{eq:EmittanceZ}
		\begin{aligned}
			\Delta\epsilon_{z\text{M}}&=C_{q}\frac{\gamma^{2}}{J_{z}}\frac{\Delta I_{5z\text{M}}}{I_{2}}\\
			&=C_{q}\frac{\gamma^{2}}{J_{z}}\frac{1}{I_{2}}\times2\int_{-\frac{L_{u\text{M}}}{2}}^{\frac{L_{u\text{M}}}{2}} \frac{\beta_{z\text{M}}}{|\rho(s)|^{3}}ds\\
			&=C_{q}\frac{\gamma^{2}}{J_{z}}\frac{1}{I_{2}}\times2\frac{\beta_{z\text{M}}}{\rho_{0\text{M}}^{3}}\frac{4}{3\pi}L_{u\text{M}},
		\end{aligned}
	\end{equation}
	where $\Delta I_{5z\text{M}}$ is the contribution of two modulators to $I_{5z}$, $\rho_{0\text{M}}$ is the bending radius at the peak magnetic field of the modulator. 
	Put in the numbers, and taking the approximation $I_{2}\approx\frac{2\pi}{\rho_{\text{ring}}}$ which means the radiation loss is mainly from dipoles in the ring, $J_{z}\approx2$, we have
	\begin{equation}\label{eq:zEmittanceM}
		\begin{aligned}
			\Delta\epsilon_{z\text{M}}[\text{nm}]
			&=8.9B_{\text{ring}}^{-1}[\text{T}]B_{0\text{M}}^{3}[\text{T}]\beta_{z\text{M}}[\text{m}]L_{u\text{M}}[\text{m}].
		\end{aligned}
	\end{equation}
}
In our example case, $B_{\text{ring}}=0.2$ T, $B_{0\text{M}}=1.13$ T, $\beta_{z\text{M}}=139\ \mu$m, $L_{u\text{M}}=1.6$ m, we have
\begin{equation}
	\Delta\epsilon_{z\text{M}}=14.3\ \text{pm},
\end{equation}
which is even larger than the desired 4 pm longitudinal emittance, and therefore is unacceptable.

The above evaluation of longitudinal emittance contribution means we need to use a weaker or shorter modulator. Since our desired longitudinal emittance is $\epsilon_{z}\lesssim 4$ pm, we need to control the contribution from two modulators to be $\Delta\epsilon_{z\text{M}}\lesssim1$~pm since the ring dipoles will also contribute longitudinal emittance with a theoretical minimum about 2~pm.  For example, we may choose to weaken the modulator field by more than a factor of two. If $\lambda_{u\text{M}}=0.15$ m and $B_{0\text{M}}=0.435$ T, $L_{u\text{M}}=1.5$ m, then the contribution of two modulators to longitudinal emittance is 
\begin{equation}
	\Delta\epsilon_{z\text{M}}=0.76\ \text{pm},
\end{equation}
which should be acceptable for a target total longitudinal emittance of 4 pm. But then to {introduce} the desired energy chirp strength, we now need $P_{L}\approx2$ GW.

Note that in this updated parameters choice, there is still one issue we need take care. In the evaluation of quantum excitation contribution of modulators to longitudinal emittance, we have implicitly assumed that the longitudinal beta function does not change inside it. This is not true strictly speaking. The undulator itself also has an $R_{56}=2N_{u}\lambda_{0}$, with $\lambda_{0}$ the fundamental resonance wavelength of the undulator and in our case is the modulation laser wavelength. And the criterion whether the thin-lens approximation applies is to evaluate whether or not $|hR_{56}|\ll1$, where $R_{56}$ is that of the undulator. Here in this updated example, we have
$
hR_{56}=h2N_{u}\lambda_L=0.9
$
for the modulator, which means that the thin-lens kick approximation actually does not apply here. So more {accurately} we should use the thick-lens map of the modulator~\cite{deng2024theoretical} to calculate the evolution of longitudinal beta function in the modulator, and then evaluate the contribution to longitudinal emittance. 


With all the subtle points carefully handled, LSF as analyzed above in principle can realize the desired nm bunch length and thus generate coherent EUV radiation. The main issue of such an EUV source is the required modulation laser power (GW level) is too high and makes the optical enhancement cavity can only work in a low duty cycle pulsed mode, thus limiting the average EUV output power.

\subsection{Generalized Longitudinal Strong Focusing}
The previous analysis of LWF and LSF leads us to consider the generalized longitudinal strong focusing (GLSF) scheme~\cite{li2023GLSF}. The basic idea of GLSF is to take advantage of the ultrasmall natural vertical emittance in a planar electron storage ring. More specifically, we will apply a partial transverse-longitudinal emittance exchange at the optical laser wavelength range to achieve efficient microbunching generation. As shown in Fig.~\ref{fig:GLSF}, the schematic setup of a GLSF ring is very similar to that of a LSF ring. But as stressed before that the energy chirp strength required in GLSF is much smaller than {that} in the LSF scheme, which means the required modulation laser power can also be smaller. A sharp reader may also notice in Fig.~\ref{fig:GLSF}, the longitudinal phase space area of beam is not conserved in the bunch compression or harmonic generation section of a GLSF ring. The fundamental physical law like Liouville's theorem of course cannot be violated in a symplectic system. The reason for this apparent “contradiction” is that GLSF invokes 4D or 6D phase space dynamics as summarized in Tab.~\ref{tab:Table1}, and what conserved {are} the eigen emittances, instead of the projected emittances.  One may also note that in the plot, the phase space rotation direction in the GLSF scheme is reversed after the radiator compared to that before the radiator, while in the LSF scheme this is not the case. In other words, in GLSF, we choose to make the upstream and downstream modulations cancel each other. In this sense, this setup is a special case of the reversible seeding scheme of SSMB~\cite{Ratner2011Reversible}. The reason of doing this is that we want to make the system transverse-longitudinal coupled only in a limited local region in the ring, the so called GLSF section, such that we can maintain $\mathcal{H}_{y}=0$ at the majority places of the ring to minimize the quantum excitation and IBS contribution to vertical emittance, thus keeping the small vertical emittance of a planar uncoupled ring. Further, this cancellation of nonlinear sinusoidal modulation waveforms will make the nonlinear dynamics of the ring easier to handle. To make the modulations perfectly cancel, we need the lattice between the upstream and down stream modulator to be an isochronous achromat. This reversible seeding setup makes the following decoupling of the system straightforward. All we need to do is to make the GLSF section an achromat, as the section from upstream modulator to downstream modulator is transparent to longitudinal dynamics. Another advantage of this reversible seeding setup is that it makes the bunch length at the modulator more flexible. It can be a short microbunned beam as shown in Fig.~\ref{fig:GLSF}.  It can also be a conventional RF-bunched beam, or even a coasting beam. Actually in our present design to be presented in Sec.~\ref{sec:application}, we actually use an RF-bunched beam in the ring. So the laser modulator in the plot is actually the RF system. Having explained the reason why we choose this reversible seeding setup for GLSF, we remind the readers that this however is not the only possible way to realize the GLSF scheme~\cite{li2023GLSF}. For example, a symmetric lattice {setup} with respect to the radiator is also possible, although the nonlinear dynamics might be challenging.

After this general introduction of GLSF scheme, we now appreciate in a more physical way why GLSF could be favored compared to LSF, in lowering the required modulation laser power{.} The key is that LSF has contribution of $\epsilon_{z}$ from both the LSF section and the ring dipoles, while GLSF has only contribution of $\epsilon_{y}$ from the GLSF section, since $\mathcal{H}_{y}$ outside the GLSF section is zero as just explained. This is the key physical argument from single-particle dynamics perspective that why GLSF may require a smaller energy modulation compared to LSF, to realize the same desired bunch length at the radiator. Actually if the longitudinal emittance in LSF is only from the quantum excitation of LSF modulators, and the vertical emittance in GLSF is only from the quantum excitation of GLSF modulators, then GLSF and LSF are actually equivalent in essence (only a factor of two different damping rate) concerning the requirement on energy modulation strength from single-particle dynamics perspective. 

Now we explain the above argument more clearly with using formulas. As we will show in the following section, more specifically Eq.~(\ref{eq:theorem1}), in GLSF at best case we have
\begin{equation}
	|h|=\frac{1}{\sqrt{\mathcal{H}_{y\text{R}}\mathcal{H}_{y\text{M}}}}=\frac{\sqrt{\epsilon_{y}/\mathcal{H}_{y\text{M}}}}{\sqrt{\epsilon_{y}\mathcal{H}_{y\text{R}}}}=\frac{\sqrt{\epsilon_{y}/\mathcal{H}_{y\text{M}}}}{\sigma_{z\text{R}}},
\end{equation}
where $\mathcal{H}_{y}=\beta_{55}^{II}$ is defined in Sec.~\ref{sec:formalism} and quantifies the contribution of vertical emittance to the bunch length. Note that we have used $\sigma_{z\text{R}}=\sqrt{\epsilon_{y}\mathcal{H}_{y\text{R}}}$, which means the bunch length at the radiator in GLSF scheme is solely determined by the beam vertical emittance.
One can appreciate the similarity of the above formula with Eq.~(\ref{eq:LSFh2}) for the case of LSF. Therefore, GLSF will be advantageous to LSF in lowering the required energy modulation strength if
\begin{equation}\label{eq:GLSFLSF}
	\frac{\epsilon_{y,\text{GLSF}}}{\mathcal{H}_{y\text{M},\text{GLSF}}}<\frac{\epsilon_{z,\text{LSF}}}{\beta_{z\text{M},\text{LSF}}}.
\end{equation}

Now we compare the two schemes in a more quantitative way. We assume that the two schemes work at the same beam energy. As we will show in Sec.~\ref{sec:emitContribution}, in a GLSF SSMB ring and if we consider only single-particle dynamics, the dominant contribution of vertical emittance is from the quantum excitation of two modulators in the GLSF section, and we have
\begin{equation}\label{eq:verticalEmitY}
	\begin{aligned}
		\epsilon_{y,\text{GLSF}}&\approx C_{q}\frac{\gamma^{2}}{J_{y}}\frac{1}{I_{2}}\times2\frac{\mathcal{H}_{y\text{M},\text{GLSF}}}{\rho_{0\text{M}}^{3}}\frac{4}{3\pi}L_{u\text{M},\text{GLSF}}.
	\end{aligned}
\end{equation}
While in LSF,  we assume that the longitudinal emittance is mainly from quantum excitation in the dipoles of the ring, and the average longitudinal beta function around the ring dipoles is the same as that at the modulators $\langle\beta_{z}\rangle\approx\beta_{z\text{M}}$. 
{Taking the approximation $I_{2}\approx\frac{2\pi}{\rho_{\text{ring}}}$ which means in both rings the radiation loss is mainly from the dipoles in the ring, $J_{z}\approx2$, $J_{x}\approx1$, and combining with Eq.~(\ref{eq:emitZLSF})}, Eq.~(\ref{eq:GLSFLSF}) then corresponds to 
\begin{equation}
	4\rho_{\text{ring},\text{GLSF}}\frac{\frac{4}{3\pi}L_{u\text{M},\text{GLSF}}}{\rho_{0\text{M},\text{GLSF}}^{3}}<\rho_{\text{ring},\text{LSF}}\frac{2\pi\rho_{\text{ring},\text{LSF}}}{\rho_{\text{ring},\text{LSF}}^{3}}.
\end{equation}
The above condition should be straightforward to fulfill in practice. For example, if the bending magnet strengths are the same in both schemes, i.e., if $\rho_{\text{ring},\text{GLSF}}=\rho_{\text{ring},\text{LSF}}=\rho_{\text{ring}}$, the above relation corresponds to
\begin{equation}
	L_{u\text{M},\text{GLSF}}<\frac{3\pi^{2}}{8}\frac{\rho_{0\text{M},\text{GLSF}}^{3}}{\rho_{\text{ring}}^{2}},
\end{equation} 
which is easy to satisfy in practice. So GLSF can be favored compared to LSF in lowering the required modulation laser power.

\subsection{Short Summary}

From the analysis in this section our tentative conclusion is: a LWF SSMB ring can be used to generate bunches with a  couple of 10 nm bunch length, thus to generate coherent visible and infrared radiation. If we want to push the bunch length to an even shorter range, the required phase slippage factor of the ring will be too small from an engineering viewpoint. A LSF SSMB ring can create bunches with a bunch length at nm level, thus to generate coherent EUV radiation. However, the required modulation laser power is in GW level, and makes the optical enhancement cavity can only work at a low duty cycle pulsed mode, thus limiting the average output EUV radiation power. At present, a GLSF SSMB ring is the most promising among these three schemes to realize nm bunch length with a smaller modulation laser power compared to LSF SSMB, thus allowing higher average power EUV radiation generation.

\section{Transverse-Longitudinal Coupling for Bunch Compression and Harmonic Generation}\label{sec:TLCTheorems}

In the following sections, we will go into more details of the GLSF scheme, more specifically we will investigate the backbone of  a GLSF SSMB storage ring, the transverse-longitudinal phase space coupling dynamics, in a systematic way. As the first step, in this section we present three theorems or inequalities that dictate such TLC-based bunch compression or harmonic generation schemes. If the initial bunch is longer than the modulation RF or laser  wavelength, then compression of bunch or microbunch can just be viewed as a harmonic generation scheme. Therefore, in this paper, we will treat bunch compression and harmonic generation as the same thing in essence. We remind the readers that the theorems presented here are the generalization of that presented in Refs.~\cite{deng2024theoretical,Deng2021NIMA} from 4D phase space to 6D phase space. The proof is basically the same. These formal mathematical relations will be useful in our later more detailed {study} for {a} GLSF SSMB light source.

\begin{figure}[tb]
	\centering 
	\includegraphics[width=1\textwidth]{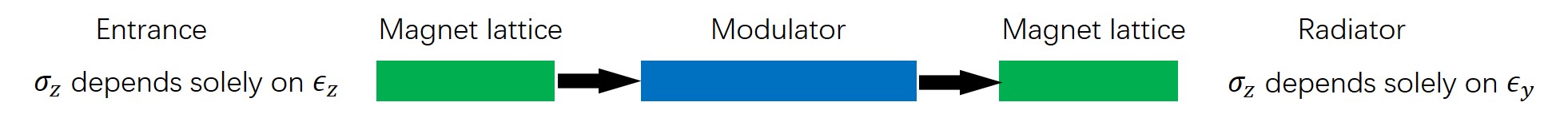}
	\caption{
		\label{fig:Chap3-TLC} 
		A schematic layout of applying TLC dynamics for bunch compression. (Figure adapted from Ref.~\cite{Deng2021NIMA}).}
\end{figure} 

\subsection{Problem Definition}\label{sec:TLCproblemdefinition}
Let us first define the problem we are trying to solve.  We assume $\epsilon_{y}$ is the small  eigen emittance we want to exploit. 
The case of using $\epsilon_{x}$ is similar.  The schematic layout of a TLC-based bunch compression section is shown in Fig.~\ref{fig:Chap3-TLC}. Suppose the beam at the entrance of the bunch compression section is $x$-$y$-$z$ decoupled, with its second moments matrix given by

{}
	\begin{equation}
		\begin{aligned}
			{\bf \Sigma}_{i}=\langle{\bf X}{\bf X}^{T}\rangle_{i}=\left(\begin{matrix}
				\epsilon_{x}\beta_{xi}&-\epsilon_{x}\alpha_{xi}&0&0&0&0\\
				-\epsilon_{x}\alpha_{xi}&\epsilon_{x}{\gamma_{xi}}&0&0&0&0\\
				0&0&\epsilon_{y}\beta_{yi}&-\epsilon_{y}\alpha_{yi}&0&0\\
				0&0&-\epsilon_{y}\alpha_{yi}&\epsilon_{y}{\gamma_{yi}}&0&0\\
				0&0&0&0&\epsilon_{z}\beta_{zi}&-\epsilon_{z}\alpha_{zi}\\
				0&0&0&0&-\epsilon_{z}\alpha_{zi}&\epsilon_{z}{\gamma_{zi}}\\
			\end{matrix}\right),
		\end{aligned}
	\end{equation}
{}
where $\alpha$, $\beta$ and $\gamma$ are the Courant-Snyder functions, the subscript ${i}$ means initial, and $\epsilon_{x}$, $\epsilon_{y}$ and $\epsilon_{z}$ are the eigen emittances of the beam corresponding to the horizontal, vertical and longitudinal mode, respectively. Note that the eigen emittances are beam invariants with respect to linear symplectic transport. For the application of TLC for bunch compression, it means that the final bunch length at the exit or radiator $\sigma_{z\text{R}}$ depends only on the vertical emittance $\epsilon_{y}$, and {neither} on the horizontal one $\epsilon_{x}$ and {nor the} longitudinal one $\epsilon_{z}$.

We divide such a bunch compression section into three parts, with their symplectic transfer matrices given by
\begin{equation}\label{eq:3Steps}
	\begin{aligned}
		&{\bf M}_{1}=\left(\begin{matrix}
			r_{11}&r_{12}&r_{13}&r_{14}&0&r_{16}\\
			r_{21}&r_{22}&r_{23}&r_{24}&0&r_{26}\\
			r_{31}&r_{32}&r_{33}&r_{34}&0&r_{36}\\
			r_{41}&r_{42}&r_{43}&r_{44}&0&r_{46}\\
			r_{51}&r_{52}&r_{53}&r_{54}&1&r_{56}\\
			0&0&0&0&0&1\\
		\end{matrix}\right),\\ 
		&{\bf M}_{2}=\text{modulation kick map},\\ 
		&{\bf M}_{3}=\left(\begin{matrix}
			R_{11}&R_{12}&R_{13}&R_{14}&0&R_{16}\\
			R_{21}&R_{22}&R_{23}&R_{24}&0&R_{26}\\
			R_{31}&R_{32}&R_{33}&R_{34}&0&R_{36}\\
			R_{41}&R_{42}&R_{43}&R_{44}&0&R_{46}\\
			R_{51}&R_{52}&R_{53}&R_{54}&1&R_{56}\\
			0&0&0&0&0&1\\
		\end{matrix}\right),\\
	\end{aligned}
\end{equation}	
where ${\bf M}_{1}$ representing ``from entrance to modulator", ${\bf M}_{2}$ representing ``modulation kick" and ${\bf M}_{3}$ representing ``modulator to radiator". 
Note that ${\bf M}_{1}$ and ${\bf M}_{3}$ are in their general thick-lens form, and does not necessarily need to be $x$-$y$ decoupled. The transfer matrix from the entrance to the radiator is then
\begin{equation}
	{{\bf O}={\bf M}_{3}{\bf M}_{2}{\bf M}_{1}.}
\end{equation}	
From the problem definition, for $\sigma_{z\text{R}}$ to be independent of $\epsilon_{x}$ and $\epsilon_{z}$, 
we need
\begin{equation}\label{eq:bunchcompressioncondition2}
	\begin{aligned}
		O_{51}&=0,\
		O_{52}=0,\
		O_{55}=0,\
		O_{56}= 0.
	\end{aligned}
\end{equation}

\subsection{Theorems and Proof}
\subsubsection{Theorems}
Given the above problem definition, and assuming the modulation kick map ${\bf M}_{2}$ is a thin-lens one, we have three theorems which dictate the relation between the modulator kick strength with the optical functions at the modulator and radiator, respectively.\\
{\bf Theorem one:} If 
\begin{equation}\label{eq:M21}
	{\bf M}_{2}=\left(\begin{matrix}
		1&0&0&0&0&0\\
		0&1&0&0&0&0\\
		0&0&1&0&0&0\\
		0&0&0&1&0&0\\
		0&0&0&0&1&0\\
		0&0&0&0&h&1\\
	\end{matrix}\right),
\end{equation}
which corresponds to the case of a normal RF or a TEM00 mode laser modulator,
then 
\begin{equation}\label{eq:theorem1}
	h^2\mathcal{H}_{y\text{M}}\mathcal{H}_{y\text{R}}\geq1,
\end{equation}
where the subscript M and R represent the place of modulator and radiator, respectively.\\
{\bf Theorem two:} If 
\begin{equation}\label{eq:M22}
	{\bf M}_{2}=\left(\begin{matrix}
		1&0&0&0&0&0\\
		0&1&0&0&0&0\\
		0&0&1&0&0&0\\
		0&0&0&1&{g}&0\\
		0&0&0&0&1&0\\
		0&0&{g}&0&0&1\\
	\end{matrix}\right),
\end{equation}
which corresponds to the case of a transverse deflecting  (in $y$-dimension) RF or a TEM01 mode laser modulator or other schemes for angular modulation,
then 
\begin{equation}\label{eq:theorem2}
	{g}^2\beta_{y\text{M}}\mathcal{H}_{y\text{R}}\geq1.
\end{equation}
{\bf Theorem three:} If 
\begin{equation}\label{eq:M23}
	{\bf M}_{2}=\left(\begin{matrix}
		1&0&0&0&0&0\\
		0&1&0&0&0&0\\
		0&0&1&0&{q}&0\\
		0&0&0&1&0&0\\
		0&0&0&0&1&0\\
		0&0&0&-{q}&0&1\\
	\end{matrix}\right),
\end{equation}
then 
\begin{equation}\label{eq:theorem3}
	{q}^2\gamma_{y\text{M}}\mathcal{H}_{y\text{R}}\geq1.
\end{equation}
{There is no commonly used single element realizing the kick map Eq.~(\ref{eq:M23}) directly. Instead, it takes the combination of several elements to realize such a map, making its application less straightforward compared to the cases correspond to Theorem one and two.}

\subsubsection{Proof}\label{sec:proof}
Here we present the details for the proof of Theorem one. The proof of the other two is just similar. From the problem definition, for $\sigma_{z\text{R}}$ to be independent of $\epsilon_{x}$ and $\epsilon_{z}$, 
we need
{}
	\begin{equation}\label{eq:TLCCondition2}
		\begin{aligned}
			O_{51}&=r_{11} R_{51}+r_{21} R_{52}+r_{31} R_{53}+r_{41} R_{54}+r_{51} \left(h R_{56}+1\right)=0,\\
			O_{52}&=r_{12} R_{51}+r_{22} R_{52}+r_{32} R_{53}+r_{42} R_{54}+r_{52} \left(h R_{56}+1\right)=0,\\
			O_{55}&= h R_{56}+1=0,\\
			O_{56}&= r_{16} R_{51}+r_{26} R_{52}+r_{36} R_{53}+r_{46} R_{54}+r_{56} \left(h R_{56}+1\right)+R_{56}=0.
		\end{aligned}
	\end{equation}
	Under the above conditions, we have
	\begin{equation}\label{eq:TransportMatrix12}
		\begin{aligned}
			&{\bf O}=\left(
			\begin{matrix}
				{\bf A}&{\bf B}&{\bf C}\\
				{\bf D}&{\bf E}&{\bf F}\\
				{\bf G}&{\bf H}&{\bf I}
			\end{matrix}
			\right),
		\end{aligned}
	\end{equation}
	with ${\bf A}\sim{\bf I}$ being $2\times2$ submatrices of ${\bf T}$, and
	\begin{equation}\label{eq:TransportMatrixSub2}
		\begin{aligned}
			&{\bf G}=\left(\begin{matrix}
				0&0\\
				r_{51}h &r_{52}h\\
			\end{matrix}\right),\\
			&{\bf H}=\left(\begin{matrix}
				r_{13} R_{51}+r_{23} R_{52}+r_{33} R_{53}+r_{43} R_{54}&r_{14} R_{51}+r_{24} R_{52}+r_{34} R_{53}+r_{44} R_{54}\\
				r_{53}h &r_{54}h\\
			\end{matrix}\right),\\
			&{\bf I}=\left(\begin{matrix}
				0&0\\
				h&r_{56}h +1\\
			\end{matrix}\right).
		\end{aligned}
	\end{equation}
	Note that ${\bf I}$ in this subsection does not mean the identity matrix. The bunch length squared at the modulator and the radiator are
	\begin{equation}\label{eq:BL2}
		\begin{aligned}
			\sigma_{z\text{M}}^{2}
			&=\epsilon_{x}\frac{\left(\beta_{xi}r_{51}-\alpha_{xi}r_{52}\right)^2+r_{52}^2}{\beta_{xi}}+\epsilon_{y}\frac{\left(\beta_{yi}r_{53}-\alpha_{yi}r_{54}\right)^2+r_{54}^2}{\beta_{yi}}+\epsilon_{z}\left(\beta_{zi}-2\alpha_{zi}r_{56}+\gamma_{zi}r_{56}^2\right)\\
			&=\epsilon_{x}\mathcal{H}_{x\text{M}}+\epsilon_{y}\mathcal{H}_{y\text{M}}+\epsilon_{z}\beta_{z\text{M}},\\
			\sigma_{z\text{R}}^{2}
			&=\epsilon_{y}\frac{\left(\beta_{yi}O_{53}-\alpha_{yi}O_{54}\right)^2+O_{54}^2}{\beta_{yi}}
			=\epsilon_{y}\mathcal{H}_{y\text{R}}.
		\end{aligned}
	\end{equation}
	According to the Cauchy-Schwarz inequality, we have
	\begin{equation}\label{eq:proof}
		\begin{aligned}
			h^2\mathcal{H}_{y\text{M}}\mathcal{H}_{y\text{R}}&=h^2\frac{\left[\left(\beta_{yi}r_{53}-\alpha_{yi}r_{54}\right)^2+r_{54}^2\right]}{\beta_{yi}}\frac{\left[\left(\beta_{yi}O_{53}-\alpha_{yi}O_{54}\right)^2+O_{54}^2\right]}{\beta_{yi}}\\
			&\geq
			\frac{h^2}{\beta_{yi}^{2}}\left[-\left(\beta_{yi}r_{53}-\alpha_{yi}r_{54}\right)O_{54}+r_{54}\left(\beta_{yi}O_{53}-\alpha_{yi}O_{54}\right)\right]^{2}\\
			&=\left(O_{53}r_{54}h-O_{54}r_{53}h\right)^{2}=\left(O_{53}O_{64}-O_{54}O_{63}\right)^{2}=|\text{Det}({\bf H})|^{2},
		\end{aligned}
	\end{equation}
	where Det() means the determinant of the matrix.
	The equality holds when
	$
	\frac{-\left(\beta_{yi}r_{53}-\alpha_{yi}r_{54}\right)}{O_{54}}=\frac{r_{54}}{\left(\beta_{yi}O_{53}-\alpha_{yi}O_{54}\right)}.
	$
	The symplecticity of ${\bf O}$ requires that ${\bf O}{\bf S}{\bf O}^{T}={\bf S}$, where ${\bf S}=\left(
	\begin{matrix}
	{\bf J}&0&0\\
	0&{\bf J}&0\\
	0&0&{\bf J}
	\end{matrix}
	\right)$ and $
	{\bf J}=\left(
	\begin{matrix}
	0&1\\
	-1&0
	\end{matrix}
	\right),$ so we have
	\begin{equation}
		\left(
		\begin{matrix}
			{\bf A}{\bf J}{\bf A}^{T}+{\bf B}{\bf J}{\bf B}^{T}+{\bf C}{\bf J}{\bf C}^{T}&{\bf A}{\bf J}{\bf D}^{T}+{\bf B}{\bf J}{\bf E}^{T}+{\bf C}{\bf J}{\bf F}^{T}&{\bf A}{\bf J}{\bf G}^{T}+{\bf B}{\bf J}{\bf H}^{T}+{\bf C}{\bf J}{\bf I}^{T}\\
			{\bf D}{\bf J}{\bf A}^{T}+{\bf E}{\bf J}{\bf B}^{T}+{\bf F}{\bf J}{\bf C}^{T}&{\bf D}{\bf J}{\bf D}^{T}+{\bf E}{\bf J}{\bf E}^{T}+{\bf F}{\bf J}{\bf F}^{T}&{\bf D}{\bf J}{\bf G}^{T}+{\bf E}{\bf J}{\bf H}^{T}+{\bf F}{\bf J}{\bf I}^{T}\\
			{\bf G}{\bf J}{\bf A}^{T}+{\bf H}{\bf J}{\bf B}^{T}+{\bf I}{\bf J}{\bf C}^{T}&{\bf G}{\bf J}{\bf D}^{T}+{\bf H}{\bf J}{\bf E}^{T}+{\bf I}{\bf J}{\bf F}^{T}&{\bf G}{\bf J}{\bf G}^{T}+{\bf H}{\bf J}{\bf H}^{T}+{\bf I}{\bf J}{\bf I}^{T}\\
		\end{matrix}
		\right)
		=
		{\bf S}.
	\end{equation}
{}
According to Eq.~(\ref{eq:TransportMatrixSub2}), we have
\begin{equation}
	{\bf G}{\bf J}{\bf G}^{T}=\left(
	\begin{matrix}
		0&0\\
		0&0
	\end{matrix}
	\right),\ {\bf I}{\bf J}{\bf I}^{T}=\left(
	\begin{matrix}
		0&0\\
		0&0
	\end{matrix}
	\right).
\end{equation} 
Therefore,
\begin{equation}
	{\bf H}{\bf J}{\bf H}^{T}={\bf J},
\end{equation}
which means ${\bf H}$ is also a symplectic matrix. So we have
$\text{Det}({\bf H})=1.
$
The theorem is thus proven.

\subsection{Dragt's Minimum Emittance Theorem}

Theorem one in Eq.~(\ref{eq:theorem1}) can also be expressed as
\begin{equation}\label{eq:BLinequality}
	|h|\geq\frac{\epsilon_{y}}{\sqrt{\epsilon_{y}\mathcal{H}_{y\text{M}}}\sqrt{\epsilon_{y}\mathcal{H}_{y\text{R}}}}=\frac{\epsilon_{y}}{\sigma_{zy\text{M}}\sigma_{z\text{R}}}.
\end{equation}
Note that in the above formula,  $\sigma_{zy\text{M}}$ means the bunch length at the modulator contributed from the vertical emittance $\epsilon_{y}$. So given a fixed $\epsilon_{y}$ and desired $\sigma_{z\text{R}}$, a smaller $h$, i.e., a smaller RF acceleration gradient or modulation laser power ($P_{L}\propto|h|^2$), means a larger $\mathcal{H}_{y\text{M}}$, thus a {larger} $\sigma_{zy\text{M}}$, is needed.   As $|h|\sigma_{z\text{M}}$ quantifies the energy spread introduced by the modulation kick, we thus also have 
\begin{equation}\label{eq:ES}
	\sigma_{z\text{R}}\sigma_{\delta\text{R}}\geq\epsilon_{y}.
\end{equation}
Similarly for Theorem two and three, we have
\begin{equation}
	\begin{aligned}
		|{g}|&\geq\frac{\epsilon_{y}}{\sigma_{y\beta\text{M}}\sigma_{z\text{R}}},
	\end{aligned}
\end{equation}
and 
\begin{equation}
	\begin{aligned}
		|{q}|&\geq\frac{\epsilon_{y}}{\sigma_{y'\beta\text{M}}\sigma_{z\text{R}}},
	\end{aligned}
\end{equation}
respectively, and also Eq.~(\ref{eq:ES}). Note that in the above formulas, the vertical beam size or divergence at the modulator contains only the vertical betatron part, i.e., that from the vertical emittance $\epsilon_{y}$.

Equation~(\ref{eq:ES}) is actually a manifestation of the classical uncertainty principle \cite{Dragt2020}, which states that 
\begin{equation}
	\begin{aligned}
		\Sigma_{11}\Sigma_{22}&\geq\epsilon_{\text{min}}^{2},\\
		\Sigma_{33}\Sigma_{44}&\geq\epsilon_{\text{min}}^{2},\\
		\Sigma_{55}\Sigma_{66}&\geq\epsilon_{\text{min}}^{2},
	\end{aligned}
\end{equation}
in which $\epsilon_{\text{min}}$ is the minimum one among the three eigen emittances $\epsilon_{I,II,III}$. In our bunch compression case, we assume that $\epsilon_{y}$ is the smaller one compared to $\epsilon_{z}$. 

Actually there is a stronger inequality compared to the classical uncertainty principle, i.e., the minimum emittance theorem \cite{Dragt2020}, which states that the projected emittance cannot be smaller than the minimum one among the three eigen emittances,
\begin{equation}
	\begin{aligned}
		\epsilon_{x,\text{pro}}^{2}=\Sigma_{11}\Sigma_{22}-\Sigma_{12}^{2}&\geq\epsilon_{\text{min}}^{2},\\
		\epsilon_{y,\text{pro}}^{2}=\Sigma_{33}\Sigma_{44}-\Sigma_{34}^{2}&\geq\epsilon_{\text{min}}^{2},\\
		\epsilon_{z,\text{pro}}^{2}=\Sigma_{55}\Sigma_{66}-\Sigma_{56}^{2}&\geq\epsilon_{\text{min}}^{2}.
	\end{aligned}
\end{equation} 

\subsection{Theorems Cast in Another Form}
As another way to appreciate the result, here we cast the theorems in a form using the generalized beta functions as introduced in Sec.~\ref{sec:formalism}. According to definition, we have
\begin{equation}
	\beta_{y}\equiv\beta_{33}^{II},\ \gamma_{y}\equiv\beta_{44}^{II},\ \mathcal{H}_{y}\equiv\beta_{55}^{II}.
\end{equation}
{\bf Theorem one:} If ${\bf M}_{2}$ is as shown in Eq.~(\ref{eq:M21}),
then 
\begin{equation}\label{eq:theorem12}
	M_{65}^2(\text{Mod})\beta_{55}^{II}(\text{Mod})\beta_{55}^{II}(\text{Rad})\geq1,
\end{equation}
where $M_{2,65}$ is the $_{65}$ matrix term of ${\bf M}_{2}$, i.e., $h$. For better visualization, in this subsection, we use brackets to denote the location, with Ent, Mod and Rad meaning entrance, modulator and radiator, respectively.\\
{\bf Theorem two:} If ${\bf M}_{2}$ is as shown in Eq.~(\ref{eq:M22}),
then 
\begin{equation}\label{eq:theorem22}
	M_{2,63}^2(\text{Mod})\beta_{33}^{II}(\text{Mod})\beta_{55}^{II}(\text{Rad})\geq1.
\end{equation}
{\bf Theorem three:} If ${\bf M}_{2}$ is as shown in Eq.~(\ref{eq:M23}),
then 
\begin{equation}\label{eq:theorem32}
	M_{2,64}^2(\text{Mod})\beta_{44}^{II}(\text{Mod})\beta_{55}^{II}(\text{Rad})\geq1.
\end{equation}
At the entrance, the generalized Twiss matrix corresponding to eigen mode $I$ is
\begin{equation}
	{\bf T}_{I}(\text{Ent})=\left(\begin{matrix}
		\beta_{xi}&-\alpha_{xi}&0&0&0&0\\
		-\alpha_{xi}&\gamma_{xi}&0&0&0&0\\
		0&0&0&0&0&0\\
		0&0&0&0&0&0\\
		0&0&0&0&0&0\\
		0&0&0&0&0&0\\
	\end{matrix}\right),
\end{equation}
and similar expressions for ${\bf T}_{II,III}(\text{Ent})$, with $x$ replaced by $y,z$ and the location of the $2\times2$ matrix shifted in the diagonal direction.
Then
\begin{equation}
	\beta_{33}^{II}(\text{Mod})=\frac{\left(\beta_{yi}r_{33}-\alpha_{yi}r_{34}\right)^2+r_{34}^2}{\beta_{yi}},
\end{equation}
\begin{equation}
	\beta_{44}^{II}(\text{Mod})=\frac{\left(\beta_{yi}r_{43}-\alpha_{yi}r_{44}\right)^2+r_{44}^2}{\beta_{yi}},
\end{equation}
\begin{equation}
	\beta_{55}^{II}(\text{Mod})=\frac{\left(\beta_{yi}r_{53}-\alpha_{yi}r_{54}\right)^2+r_{54}^2}{\beta_{yi}},
\end{equation}
\begin{equation}
	\beta_{55}^{I}(\text{Rad})=\frac{\left(\beta_{xi}O_{51}-\alpha_{xi}O_{52}\right)^2+O_{52}^2}{\beta_{xi}},
\end{equation}
\begin{equation}
	\beta_{55}^{II}(\text{Rad})=\frac{\left(\beta_{yi}O_{53}-\alpha_{yi}O_{54}\right)^2+O_{54}^2}{\beta_{yi}},
\end{equation}
\begin{equation}
	\beta_{55}^{III}(\text{Rad})=\frac{\left(\beta_{zi}O_{55}-\alpha_{zi}O_{56}\right)^2+O_{56}^2}{\beta_{zi}}.
\end{equation}
For $\sigma_{z\text{R}}$ to be independent of $\epsilon_{x}$ and $\epsilon_{z}$, we need $\beta_{55}^{I}(\text{Rad})=0$ and $\beta_{55}^{III}(\text{Rad})=0$, which then lead to Eq.~(\ref{eq:bunchcompressioncondition2}). And the following proof procedures are the same as that shown in the above Sec.~\ref{sec:proof}.

\section{Energy Modulation-Based Coupling Schemes}\label{sec:TLCE}

After introducing the three formal theorems, now we conduct more detailed analysis of the TLC-based bunch compression or microbunching generation schemes. We group these schemes into two categories, i.e., energy modulation-based and angular modulation-based schemes. They corresponds to the case of Theorem One and Two presented in last section. In this section, we focus on energy modulation-based schemes, and next section is dedicated to angular modulation-based schemes. The physical realization corresponds to the case of Theorem Three is not that straightforward compared to the cases of Theorem One and Two, and we do not expand its discussion in this paper.

\subsection{Form Function and Bunching Factor}\label{sec:BF}

\subsubsection{General Formula}
For coherent radiation generation, a parameter of vital important is the bunching factor of the electron beam. Here first we derive the bunching factor of {the} energy-modulation based TLC microbunching schemes. The mathematical model is formulated as follows.\\
6D particle state vector: 
\begin{equation}
	{\bf X}\equiv\left(
	\begin{matrix}
		x&
		x'& 
		y&
		y'&
		z&
		\delta
	\end{matrix}
	\right)^{T}.
\end{equation}
6D spectral vector: 
\begin{equation}
	{\bf K}\equiv\left(
	\begin{matrix}
		k_{x}&
		k_{x'}&
		k_{y}&
		k_{y'}&
		k_{z}&
		k_{\delta}
	\end{matrix}
	\right).
\end{equation}
Normalized particle density distribution in phase space $\psi({\bf X})$: 
\begin{equation}
	\int\psi({\bf X})d{\bf X}=1,\ \psi({\bf X})\geq0,
\end{equation}
where $\int d{\bf X}$ means $\int_{-\infty}^{\infty}\int_{-\infty}^{\infty}\int_{-\infty}^{\infty}\int_{-\infty}^{\infty}\int_{-\infty}^{\infty}\int_{-\infty}^{\infty}dxdx'dy\\dy'dzd\delta$. 
Here we introduce the form function (FF) of beam as:
\begin{equation}
	\begin{aligned}
		&\mathcal{F}({\bf K})\equiv\int\psi({\bf X})e^{-i{\bf K}{\bf X}}d{\bf X}.
	\end{aligned}
\end{equation}
$\psi({\bf X})$ and $\mathcal{F}({\bf K})$ then forms a Fourier transform pair
\begin{equation}
	\psi({\bf X})=\frac{1}{2\pi}\int \mathcal{F}({\bf K})e^{i{\bf K}{\bf X}}d{\bf K},
\end{equation}
where $\int d{\bf K}$ means $\int_{-\infty}^{\infty}\int_{-\infty}^{\infty}\int_{-\infty}^{\infty}\int_{-\infty}^{\infty}\int_{-\infty}^{\infty}\int_{-\infty}^{\infty}dk_xdk_{x'}\\dk_ydk_{y'}dk_zdk_\delta$. This form function is another complete description of beam distribution and can offer complementary insight for beam dynamics study. More details on this respect will be reported elsewhere.

The classical 1D bunching factor or form factor used in literature is a specific point in our defined FF, i.e., with ${\bf K}=(0,0,0,0,k_{z},0)$,
\begin{equation}\label{eq:BF}
	b(k_{z})=\mathcal{F}(0,0,0,0,k_{z},0)=\int_{-\infty}^{\infty}\psi(z)e^{-ik_{z}z}dz,
\end{equation}
where $\psi(z)=\int_{-\infty}^{\infty}\int_{-\infty}^{\infty}\int_{-\infty}^{\infty}\int_{-\infty}^{\infty}\int_{-\infty}^{\infty}\psi({\bf X})dxdx'dydy'd\delta$ is the normalized longitudinal distribution of particles. Here in this paper, we will use $\mathcal{F}$ to denote the 6D form function, and $b$ the classical 1D bunching factor.

Now we derive the form function and bunching factor for a single-stage energy modulation based microbunching schemes.  {A lumped description of the laser-induced energy modulation can be written as}:
\begin{equation}
	\delta=\delta+A\sin(k_{L}z),
\end{equation}
with $k_{L}=\frac{2\pi}{\lambda_{L}}$ the laser wavenumber and $A$ is the modulation strength. After the modulation, the particle state vector evolves according to:
\begin{equation}
	\begin{aligned}
		{{\bf X}_{f}={\bf R}{\bf X}_{i}},
	\end{aligned}
\end{equation}
where ${\bf R}$ is the linear symplectic transfer matrix of the magnet lattice, which could be a single-pass one like a linear accelerator or a multi-pass one like a storage ring. In this paper we only consider the case that the magnet lattice is linear. 
Denote: 
\begin{equation}\label{eq:Mp}
	\begin{aligned}
		{\bf A}&\equiv(0,0,0,0,0,A)^{T},\\
		{\bf U}_{p}&\equiv\left(
		\begin{matrix}
			0&
			0&
			0&
			0&
			pk_{L}&
			0
		\end{matrix}
		\right),\\
		{\bf M}_{p}&\equiv{\bf K}{\bf R}-{\bf U}_{p},
	\end{aligned}
\end{equation}
{with $p$ being an integer.
	Then the final FF is
	\begin{equation}
		\begin{aligned}
			\mathcal{F}({\bf K})&=\int\psi_{f}({\bf X})e^{-i{\bf K}{\bf X}}d{\bf X}\\
			&=\int\psi_{m+}({\bf X})e^{-i{\bf KR}{\bf X}}d{\bf X}\\
			&=\int\psi_{0}({\bf X})e^{-i\left({\bf KR}{\bf X}+{\bf KRA}\sin(k_{L}z)\right)}d{\bf X}\\
			&=\sum_{p=-\infty}^{\infty}J_{p}\left(-{\bf KRA}\right)\int\psi_{0}({\bf X})e^{-i\left({\bf KR}{\bf X}-pk_{L}z\right)}d{\bf X}\\
			&=\sum_{p=-\infty}^{\infty}J_{p}\left(-{\bf KRA}\right)\int\psi_{0}({\bf X})e^{-i{\bf M}_{p}{\bf X}}d{\bf X}\\
			&=\sum_{p=-\infty}^{\infty}J_{p}\left(-{\bf KRA}\right)\mathcal{F}_{0}({\bf M}_{p}),
		\end{aligned}
	\end{equation}
	where $\psi_{0}({\bf X})$, $\psi_{m+}({\bf X})$, and $\psi_{f}({\bf X})$ mean the beam distribution at the beginning, right after the energy modulation and the final point, respectively. $J_{p}$ is the $p$-th order Bessel function of the first kind. Jacobi-Anger identity $e^{ix\sin y}=\sum_{n=-\infty}^{\infty}e^{iny}J_{n}[x]$ has been used in the above derivation. Note that we have also used the fact that for a symplectic matrix ${\bf R}$ we have $\text{Det}\left({\bf R}\right)=1$. $\mathcal{F}_{0}({\bf K})$ the initial FF.}

The above formula is general and applies for arbitrary initial beam distribution.  If the initial beam distribution is Gaussian in 6D phase space:
\begin{equation}\label{eq:Gaussian}
	\begin{aligned}
		\psi_{0}({\bf X})&=\frac{1}{(2\pi)^{3}\sqrt{\text{Det}({\bf \Sigma_{0}})}}\text{exp}\left(-\frac{1}{2}{\bf X}^{T}{\bf \Sigma_{0}^{-1}}{\bf X}\right)
	\end{aligned}
\end{equation}
with ${\bf \Sigma}_{0}$ being the initial second moments of the beam, the initial FF is then 
\begin{equation}\label{eq:FFGaussian}
	\mathcal{F}_{0}({\bf K})=\text{exp}\left(-\frac{{\bf K}{\bf \Sigma}_{0}{\bf K}^{T}}{2}\right).
\end{equation}
The final FF is then
\begin{equation}\label{eq:BFGeneral}
	\begin{aligned}
		\mathcal{F}({\bf K})
		&=\sum_{p=-\infty}^{\infty}J_{p}\left(-{\bf KRA}\right)\text{exp}\left(-\frac{{\bf M}_{p}{\bf \Sigma}_{0}{\bf M}_{p}^{T}}{2}\right),
	\end{aligned}
\end{equation}

\subsubsection{HGHG}

In many applications, the classical 1D bunching factor suffices. With
$
{\bf K}=\left(
\begin{matrix}
0&
0&
0&
0&
k_{z}&
0
\end{matrix}
\right)$,
then
\begin{equation}
	\begin{aligned}
		{\bf KRA}&=k_{z}R_{56}A,\\
		{\bf M}_p&=k_{z}\left(
		\begin{matrix}
			R_{51}&
			R_{52}&
			R_{53}&
			R_{54}&
			R_{55}-\frac{pk_{L}}{k_{z}}&
			R_{56}
		\end{matrix}
		\right).
	\end{aligned}
\end{equation}
If further $R_{51}=0$, $R_{52}=0$, $R_{53}=0$, $R_{54}=0$ and $R_{55}=1$, and the initial beam is transverse-longitudinal decoupled and upright in the longitudinal phase space, which corresponds to the case of high-gain harmonic-generation (HGHG)~\cite{Yu1991}, then 
\begin{equation}
	\begin{aligned}
		b(k_{z})&=\sum_{p=-\infty}^{\infty}J_{p}\left(-k_{z}R_{56}A\right)\\
		&\text{exp}\left(-\frac{k_{z}^{2}}{2}\left[\left(1-\frac{pk_{L}}{k_{z}}\right)^{2}\sigma_{z0}^{2}+\left(R_{56}\sigma_{\delta0}\right)^{2}\right]\right),
	\end{aligned}
\end{equation}
where $\sigma_{z0}$ and $\sigma_{\delta0}$ are the initial RMS bunch length and energy spread, respectively.
If the initial bunch length is much longer than the laser wavelength, i.e., $k_{L}\sigma_{z0}\gg1$, the above exponential terms will be non-zero only when $k_{z}=pk_{L}$, which means there is only bunching at the laser  harmonics. IN this case, we have the bunching factor at the $n$-th ($n$ being integer) laser harmonic 
\begin{equation}\label{eq:HGHG}
	\begin{aligned}
		b_{n}&=b(k_{z}=nk_{L})\\
		&=J_{n}(-nk_{L}{R}_{56}A)\text{exp}\left[-\frac{\left(nk_{L}R_{56}\sigma_{\delta0}\right)^{2}}{2}\right].
	\end{aligned}
\end{equation}
For $n>4$, the maximal value of the Bessel function
$J_{n}$ is about $0.67/n^{1/3}$ and is achieved when its
argument is equal to $n+0.81n^{1/3}$. For large $n$, this argument corresponds to $k_{L}{R}_{56}A\sim1$. Then to make sure the exponential term in Eq.~(\ref{eq:HGHG}) not too small,  we need $A\sim n\sigma_{\delta0}$. So generally if we want to realize $n$-th harmonic bunching in HGHG, we need an energy modulation strength a factor of $n$ larger than the initial energy spread. 


\subsubsection{TLC-based Microbunching}


Now let us consider the case of nonzero {$R_{51,52,53,54}$}, which corresponds to transverse-longitudinal coupling (TLC)-based microbunching. {Below} we use $y$-$z$ coupling as an example for the analysis and ignore the $x$-dimension. The analysis for $x$-$z$ coupling is similar. If
$
{\bf K}\equiv\left(
\begin{matrix}
0&
0&
0&
0&
k_{z}&
0
\end{matrix}
\right)$, $R_{51}=0$, $R_{52}=0$, $R_{55}=1$,  $R_{66}=1$, then 
\begin{equation}
	\begin{aligned}
		{\bf KRA}&=k_{z}R_{56}A,\\
		{\bf M}_p&=k_{z}\left(
		\begin{matrix}
			0&
			0&
			R_{53}&
			R_{54}&
			1-\frac{pk_{L}}{k_{z}}&
			R_{56}
		\end{matrix}
		\right).
	\end{aligned}
\end{equation}
Using the real and imaginary generalized beta functions and Twiss matrices introduced in Sec.~\ref{sec:formalism}, the bunching factor at the $n$-th laser harmonic can be expressed as
{}
	\begin{equation}
		\begin{aligned}
			b_{n}&=\sum_{p=-\infty}^{\infty}J_{p}\left(-nk_{L}R_{56}A\right)\text{exp}\left[-\frac{\left(\epsilon_{II}{\bf M}_{p}{\bf T}_{II}{\bf M}_{p}^{T}+\epsilon_{III}{\bf M}_{p}{\bf T}_{III}{\bf M}_{p}^{T}\right)}{2} \right],
		\end{aligned}
	\end{equation}
	where ${\bf T}_{II,III}$ are the real generalized Twiss matrices right before the modulation.
	We then require
	\begin{equation}
		\begin{aligned}
			{\bf M}_{p}{\bf T}_{III}{\bf M}_{p}^{T}\bigg|_{p=n}&=(nk_{L})^{2}\left({\bf R}{\bf T}_{III}{\bf R}^{T}-2{\bf R}{\bf T}_{III}+{\bf T}_{III}\right)_{55}=0.
		\end{aligned}
	\end{equation}
	{The physical reason of this requirement is that for $p=n$, we want the longitudinal emittance does not contribute to the bunching factor. By doing this, the bunching factor will be mainly determined by the vertical emittance, which is assumed to be a small value. Therefore, we can realize high harmonic bunching with a shallow energy modulation strength $A$.}
	The above relation can be written more specifically using generalized beta functions and the matrix terms of ${\bf R}$ as
	\begin{equation}\label{eq:ADMcondition}
		R_{53}^2 \beta _{33}^{III}+R_{54}^2 \beta _{44}^{III}+R_{56}^2 \beta _{66}^{III}+2 R_{53}R_{54} \beta _{34}^{III}+2 R_{53}R_{56} \beta _{36}^{III}+2 R_{54} R_{56} \beta _{46}^{III}=0.
	\end{equation}
	If the generalized Twiss matrices of the eigenmode $II$ and $III$ right before the modulation are given by Eq.~(\ref{eq:Twiss}),
	and further assuming that at the modulation point we have $D_{x}=0$ and $D_{x}'=0$, then Eq.~(\ref{eq:ADMcondition}) can be casted into
	\begin{equation}
		R_{53} D_{y}+R_{54} D_{y}'+R_{56}=0.
	\end{equation}
	This relation means the final coordinate $z$ does not depend on the initial energy deviation $\delta$ {in linear approximation}.
	Under the above condition, we have
	\begin{equation}
		\begin{aligned}
			{\bf M}_{p}{\bf T}_{III}{\bf M}_{p}^{T}
			&=(nk_{L})^{2}\left(1-\frac{p}{n}\right)^{2} \beta _{z\text{M}},
		\end{aligned}
	\end{equation}
	and
	\begin{equation}
		\begin{aligned}
			&{\bf M}_{p}{\bf T}_{II}{\bf M}_{p}^{T}
			=(nk_{L})^{2}\left[\mathcal{H}_{y\text{R}}+\left(1-\frac{p}{n}\right)^{2}\mathcal{H}_{y\text{M}}\right.\\
			&\left.\ \ \ \ +2\left(1-\frac{p}{n}\right)\left(\gamma_{y}R_{54}D_{y}-\alpha_{y}R_{53}D_{y}+\alpha_{y}R_{54}D_{y}'-\beta_{y}R_{53}D_{y}'\right)\right].
		\end{aligned}
	\end{equation}
	Then the bunching factor at the $n$-th laser harmonic at the final point {which in our context means the radiator} is
	\begin{equation}\label{eq:BFTLC}
		\begin{aligned}
			b_{n}&=\sum_{p=-\infty}^{\infty}J_{p}\left(-nk_{L}R_{56}A\right)\text{exp}\left[-\frac{(nk_{L})^{2}}{2} \epsilon_{y}\mathcal{H}_{y\text{R}} \right]\text{exp}\left[-\frac{k_{L}^{2}}{2}\left(n-p\right)^{2} \left(\epsilon_{y}\mathcal{H}_{y\text{M}}+\epsilon_{z}\beta _{z\text{M}}\right) \right]\\
			&\ \ \ \ \text{exp}\left[-\frac{k_{L}^{2}}{2}2n\left(n-p\right)\epsilon_{y}\left(\gamma_{y}R_{54}D_{y}-\alpha_{y}R_{53}D_{y}+\alpha_{y}R_{54}D_{y}'-\beta_{y}R_{53}D_{y}'\right) \right].\\
		\end{aligned}
	\end{equation}
{}
Here we remind the readers that there is an extra term in Eq.~(\ref{eq:BFTLC}) compared to the result in our previous publications in Refs.~\cite{deng2024theoretical,Deng2021NIMA}, i.e.,
\begin{equation}\label{eq:BFTLCOld}
	\begin{aligned}
		b_{n}&=\sum_{p=-\infty}^{\infty}J_{p}\left(-nk_{L}R_{56}A\right)\text{exp}\left[-\frac{(nk_{L})^{2}}{2} \epsilon_{y}\mathcal{H}_{y\text{R}} \right]\\
		&\ \ \ \ \text{exp}\left[-\frac{k_{L}^{2}}{2}\left(n-p\right)^{2} \left(\epsilon_{y}\mathcal{H}_{y\text{M}}+\epsilon_{z}\beta _{z\text{M}}\right) \right].
	\end{aligned}
\end{equation}
We conclude that the Eq.~(\ref{eq:BFTLC}) here  is rigorously more accurate. 

{If the modulation waveform is linear, according to Eq.~(\ref{eq:BL2}), we have the RMS bunch length at the modulator and radiator given by
	\begin{equation}
		\begin{aligned}
			\sigma_{z\text{M}}^{2}&=\epsilon_{z}\beta _{z\text{M}}+\epsilon_{y}\mathcal{H}_{y\text{M}},\\
			\sigma_{z\text{R}}^{2}&=\epsilon_{y}\mathcal{H}_{y\text{R}}.
		\end{aligned}
	\end{equation}
	So in this paper, we call $\sigma_{z\text{R}}=\sqrt{\epsilon_{y}\mathcal{H}_{y\text{R}}}$ the linear bunch length at the radiator.} We have proven in last section there is a fundamental inequality dictating the energy chirp strength $h$ and $\mathcal{H}_{y}$ at the modulator and radiator, respectively, i.e., $h^{2}\mathcal{H}_{y\text{M}}\mathcal{H}_{y\text{R}}\geq1$. Basically, given the vertical emittance and desired $\sigma_{z\text{R}}$, to lower the energy chirp strength, we need to lengthen the bunch at the modulator. If the initial bunch is shorter than the modulation laser wavelength, {considering the actual laser modulation waveform is sinusoidal,} then according to Eq.~(\ref{eq:BFTLC}), a bunch lengthening at the modulator means a bunching factor drop at the radiator as can be seen from the second exponential term. For more discussion on this point, {the readers are referred to} Refs.~\cite{deng2024theoretical, Deng2021NIMA}.

When $k_{L}^{2}\left(\epsilon_{y}\mathcal{H}_{y\text{M}}+\epsilon_{z}\beta _{z\text{M}}\right)\gg1$ which means the bunch length at the modulation point is much longer than the laser wavelength, then there is only one term non-vanishing in the above summation, i.e., the term with $p=n$. Then 
\begin{equation}\label{eq:BFADM}
	\begin{aligned}
		b_{n}&=J_{n}\left(-nk_{L}R_{56}A\right)\text{exp}\left[-\frac{(nk_{L})^{2}}{2}\epsilon_{y}\mathcal{H}_{y\text{R}} \right].
	\end{aligned}
\end{equation}


{Here we make a short comment that our derivation of bunching factor here, and also most of that found in literature,  neglects the collective interactions between the electron beam and the co-propagation electromagnetic field. Such a collective interaction may disturb the modulation performance. Although not much work on this subject, the interested readers may refer to a recent relevant work~\cite{TsaiNIMA2023}.}


\subsection{Modulation Strength}

After deriving the bunching factor, now we derive the formula of modulation strength, given the laser, electron and undulator parameters. This is a necessary work for quantitative analysis and comparison.

\subsubsection{A Normally Incident Laser}\label{sec:TEM00}
The most common method of imprinting energy modulation on an electron beam at the laser wavelength is to use a TEM00 mode laser to resonate with the electrons in an undulator. Below we use a planar undulator as the modulator. A helical undulator can also be applied for energy modulation, but since we want to preserve the ultrasmall vertical emittance, we need to avoid $x$-$y$ coupling as much as possible, and thus a planar undulator might be preferred.    
The electromagnetic field of a TEM00 mode Gaussian laser polarized in the horizontal plane  is~\cite{Chao2023FocusedLaser}
{}
	\begin{equation}\label{eq:TEM00}
		\left(
		\begin{matrix}
			E_{x}\\
			E_{y}\\
			E_{z}\\
			cB_{x}\\
			cB_{y}\\
			cB_{z}
		\end{matrix}
		\right)=
		E_{x0}e^{ik_{L}{s}-i\omega_{L} t+i\phi_{0}}\left(-iZ_{R}Q\right)\text{exp}\left[i\frac{k_{L}Q}{2}\left(x^{2}+y^{2}\right)\right]
		\left(
		\begin{matrix}
			1\\
			0\\
			-Qx\\
			-Q^{2}xy\\
			Q^{2}x^{2}-\frac{iQ}{k_{L}}+1\\
			-Qy
		\end{matrix}
		\right),
	\end{equation}
{}
where {$s$ is the longitudinal global path length variable, $t$ is the time variable}, $k_{L}={2\pi}/{\lambda_{L}}$ the wavenumber and $\lambda_{L}$ is the wavelength of the laser, $c$ is the speed of light in free space, $\omega_{L}=k_{L}c$, $Z_{R}={\pi w_{0}^{2}}/{\lambda_{L}}$ is the Rayleigh length, $w_{0}$ the beam waist radius, and 
\begin{equation}\label{eq:Q}
	Q=\frac{i}{Z_{R}(1+\frac{s}{Z_{R}})}
\end{equation}
{with $i$ here being the imaginary unit.}
{We remind the readers that it has been implicitly assumed that all the fields will take the real part of their complex expressions.}

The relationship between the peak electric field $E_{x0}$ and the laser peak power $P_{L}$ is given by
\begin{equation}\label{eq:laserE}
	P_{L}=\frac{E_{x0}^{2}Z_{R}\lambda_{L}}{4Z_{0}},
\end{equation}
in which $Z_{0}=376.73\ \Omega$ is the impedance of free space. 
The prescribed wiggling motion of electron in a planar undulator is
\begin{equation}\label{eq:undulatorxs}
	x(s)=\frac{K}{\gamma k_{u}}\sin(k_{u}s),
\end{equation}
with $\gamma$ the Lorentz factor, 
\begin{equation}\label{eq:K}
	K=\frac{eB_{0}\lambda_{u}}{2\pi m_{e}c}=0.934\cdot B_{0}[\text{T}]\cdot\lambda_{u}[\text{cm}]
\end{equation}  
the dimensionless undulator parameter, $B_{0}$ the peak magnetic field, $\lambda_{u}$ the undulator period, and $k_{u}={2\pi}/{\lambda_{u}}$ the undulator wavenumber.
The resonant condition of laser-electron interaction inside a planar undulator is 
\begin{equation}
	\lambda_{L}=\frac{1+\frac{K^{2}}{2}}{2\gamma^{2}}\lambda_{u}.
\end{equation}
From the prescribed motion we can calculate the electron horizontal and longitudinal velocity 
\begin{equation}\label{eq:prescribeMotion}
	\begin{aligned}
		v_{x}(s)&\approx\frac{\beta cK}{\gamma}\cos(k_{u}s),\\
		v_{z}(s)&=\sqrt{v^{2}-v_{x}^{2}}\
		\approx \bar{v}_{z} -\frac{cK^{2}}{4\gamma^{2}} \cos(2k_{u}s),
	\end{aligned}
\end{equation}
{with $\beta=\sqrt{1-\frac{1}{\gamma^{2}}}$,} and
\begin{equation}
	\bar{v}_{z}=c\left(1-\frac{1+K^2/2}{2\gamma^{2}}\right)
\end{equation}
the average longitudinal velocity of electron in the undulator.
Therefore, we have the longitudinal path length of electron as a function of time given by
\begin{equation}
	\begin{aligned}
		{s}(t)&\approx\bar{v}_{z}t-\frac{K^{2}}{8\gamma^{2}k_{u}}\sin(2k_{u}\bar{v}_{z}t).
	\end{aligned}
\end{equation}
Then 
\begin{equation}
	k_{L}s-\omega_{L} t\approx-k_{u}s-\chi\sin(2k_{u}s),
\end{equation}
where 
\begin{equation}
	\chi=\frac{K^{2}k_{L}}{8\gamma^{2}k_{u}}=\frac{K^{2}}{4+2K^{2}}.
\end{equation}
Note that since the longitudinal coordinate of electron will affect the laser phase observed, so we need to calculate its precision to the order of $\frac{1}{\gamma^{2}}$. While for the horizontal coordinate $x$, we only need to calculate it to the order of $\frac{1}{\gamma}$. In the following, we will adopt the approximation $\beta\approx1$ since we are interested in relativistic cases.

\begin{figure}[tb]
	\centering 
	\includegraphics[width=0.7\columnwidth]{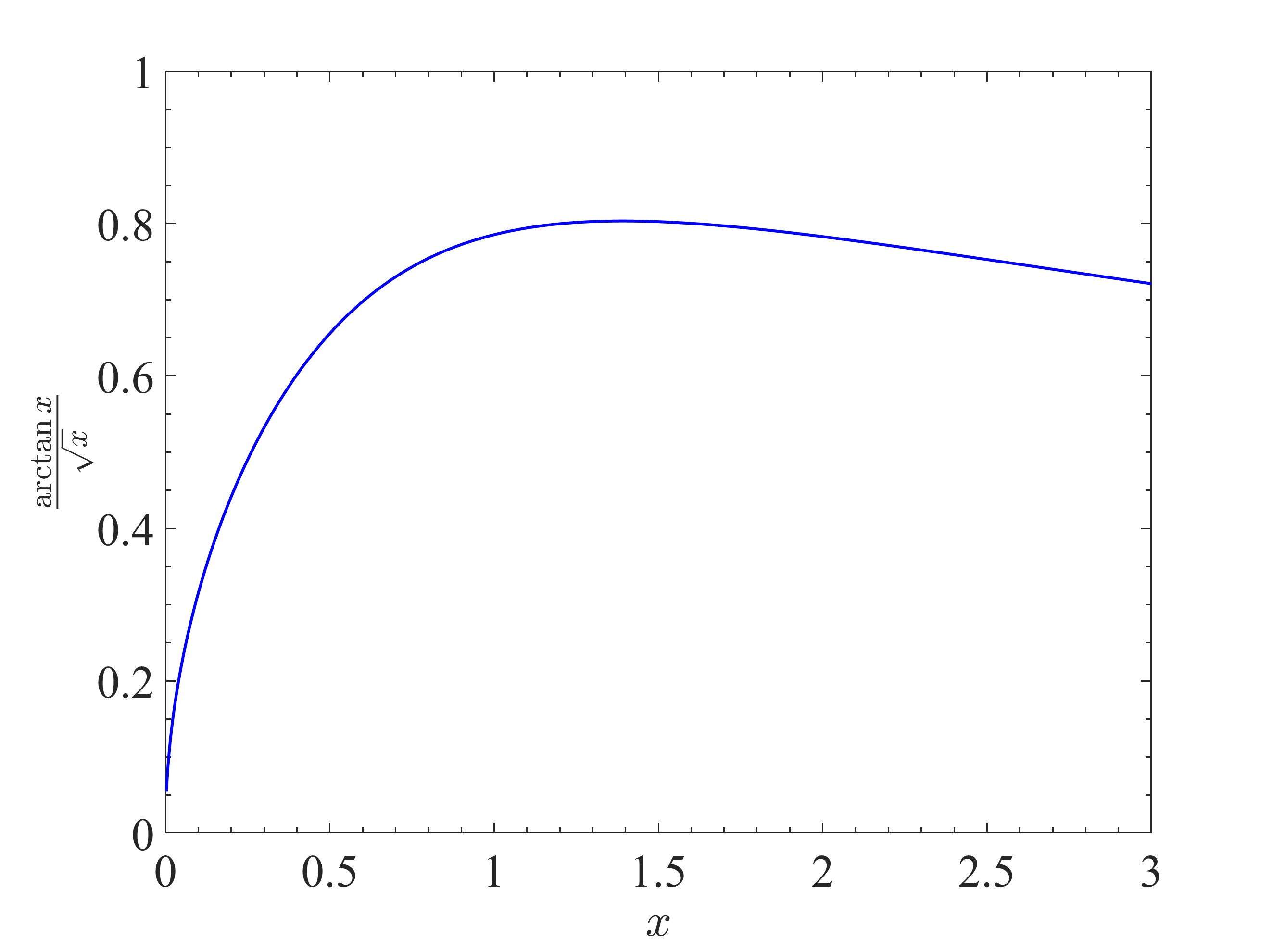}
	\caption{
		\label{fig:PlanarUndulatorEnergyModulation} 
		$f(x)=\frac{\tan^{-1}(x)}{\sqrt{x}}$ vs. $x$. }
\end{figure}

Given the electron prescribed motion and laser electric field, the laser and electron  exchange energy according to
\begin{equation}\label{eq:LEinteraction}
	\frac{d\mathcal{W}}{dt}=-\left(ev_{x}E_{x}+ev_{z}E_{z}\right),
\end{equation}
{where $\mathcal{W}$ is the work done on the electron by the laser, or the energy transfer from the laser to the electron. Note that in this manuscript $e$ represents the elementary charge and is positive.} 
Assuming that the laser beam waist is in the middle of the undulator, whose length is $L_{u}$, and when electron transverse coordinates are much smaller than the laser beam waist $x,y\ll w_{0}$, which is true in most of our interested cases, we drop the $\text{exp}\left[i\frac{k_{L}Q}{2}(x^2+y^2)\right]$ in the laser electric field. Further, when the transverse displacement of electron is much smaller than the Rayleigh length $x\ll Z_{R}$ which is also satisfied in our interested cases, we can also drop the contribution from $E_{z}$ on the energy modulation. {Assuming the relative phase of laser field to electron horizontal velocity $v_{x}$ at the undulator center is $\phi_{0}$}, the integrated modulation voltage induced by the laser on the electron beam in a planar undulator is then
{}
	\begin{equation}
		\begin{aligned}
			V_{L}&\approx\text{Re}\left[\int_{-\frac{L_{u}}{2}}^{\frac{L_{u}}{2}}v_{x}E_{x}\frac{ds}{c}\right]
			\approx E_{x0}\frac{K}{\gamma}\text{Re}\left[e^{i\phi_{0}}\int_{-\frac{L_{u}}{2}}^{\frac{L_{u}}{2}}\frac{1}{1+i\frac{s}{Z_{R}}}\sum_{n=-\infty}^{\infty}J_{n}(-\chi)e^{in2k_{u}s}\frac{1+e^{-i2k_{u}s}}{2}ds\right],
		\end{aligned}
	\end{equation}
{}
where Re() means taking the real component of the complex number.
When $L_{u}\gg\lambda_{u}$, which means the undulator period number $N_{u}\gg1$, in the above integration, only the term with $n=0$ and $n=1$ will give notable non-vanishing value. Denote $[JJ]\equiv J_{0}(\chi)-J_{1}(\chi)$, we then have 
\begin{equation}
	\begin{aligned}
		V_{L}
		&= E_{x0}\frac{K}{\gamma}\frac{[JJ]}{2}\text{Re}\left[e^{i\phi_{0}}\int_{-\frac{L_{u}}{2}}^{\frac{L_{u}}{2}}\frac{1-i\frac{s}{Z_{R}}}{1+\left(\frac{s}{Z_{R}}\right)^{2}}ds\right]\\
		&= E_{x0}\frac{K}{\gamma}[JJ]Z_{R}\tan^{-1}\left(\frac{L_{u}}{2Z_{R}}\right)\cos\phi_{0}.
	\end{aligned}
\end{equation}
We want the energy modulation strength as large as possible, so we choose $\phi_{0}=0$. Put in the expression of peak electric field from Eq.~(\ref{eq:laserE}), 
the linear energy chirp strength around the zero-crossing phase is therefore
\begin{equation}\label{eq:TEM00h}
	\begin{aligned}
		h&=\frac{eV_{L}}{E_{0}}k_{L}\\
		&=\frac{ek_{L}K[JJ]}{\gamma^{2}m_{e}c^{2}}\sqrt{\frac{2P_{L}Z_{0}}{\lambda_{L}}}\frac{\tan^{-1}\left(\frac{L_{u}}{2Z_{R}}\right)}{\sqrt{\frac{L_{u}}{2Z_{R}}}}\sqrt{L_{u}},
	\end{aligned}
\end{equation}
where $E_{0}$ is the electron energy and $m_{e}$ is the mass of electron. Once the modulator length is given, we can optimize the laser Rayleigh length to maximize the energy modulation. Figure~\ref{fig:PlanarUndulatorEnergyModulation} is a plot of $f(x)=\frac{\tan^{-1}(x)}{\sqrt{x}}$ as a function of $x$. The maximum value of $f(x)$ is 0.8034 and is realized when $x=1.392$. So when $Z_{R}=\frac{L_{u}}{2\times1.392}=0.359L_{u}$, the energy modulation reaches the maximum value. Note that when $Z_{R}$ is within a small range  close to the optimal value, the impact of Rayleigh length on energy modulation strength is not very sensitive. Therefore, for easy of remembering, the optimal condition can be expressed as
\begin{equation}
	Z_{R}\approx\frac{L_{u}}{3}.
\end{equation}
From Eq.~(\ref{eq:TEM00h}), we can see that under the optimal condition, we have $h\propto\sqrt{P_{L}L_{u}}$. Using Eq.~(\ref{eq:TEM00h}) we can also do some example calculation to get a more concrete feeling. If $E_{0}=600$ MeV, $\lambda_{L}=1064$~nm,  $\lambda_{u}=8$ cm ($B_{0}=1.13$ T), $L_{u}=0.8$ m ($N_{u}=10$),  $Z_{R}=0.359L_{u}$, then the induced energy chirp strength with $P_{L}=1$~MW is $h=955\ \text{m}^{-1}$.

\subsubsection{Dual-Tilted-Laser for Energy Modulation}

Now with a hope to increase the energy modulation strength with a given laser power, we may use a configuration of crossing two lasers for energy modulation. The basic idea is that if two crossing lasers can double the energy modulation strength of that of a single laser, then the effect is like that induced by single laser with a laser power four times larger. Our calculation shows that indeed dual-tilted-laser (DTL) can induce a larger energy modulation compared to that of a single laser, but the issue is that the required crossing angle (less than 2 mrad) is too small from an engineering viewpoint.

Now we present the analysis. First we consider the case of two lasers crossing in the $y$-$z$ plane. The laser field of an oblique TEM00 laser is given by first replacing the physical coordinate with the rotated coordinates
{}
	\begin{equation}
		\begin{aligned}
			&x\rightarrow x,\ y\rightarrow y_{1}=y\cos\theta+s\sin\theta,\ s\rightarrow s_{1}=-y\sin\theta+s\cos\theta,\ t\rightarrow t,\\
		\end{aligned}
	\end{equation}
	{with $\theta$ in this section and below the tilted angle of the incident laser instead of the dipole bending angle in Sec.~\ref{sec:TME}, We hope their difference is clear from context.}
	The resulting field expression according to Eq.~(\ref{eq:TEM00}) is
	\begin{equation}
		\left(
		\begin{matrix}
			E_{x}\\
			E_{y}\\
			E_{z}\\
			cB_{x}\\
			cB_{y}\\
			cB_{z}
		\end{matrix}
		\right)_{\text{rot}}=
		E_{x0}e^{ik_{L}s_{1}-i\omega_{L} t}\frac{\text{exp}\left[-\frac{k_{L}(x^{2}+y_{1}^{2})}{2(Z_{R}+is_{1})}\right]}{1+i\frac{s_{1}}{Z_{R}}}
		\left(
		\begin{matrix}
			1\\
			0\\
			-i\frac{x}{Z_{R}+is_{1}}\\
			\frac{xy_{1}}{\left(Z_{R}+is_{1}\right)^{2}}\\
			-\frac{x^2}{\left(Z_{R}+is_{1}\right)^{2}}-\frac{1}{k_{L}\left(Z_{R}+is_{1}\right)}+1\\
			-i\frac{y_{1}}{Z_{R}+is_{1}}
		\end{matrix}
		\right).
	\end{equation}
	Note however, in the above expression, $x,y,s$ of the electromagnetic fields are defined according to the oblique laser propagating direction. To get the expression back in the original coordinate system, i.e., undulator axis as the $z$ axis, we need to rotate the laser field as, 
	\begin{equation}
		\begin{aligned}
			&E_{x}\rightarrow E_{x\text{rot}},\ E_{y}\rightarrow E_{y}=E_{y\text{rot}}\cos\theta-E_{z\text{rot}}\sin\theta,\ E_{z}\rightarrow E_{z}=E_{y\text{rot}}\sin\theta+E_{z\text{rot}}\cos\theta,\ t\rightarrow t,
		\end{aligned}
	\end{equation}
	and for the electric field the result is
	\begin{equation}
		\left(
		\begin{matrix}
			E_{x}\\
			E_{y}\\
			E_{z}\\
		\end{matrix}
		\right)_{\text{unrot}}=E_{x0}e^{ik_{L}s_{1}-i\omega_{L} t}\frac{\text{exp}\left[-\frac{k_{L}(x^{2}+y_{1}^{2})}{2(Z_{R}+is_{1})}\right]}{1+i\frac{s_{1}}{Z_{R}}}
		\left(
		\begin{matrix}
			1\\
			-i\frac{x}{Z_{R}+is_{1}}\sin\theta\\
			-i\frac{x}{Z_{R}+is_{1}}\cos\theta\\
		\end{matrix}
		\right).
	\end{equation}
	Assuming that the two crossing lasers are in-phase and have the same amplitude. In addition, we assume that $\theta_{2}=-\theta_{1}=-\theta$, and the two lasers have the same Rayleigh {length}. Then the superimposed field is
	\begin{equation}\label{eq:DTLField}
		\begin{aligned}
			\left(
			\begin{matrix}
				E_{x}\\
				E_{y}\\
				E_{z}\\
			\end{matrix}
			\right)_{\text{unrot}}&=
			E_{x0}e^{ik_{L}\left(-y\sin\theta+s\cos\theta\right)-i\omega_{L} t}\frac{\text{exp}\left[-\frac{k_{L}(x^{2}+\left(y\cos\theta+s\sin\theta\right)^{2})}{2(Z_{R}+i\left(-y\sin\theta+s\cos\theta\right))}\right]}{1+i\frac{\left(-y\sin\theta+s\cos\theta\right)}{Z_{R}}}
			\left(
			\begin{matrix}
				1\\
				-i\frac{x}{Z_{R}+i\left(-y\sin\theta+s\cos\theta\right)}\sin\theta\\
				-i\frac{x}{Z_{R}+i\left(-y\sin\theta+s\cos\theta\right)}\cos\theta\\
			\end{matrix}
			\right)\\
			&\ \ \ \ +E_{x0}e^{ik_{L}\left(y\sin\theta+s\cos\theta\right)-i\omega_{L} t}\frac{\text{exp}\left[-\frac{k_{L}(x^{2}+\left(y\cos\theta-s\sin\theta\right)^{2})}{2(Z_{R}+i\left(y\sin\theta+s\cos\theta\right))}\right]}{1+i\frac{\left(y\sin\theta+s\cos\theta\right)}{Z_{R}}}
			\left(
			\begin{matrix}
				1\\
				i\frac{x}{Z_{R}+i\left(y\sin\theta+s\cos\theta\right)}\sin\theta\\
				-i\frac{x}{Z_{R}+i\left(y\sin\theta+s\cos\theta\right)}\cos\theta\\
			\end{matrix}
			\right).
		\end{aligned}
	\end{equation}
{}
As before here we focus on the impact of $E_{x}$ on laser-electron interaction, and ignore the contribution from $E_{z}$. When $\theta$ is very small, the superimposed $E_{x}$ can be approximated as
\begin{equation}
	\begin{aligned}
		E_{x}&=2E_{x0}\frac{\exp \left[i k_L s \cos \theta -i \omega t  -\frac{k_L \left(x^2+s^2 \sin ^2\theta\right)}{2 \left(Z_R+i s \cos \theta\right)}\right]}{1+\frac{i s \cos \theta}{Z_R}}\\
		&\approx2E_{x0}e^{ik_{L}s\cos\theta-i\omega_{L}t}\frac{\exp \left[-\frac{k_L s^2 \theta^{2} }{2 \left(Z_R+i s \right)} \right]}{1+i\frac{s}{Z_{R}}}.
	\end{aligned}
\end{equation}
Note that in the final approximated expression, we have kept $s\cos\theta$ in the laser phase term. The reason is that the laser phase is of key importance in laser-electron interaction and the accuracy requirement is high. In addition, we have also kept the $s^{2}\theta^{2}$ in the intensity decay term, this is because $L_{u}\theta$ may not be small compared to the laser beam wait radius $w_{0}$.  The expression of $E_{x}$ for the case of crossing in $x$-$z$ plane is similar. So the difference of crossing in $x$-$z$ plane and $y$-$z$ plane is not much in inducing energy modulation.

\begin{figure}[tb]
	\centering
	\includegraphics[width=0.7\columnwidth]{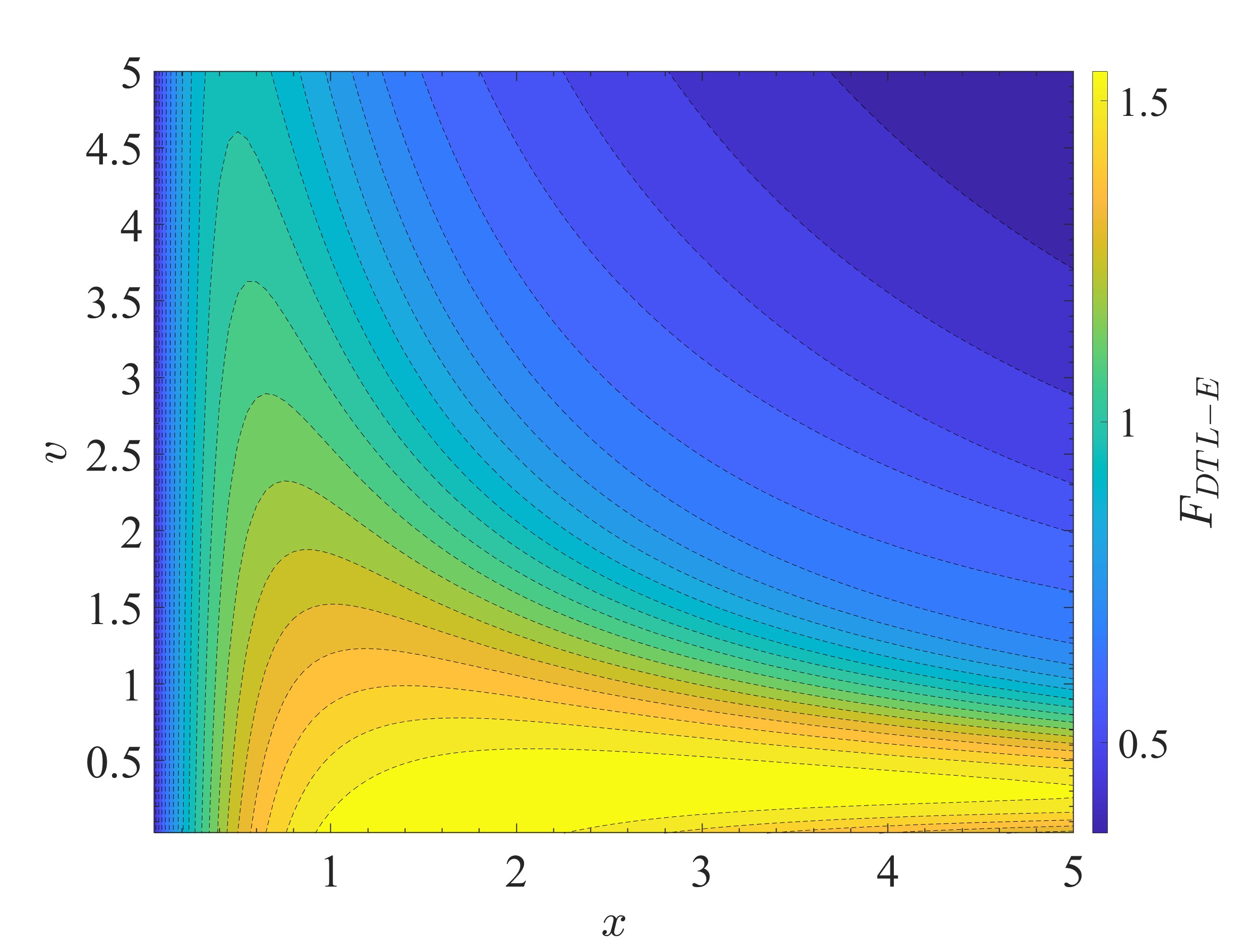}
	\caption{Contour plot of $F_{\text{DTL-E}}(x,v)$ given by Eq.~(\ref{eq:DTLEDeng}).  }
	\label{fig:dtlefdeng}
\end{figure}

For effective laser-electron interaction, the off-axis resonance condition now is
\begin{equation}
	c\frac{\lambda_{u}}{\bar{v}_{z}}-\lambda_{u}\cos\theta=\lambda_{L},
\end{equation}
or
\begin{equation}\label{eq:offaxis}
	\begin{aligned}
		\lambda_{L}&=\left(\frac{1}{1-\frac{1+\frac{K^{2}}{2}}{2\gamma^{2}}}-\cos\theta\right)\lambda_{u}\\
		&\approx\frac{1+\frac{K^{2}}{2}+\gamma^{2}\theta^{2}}{2\gamma^{2}}\lambda_{u}.
	\end{aligned}
\end{equation}
Then 
\begin{equation}
	\begin{aligned}
		k_{L}s-\omega_{L} t\approx-k_{u}s-\chi\sin(2k_{u}s).
	\end{aligned}
\end{equation}
Note that $\chi$ now  depends on $\theta$, more specifically,
\begin{equation}
	\chi=\frac{K^{2}k_{L}}{8\gamma^{2}k_{u}}\approx\frac{K^{2}}{4+2K^{2}+4\gamma^{2}\theta^{2}}.
\end{equation} 
Assume that the laser beam waists are in the middle of the undulator, whose length is $L_{u}$, and denote
\begin{equation}\label{eq:v}
	v\equiv\frac{k_{L}Z_{R}\theta^{2}}{2 }=\left(\frac{Z_{R}\theta}{w_{0}}\right)^{2},
\end{equation}
the integrated modulation voltage induced by the DTL in a planar undulator is then
{}
	\begin{equation}
		\begin{aligned}
			V_{L}&\approx\text{Re}\left[\int_{-\frac{L_{u}}{2}}^{\frac{L_{u}}{2}}v_{x}E_{x}\frac{ds}{c}\right]\\
			&\approx E_{x0}\frac{K[JJ]}{\gamma}Z_{R}\int_{-\frac{L_{u}}{2Z_{R}}}^{\frac{L_{u}}{2Z_{R}}}\frac{ \exp \left(-v\frac{{u}^{2}}{1+{u}^{2}} \right)}{1+{u}^{2}}\left[\cos\left(v\frac{{u}^{3}}{1+{u}^{2}}\right)+{u}\sin\left(v\frac{{u}^{3}}{1+{u}^{2}}\right)\right] d{u}\cos\phi_{0}.
		\end{aligned}
	\end{equation}
	To maximize the energy modulation, we choose $\phi_{0}=0$. 
	Put in the expression of $E_{x0}$ from Eq.~(\ref{eq:laserE}), the linear energy chirp strength around the zero-crossing phase is therefore
	\begin{equation}\label{eq:TEM00DTL}
		h=\frac{eV_{L}}{E_{0}}k_{L}=\frac{ek_{L}K[JJ]}{\gamma^{2}m_{e}c^{2}}\sqrt{\frac{2P_{L}Z_{0}}{\lambda_{L}}}F_{\text{DTL-E}}\left(\frac{L_{u}}{2Z_{R}},\frac{k_{L}Z_{R}\theta^{2}}{2 }\right)\sqrt{L_{u}},
	\end{equation}
	where
	\begin{equation}\label{eq:DTLEDeng}
		F_{\text{DTL-E}}(x,v)=\frac{1}{\sqrt{x}}\int_{-x}^{x}\frac{ \exp \left(-v\frac{{u}^{2}}{1+{u}^{2}} \right)}{1+{u}^{2}}\left[\cos\left(v\frac{{u}^{3}}{1+{u}^{2}}\right)+{u}\sin\left(v\frac{{u}^{3}}{1+{u}^{2}}\right)\right] d{u}.
	\end{equation}
{}
A flat contour plot of $F_{\text{DTL-E}}(x,v)$ is given in Fig.~\ref{fig:dtlefdeng}.

\begin{figure}[H]
	\centering 
	\includegraphics[width=0.7\columnwidth]{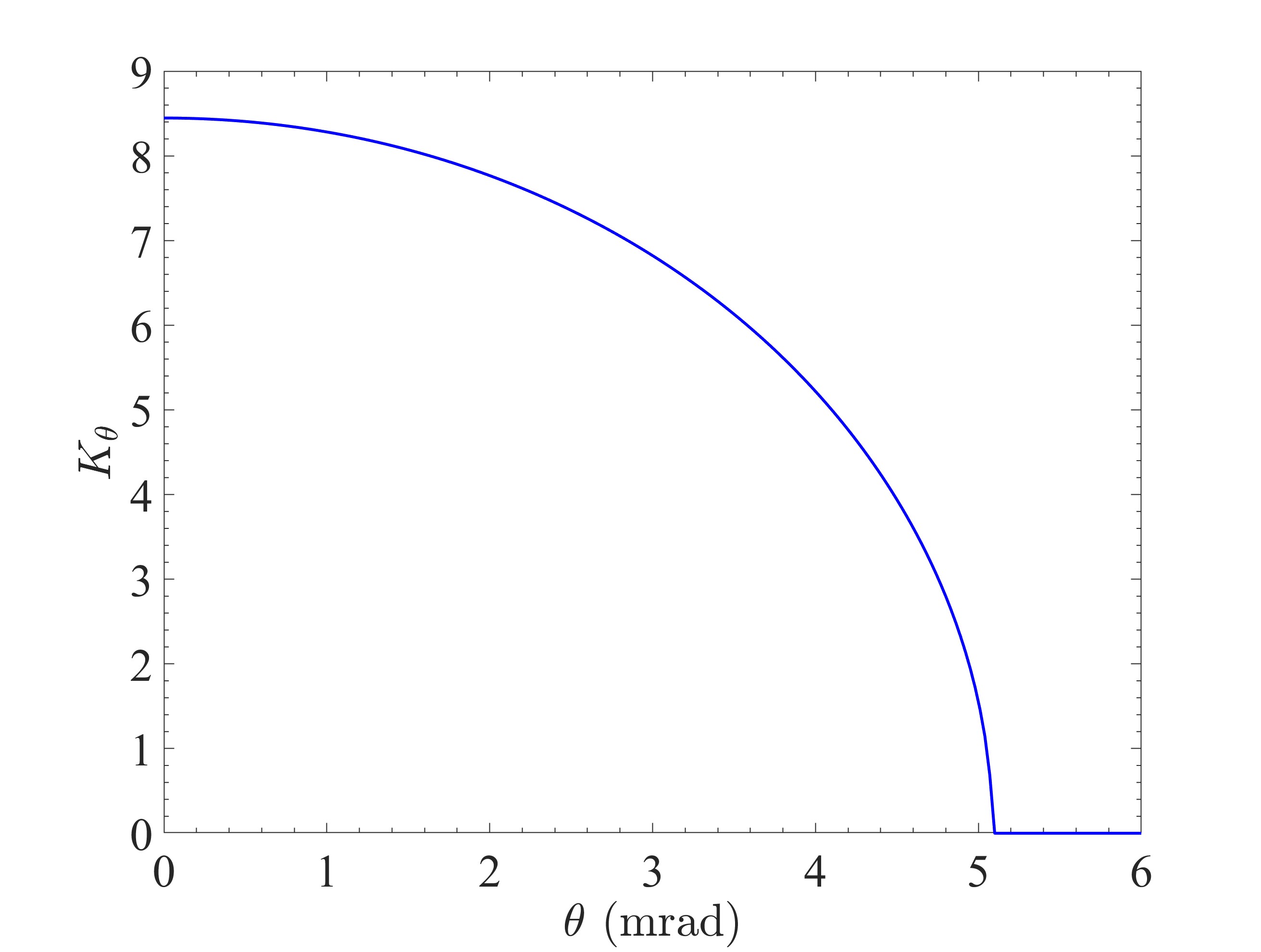}
	\caption{
		\label{fig:KThetaLmabdaUFixed} 
		Undulator parameter $K_{\theta}$ vs. $\theta$ with $\lambda_{u}$ kept fixed. Parameters used: $E_{0}=600$ MeV, $\lambda_{L}=1064$ nm, $\lambda_{u}=8$ cm. }
\end{figure}

\begin{figure}[H]
	\centering 
	\includegraphics[width=0.7\columnwidth]{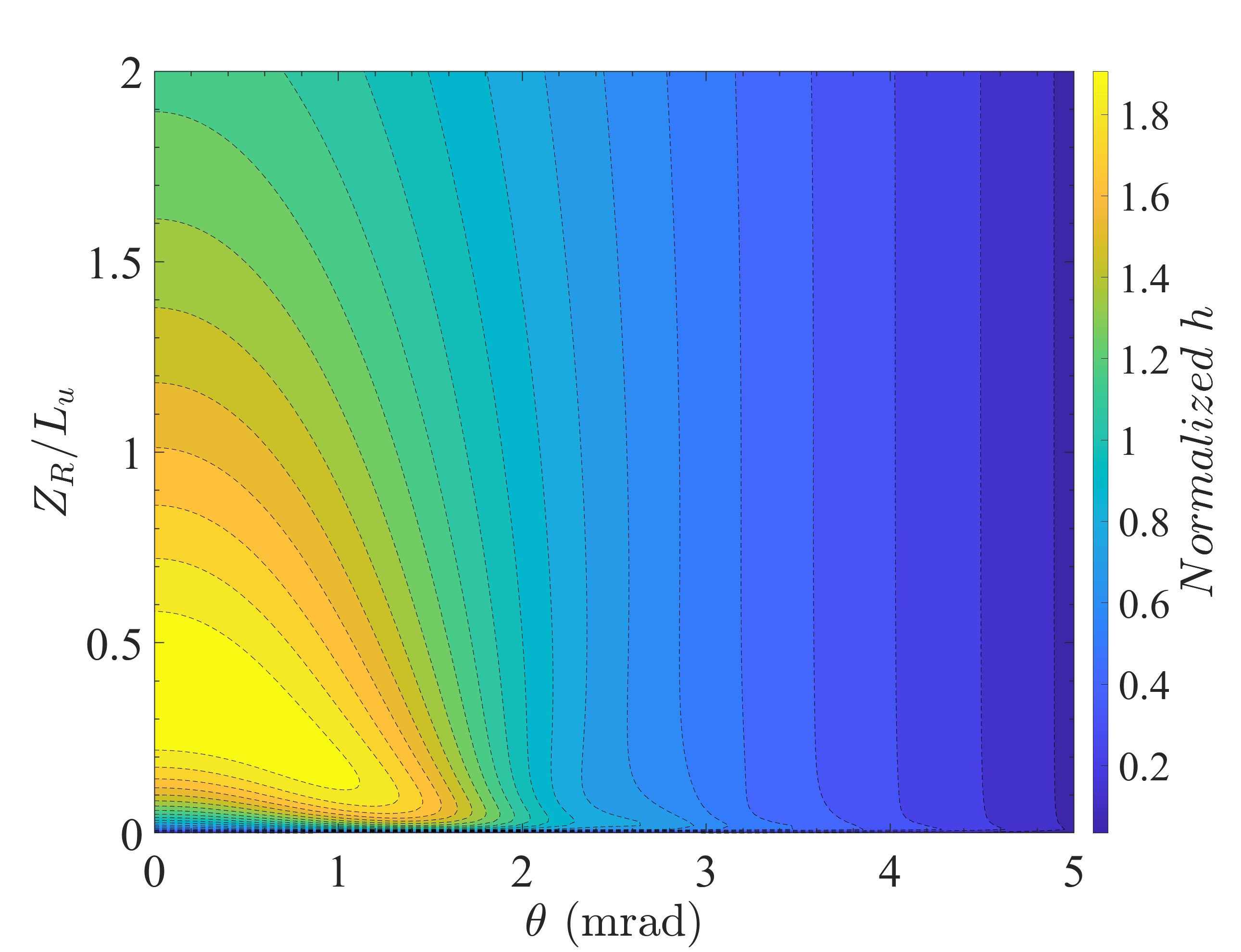}
	\caption{
		\label{fig:DTLEDeng} 
		Energy chirp strength  $h$ normalized by the largest energy chirp induced by a single normally incident laser,  vs. $\theta$ and $Z_{R}/L_{u}$. Keep $\lambda_{u}$ fixed when changing $\theta$. Parameters used: $E_{0}=600$ MeV, $\lambda_{L}=1064$ nm, $\lambda_{u}=0.08$ m, $N_{u}=10$, $L_{u}=0.8$ m.  }
\end{figure}

Now we can use the derived formula to calculate the energy chirp strength induced by DTL. First we consider the case of keeping $\lambda_{u}$ fixed, when changing $\theta$. If we keep $\lambda_{u}$ fixed when changing $\theta$, then the off-axis resonant condition leads to the undulator parameter as a function of $\theta$ given by
\begin{equation}
	K_\theta=\sqrt{2}\sqrt{2\gamma^{2}\left(1-\frac{1}{\frac{\lambda_{L}}{\lambda_{u}}+\cos\theta}\right)-1}.
\end{equation}
With the increase of $\theta$, $K_{\theta}$ will decrease. Note that in this case $K_{\theta}\propto B_{0\theta}$, so the magnetic field strength $B_{0\theta}$ also decreases with the increase of $\theta$. An example calculation of $K_{\theta}$ vs. $\theta$ is given in Fig.~\ref{fig:KThetaLmabdaUFixed}. The corresponding contour plot of the energy chirp strength normalized by the largest energy chirp induced by a single normally incident laser, vs. $\theta$ and $Z_{R}/L_{u}$, is given in Fig.~\ref{fig:DTLEDeng}.

\begin{figure}[H]
	\centering 
	\includegraphics[width=0.7\columnwidth]{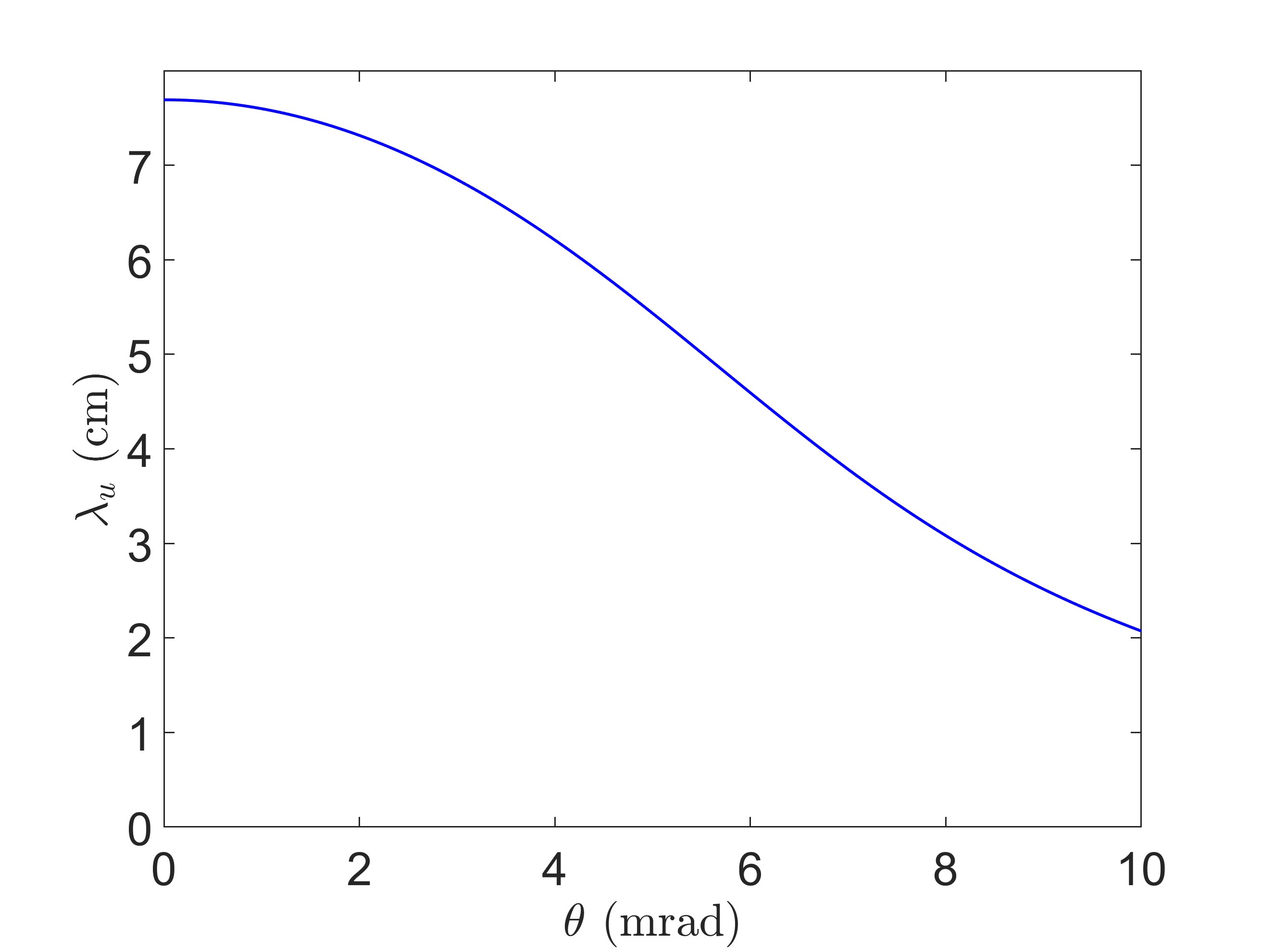}
	\caption{
		\label{fig:lambdaUVSTheta2} 
		$\lambda_{u}$ vs. $\theta$, with $B_{0}$ kept fixed. Parameters used: $E_{0}=600$~MeV, $\lambda_{L}=1064$~nm, $B_{0}=1.2$ T.  }
\end{figure}

\begin{figure}[H]
	\centering 
	\includegraphics[width=0.7\columnwidth]{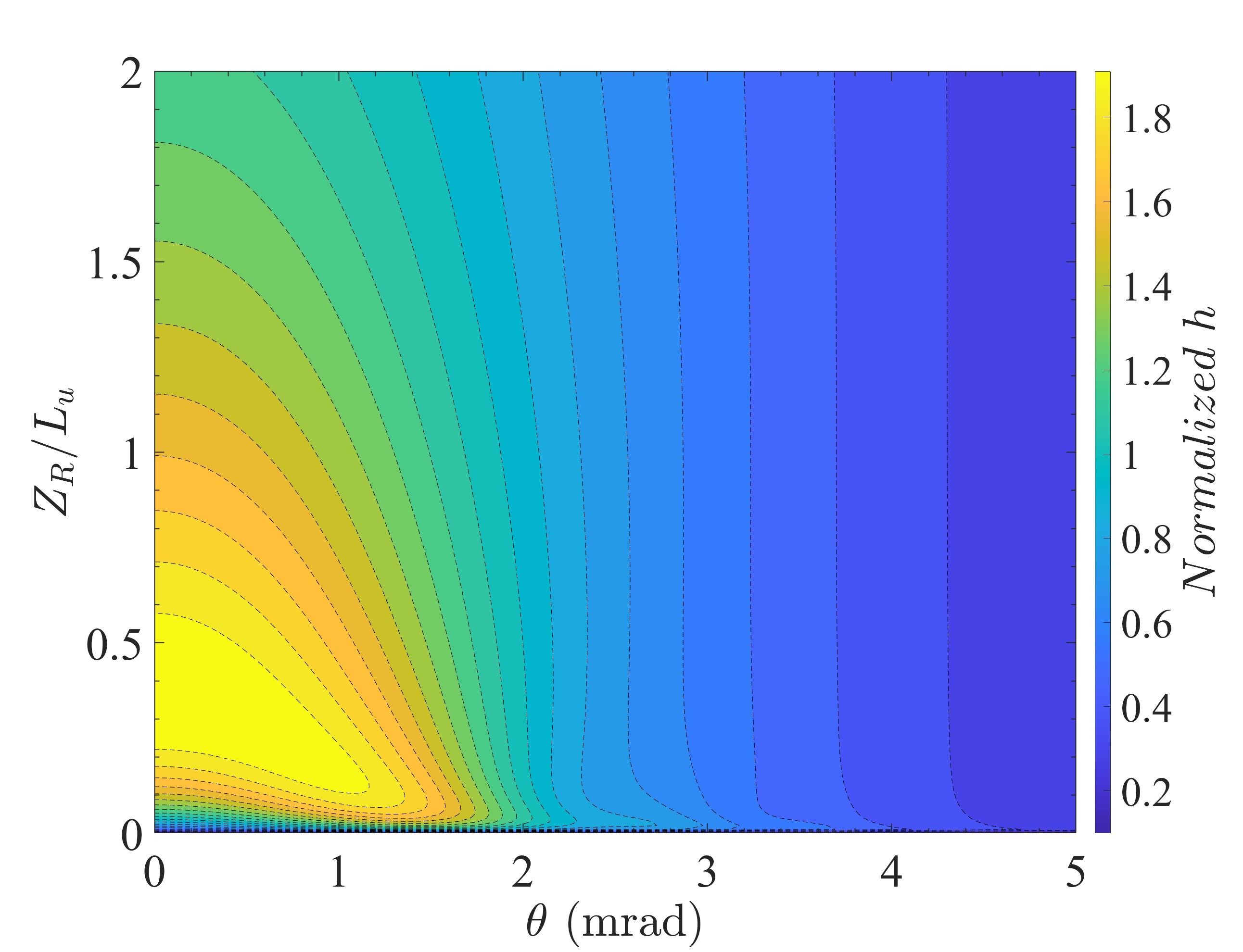}
	\caption{
		\label{fig:DTLEDengB0Fxied} 
		Energy chirp strength  $h$ normalized by the largest energy chirp induced by a single normally incident laser,  vs. $\theta$ and $Z_{R}/L_{u}$. Keep $B_{0}$ fixed when changing $\theta$. Parameters used: $E_{0}=600$ MeV, $\lambda_{L}=1064$ nm, $B_{0}=1.2$ T, $L_{u}=0.8$ m.  }
\end{figure}

The previous calculation assumes $\lambda_{u}$ is kept unchanged when we adjust the incident angle $\theta$. This will result in a limited region of $\theta$ to fulfill resonant condition as shown in Fig.~\ref{fig:KThetaLmabdaUFixed}. Now we conduct the calculation by assuming the peak magnetic field $B_{0}$ unchanged when we adjust $\theta$. 
Put the expression of undulator parameter Eq.~(\ref{eq:K}) in the off-axis resonant condition Eq.~(\ref{eq:offaxis}), we have
\begin{equation}
	\frac{1}{2}\left(\frac{eB_{0}}{2\pi m_{e}c}\right)^{2}\lambda_{u}^{3}+(1+\gamma^{2}\theta^{2})\lambda_{u}-2\gamma^{2}\lambda_{L}=0,
\end{equation}
from which we get
{
	\begin{equation}
		\begin{aligned}
			\lambda_{u}=&\frac{2 \pi ^{2/3} \sqrt[3]{\mathcal{D}}}{3^{2/3} B_0^2 e^2}-\frac{4 \pi ^{4/3} c^2 \left(\gamma ^2 \theta ^2+1\right) m_e^2}{\sqrt[3]{3} \sqrt[3]{\mathcal{D}}},
		\end{aligned}
	\end{equation}
	with
	\begin{equation}
		\begin{aligned}
			\mathcal{D}&=\sqrt{3B_0^6 c^4 e^6 m_e^4 \left(27 B_0^2 \gamma ^4 e^2 \lambda _L^2+8 \pi ^2 c^2 \left(\gamma ^2 \theta ^2+1\right)^3 m_e^2\right)},\\
			&+9 B_0^4 c^2 \gamma ^2 e^4 m_e^2 \lambda _L.
		\end{aligned}
	\end{equation}
}
For example, if $\lambda_{L}=1064$ nm and $B_{0}=1.2$ T, then $\lambda_{u}$ as a function of $\theta$ for the case of $E_{0}=600$~MeV is shown in Fig.~\ref{fig:lambdaUVSTheta2}. Note that in this case $K_{\theta}\propto\lambda_{u}$ which depends on $\theta$, so the undulator parameter also decreases with the increase of $\theta$. But the decrease is not that fast compared to that presented in Fig.~\ref{fig:KThetaLmabdaUFixed}.  The corresponding contour plot of the energy chirp strength induced by DTL normalized by the largest energy chirp induced by a single normally incident laser,  vs. $\theta$ and $Z_{R}/L_{u}$, is given in Fig.~\ref{fig:DTLEDengB0Fxied}.  

As can be seen from the calculation results in Figs.~\ref{fig:DTLEDeng} and \ref{fig:DTLEDengB0Fxied}, DTL indeed can induce a larger energy modulation compared to a single normally incident laser. But the required crossing angle (less than 2 mrad) is too small for engineering. So the usual setup of a single normally incident TEM00 mode laser is still the preferred choice in practical application.


\subsection{Realization Examples}

After the derivation of bunching factor and laser-induced modulation strengths, finally in this section we give some realization examples of microcbunching schemes belonging to what we have analyzed. FEL seeding technique like phase-merging enhanced harmonic generation
(PEHG) \cite{Deng2013,Feng2014},  and angular dispersion-induced
microbunching (ADM) \cite{Feng2017} can be viewed as specific examples of our general definition of TLC-based microbunching schemes in Theorem One. More detailed discussion in this respect has been presented before in Ref.~\cite{Deng2021NIMA}.

Here we make a short comment about the relation between FEL seeding techniques like HGHG, PEHG, ADM and the storage ring schemes like LSF, GLSF discussed in this paper. One is single-pass, and the other is multi-pass. One invokes matrix multiplication or nonlinear transfer map once, and the other invokes eigen analysis or normal form analysis of the one-turn map. What they share is the bunch compression or harmonic generation mechanism. The relation between HGHG and LSF is similar to the relation between PEHG/ADM and GLSF.

\section{Angular Modulation-Based Coupling Schemes}\label{sec:TLCA}

\subsection{Form Function and Bunching Factor}

After investigating energy modulation-based TLC microbunching schemes, now in this section we discuss angular modulation-based ones. The problem definition is similar to that in the energy modulation-based schemes, only replacing the energy modulation by angular modulation. We use $y'$ modulation as an example, since we will take advantage of the ultrasmall vertical emittance in a planar ring as explained before.
The lumped laser-induced angular modulation is modeled as:
\begin{equation}
	\begin{aligned}
		y'&=y'+A\sin(k_{L}z),\\
		\delta&=\delta+Ak_{L}y\cos(k_{L}z).
	\end{aligned}
\end{equation}
Note that we need to add the second equation above to make the above modulation symplectic. Such an angular modulation process also satisfies the Panofsky-Wenzel theorem~\cite{Chao1993}
\begin{equation}
	\frac{\partial\Delta y'}{\partial z}=\frac{\partial\Delta \delta}{\partial y}.
\end{equation}
Following the derivations in the section of energy modulation-based schemes, the final form function in this case is 
{}
	{
		\begin{equation}
			\begin{aligned}
				\mathcal{F}({\bf K})
				&=\int\psi_{f}({\bf X})e^{-i{\bf K}{\bf X}}d{\bf X}\\
				&=\int\psi_{m+}({\bf X})e^{-i{\bf KR}{\bf X}}d{\bf X}\\
				&=\int\psi_{0}({\bf X})e^{-i\left({\bf KR}{\bf X}+{\bf K}{\bf R}_{i4}A\sin(k_{L}z)+{\bf K}{\bf R}_{i6}Ak_{L}y\cos(k_{L}z)\right)}d{\bf X}\\
				&=\sum_{p_{1}=-\infty}^{\infty}J_{p_{1}}\left(-{\bf K}{\bf R}_{i4}A\right)\int\psi_{0}({\bf X})e^{-i\left({\bf KR}{\bf X}-p_{1}k_{L}z+{\bf K}{\bf R}_{i6}Ak_{L}y\cos(k_{L}z)\right)}d{\bf X}\\
				&=\sum_{p_{1}=-\infty}^{\infty}J_{p_{1}}\left(-{\bf K}{\bf R}_{i4}A\right)\int\psi_{0}({\bf X})\sum_{p_{2}=-\infty}^{\infty}i^{p_{2}}J_{p_{2}}\left(-{\bf K}{\bf R}_{i6}Ak_{L}y\right)e^{-i\left[{\bf KR}{\bf X}-(p_{1}+p_{2})k_{L}z\right]}d{\bf X}.
			\end{aligned}
		\end{equation}
		The integration of the above result is given by generalized hypergeometric function. If
		$
		{\bf K}=\left(
		\begin{matrix}
		0&
		0&
		0&
		0&
		k_{z}&
		0
		\end{matrix}
		\right)
		$
		and $R_{56}=0$, then ${\bf K}{\bf R}_{i6}=0$, and the initial beam distribution is Gaussian as given in Eq.~(\ref{eq:Gaussian}), then the 1D bunching factor is
		\begin{equation}
			\begin{aligned}
				b(k_{z})
				&=\sum_{p=-\infty}^{\infty}J_{p}\left(-{\bf K}{\bf R}_{i4}A\right)\int\psi_{0}({\bf X})e^{-i\left({\bf KR}{\bf X}-pk_{L}z\right)}d{\bf X}\\
				&=\sum_{p=-\infty}^{\infty}J_{p}\left(-k_{z}R_{54}A\right)\text{exp}\left(-\frac{{\bf M}_{p}{\bf \Sigma}_{0}{\bf M}_{p}^{T}}{2}\right),
			\end{aligned}
		\end{equation}
		with ${\bf M}_{p}$ given by Eq.~(\ref{eq:Mp}).}
	
	To appreciate the physical principle, instead of a general mathematical analysis, we use one specific case as an example. 
	If $R_{51}=0$, $R_{52}=0$, $R_{55}=1$, $R_{56}=0$, and the initial beam is transverse-longitudinal decoupled and has an upright distribution in the longitudinal phase space, then 
	\begin{equation}
		\begin{aligned}
			b(k_{z})&=\sum_{p=-\infty}^{\infty}J_{p}\left(-k_{z}R_{54}A\right)\text{exp}\left(-\frac{k_{z}^{2}}{2}\left[\epsilon_{y}\left(R_{53}^{2}\beta_{y}-2R_{53}R_{54}\alpha_{y}+R_{54}^{2}{\gamma_{y}}\right)+\left(1-\frac{pk_{L}}{k_{z}}\right)^{2}\sigma_{z0}^{2}\right]\right),
		\end{aligned}
	\end{equation}
{}
with $\alpha_{y},\beta_{y},\gamma_{y}$ the Courant-Snyder functions before the modulation, and $\sigma_{\delta 0}$ the initial RMS energy spread.
If the initial bunch length is much longer than the laser wavelength, the above exponential terms will be non-zero only when $1-\frac{pk_{L}}{k_{z}}=0$, which means there is only bunching at the laser harmonics. In this case, we have the bunching factor at the $n$-th laser harmonic to be
\begin{equation}\label{eq:BFTLCA}
	\begin{aligned}
		b_{n}&=b(k_{z}=nk_{L})\\
		&=J_{n}(-nk_{L}{R}_{54}A)\text{exp}\left[-\frac{\left(nk_{L}\right)^{2}}{2}\epsilon_{y}\mathcal{H}_{y\text{R}}\right],
	\end{aligned}
\end{equation}
where $\mathcal{H}_{y\text{R}}=R_{53}^{2}\beta_{y}-2R_{53}R_{54}\alpha_{y}+R_{54}^{2}{\gamma_{y}}$. One can appreciate the similarity of the above result with the bunching factor of energy modulation-based TLC microbunching schemes, i.e., Eq.~(\ref{eq:BFADM}). The $R_{54}$ here plays the role of $R_{56}$ there. If further $R_{53}=0$, then Eq.~(\ref{eq:BFTLCA}) reduces to
\begin{equation}
	\begin{aligned}
		b_{n}
		&=J_{n}(-nk_{L}{R}_{54}A)\text{exp}\left[-\frac{\left(nk_{L}R_{54}\sigma_{y'}\right)^{2}}{2}\right],
	\end{aligned}
\end{equation}
where $\sigma_{y'}$ is the initial beam angular divergence. 
So $R_{54}$ and $\sigma_{y'}$ in this scheme play the role of $R_{56}$ and $\sigma_{\delta}$ in HGHG as shown in Eq.~(\ref{eq:HGHG}), respectively.  In a planar uncoupled ring, the natural vertical emittance is quite small, thus also $\sigma_{y'}$. Therefore, using this scheme we can realize a high harmonic bunching in a storage ring, for example to generate ultrashort soft X-ray pulse.

As can be seen from our analysis, both the energy modulation-based and the angular modulation-based TLC microbunching schemes share the same spirit, i.e., to take advantage of the small transverse emittance, the vertical emittance in our case, to generate microbunching with a shallow modulation strength. These TLC-based microbunching schemes can be viewed as partial transverse-longitudinal emittance exchange at the optical laser wavelength range. They do not necessarily need to be a complete emittance exchange since for microbunching, the most important coordinate is $z$, and $\delta$ is relatively less important.  As we will show soon, although the spirit is the same, given the same level of modulation laser power, the physical realization of energy modulation-based TLC microbunching schemes turn out to be more effective for our SSMB application compared to angular modulation-based schemes.



\subsection{Modulation Strength}

\subsubsection{TEM01 Mode Laser-Induced Angular Modulation}
After deriving the bunching factor, now we derive the laser-induced angular modulation strength for quantitative evaluation. We start with the usual angular modulation proposal by applying a TEM01 mode laser in an undulator~\cite{Xiang2010}.  The electric field of a Hermite-Gaussian TEM01 mode laser polarized in the horizontal plane is~\cite{Chao2023FocusedLaser}
{}
	\begin{equation}\label{eq:TEM01}
		\left(
		\begin{matrix}
			E_{x}\\
			E_{y}\\
			E_{z}\\
			cB_{x}\\
			cB_{y}\\
			cB_{z}
		\end{matrix}
		\right)=
		E_{x0}e^{ik_{L}s-i\omega_{L} t}\left(-iZ_{R}Q\right)^{2}\text{exp}\left[i\frac{k_{L}Q}{2}\left(x^{2}+y^{2}\right)\right]\left(\frac{2\sqrt{2}}{w_{0}}\right)
		\left(
		\begin{matrix}
			y\\
			0\\
			-Qxy\\
			i\frac{Qx}{k_{L}}-Q^{2}xy^{2}\\
			\left(Q^{2}x^{2}-\frac{iQ}{k_{L}}+1\right)y\\
			\frac{i}{k_{L}}-Qy^{2}
		\end{matrix}
		\right).
	\end{equation}
{}
The relation between $E_{x0}$ and the laser peak power for a TEM01 mode laser is given by
\begin{equation}\label{eq:ExoTEM01}
	P_{L}=\frac{E_{x0}^{2}Z_{R}\lambda_{L}}{2Z_{0}}.
\end{equation}
Note there is a factor of two difference in the above laser power formula compared to the case of a TEM00 mode laser.
The electron wiggles in a horizontal planar undulator according to Eq.~(\ref{eq:undulatorxs}),
and the laser-electron  exchanges energy according to Eq.~(\ref{eq:LEinteraction}).
{Making the same assumption} and following similar procedures as before, we can get  the integrated modulation voltage induced by a TEM01 mode laser in the undulator
\begin{equation}
	\begin{aligned}
		V_{L}
		&= \text{Re}\left[E_{x0}\frac{K}{\gamma}e^{i\phi_{0}}\frac{[JJ]}{2}\frac{2\sqrt{2}}{w_{0}}y\int_{-\frac{L_{u}}{2}}^{\frac{L_{u}}{2}}\frac{1}{\left(1+i\frac{s}{Z_{R}}\right)^{2}}ds\right]\\
		&= E_{x0}\frac{K[JJ]}{\gamma}\frac{2\sqrt{2}}{w_{0}}yZ_{R}\frac{\frac{L_{u}}{2Z_{R}}}{1+\left(\frac{L_{u}}{2Z_{R}}\right)^{2}}\cos\phi_{0}.\\
	\end{aligned}
\end{equation}
We will choose $\phi_{0}=0$ to maximize the modulation voltage. Put in the expression of $E_{x0}$ from Eq.~(\ref{eq:ExoTEM01}) and $w_{0}=\sqrt{\frac{Z_{R}\lambda_{L}}{\pi}}$ we have
\begin{equation}
	\begin{aligned}
		V_{L}&=\frac{4K[JJ]}{\gamma}\frac{\sqrt{\pi P_{L}Z_{0}}}{\lambda_{L}}\frac{\frac{L_{u}}{2Z_{R}}}{1+\left(\frac{L_{u}}{2Z_{R}}\right)^{2}}y.
	\end{aligned}
\end{equation}
The induced energy modulation strength with respect to $y$ around the zero-crossing phase is then
\begin{equation}\label{eq:angular}
	{g}=\frac{\partial\left(\frac{eV_{L}}{E_{0}}\right)}{\partial{y}}=\frac{2ek_{L}K[JJ]}{\gamma^{2}m_{e}c^{2}}\sqrt{\frac{P_{L}Z_{0}}{\pi} }\frac{\frac{L_{u}}{2Z_{R}}}{1+\left(\frac{L_{u}}{2Z_{R}}\right)^{2}}.
\end{equation}
The symplecticity of the dynamical system will require that this formula also gives the linear angular chirp strength around the zero-crossing phase.
It is interesting to note that, given the laser power, the modulation kick strength depends on the ratio between $Z_{R}$ and $L_{u}$, {instead of} their absolute values. 

One may wonder that when $\frac{L_{u}}{2Z_{R}}$ is fixed, the induced angular chirp strength is independent of the modulator length $L_{u}$. While in a TEM00 mode laser modulator, as given in Eq.~(\ref{eq:TEM00h}), we have the energy modulation strength proportional to $\sqrt{L_{u}}$. Mathematically this is because in the expression of TEM01 mode laser field, there is a term $\left(-iZ_{R}Q\right)^{2}$, while in TEM00 mode laser, this term is $\left(-iZ_{R}Q\right)$. Physically it means the on-axis power density of a TEM01 mode laser decays faster, compared to that of a TEM00 mode laser, when we go away from the laser waist. This may not be surprising if we keep in mind that the intensity peaks of a TEM01 mode laser are not on-axis. 

According to Eq.~(\ref{eq:angular}), the maximal modulation is realized when $Z_{R}=\frac{L_{u}}{2}$ and the value is 
\begin{equation}
	{g}_{\text{max}}=\frac{\partial\left(\frac{eV_{L}}{E_{0}}\right)}{\partial{y}}=\frac{ek_{L}K[JJ]}{\gamma^{2}m_{e}c^{2}}\sqrt{\frac{P_{L}Z_{0}}{\pi} }.
\end{equation}
For example, if $E_{0}=600$ MeV, $\lambda_{L}=1064$~nm,  $\lambda_{u}=8$~cm ($B_{0}=1.13$ T), $L_{u}=0.8$~m ($N_{u}=10$),  $Z_{R}=\frac{L_{u}}{2}$, then for $P_{L}=1$~MW the induced angular chirp strength is ${g}=0.55\ \text{m}^{-1}$. As a comparison, the energy chirp strength induced by such a 1 MW TEM00 laser modulator as evaluated before is $h=955\ \text{m}^{-1}$. So generally, a TEM01 mode laser is not effective in imprinting angular modulation. Actually we will see in the next section, even a dual-tilted-laser setup is still not effective enough for our application.

\subsubsection{Dual-Titled-Laser-Induced Angular Modulation}

Another way to imprint angular modulation on the electron beam is using a titled incident TEM00 mode laser to modulate the beam in an undulator. To further lower the required laser power, we can apply dual tilted laser with a crossing configuration~\cite{Wang2020,Feng2022}. Here in this paper, we focus on the angular modulation scheme based on the dual-tilted-laser (DTL) setup. Note that if we want to use DTL for energy modulation, the two lasers should {be} in phase {to make the two laser induced energy modulations add}. While for angular modulation, they should be $\pi$-phase shifted with respect to each other. {This is because for angular modulation, the particle on the reference orbit should get zero energy kick. Only when the particle transverse coordinate is nonzero will it get an energy kick. So the two laser induced energy modulation should cancel on axis.}

\begin{figure}[tb]
	\centering 
	\includegraphics[width=0.7\columnwidth]{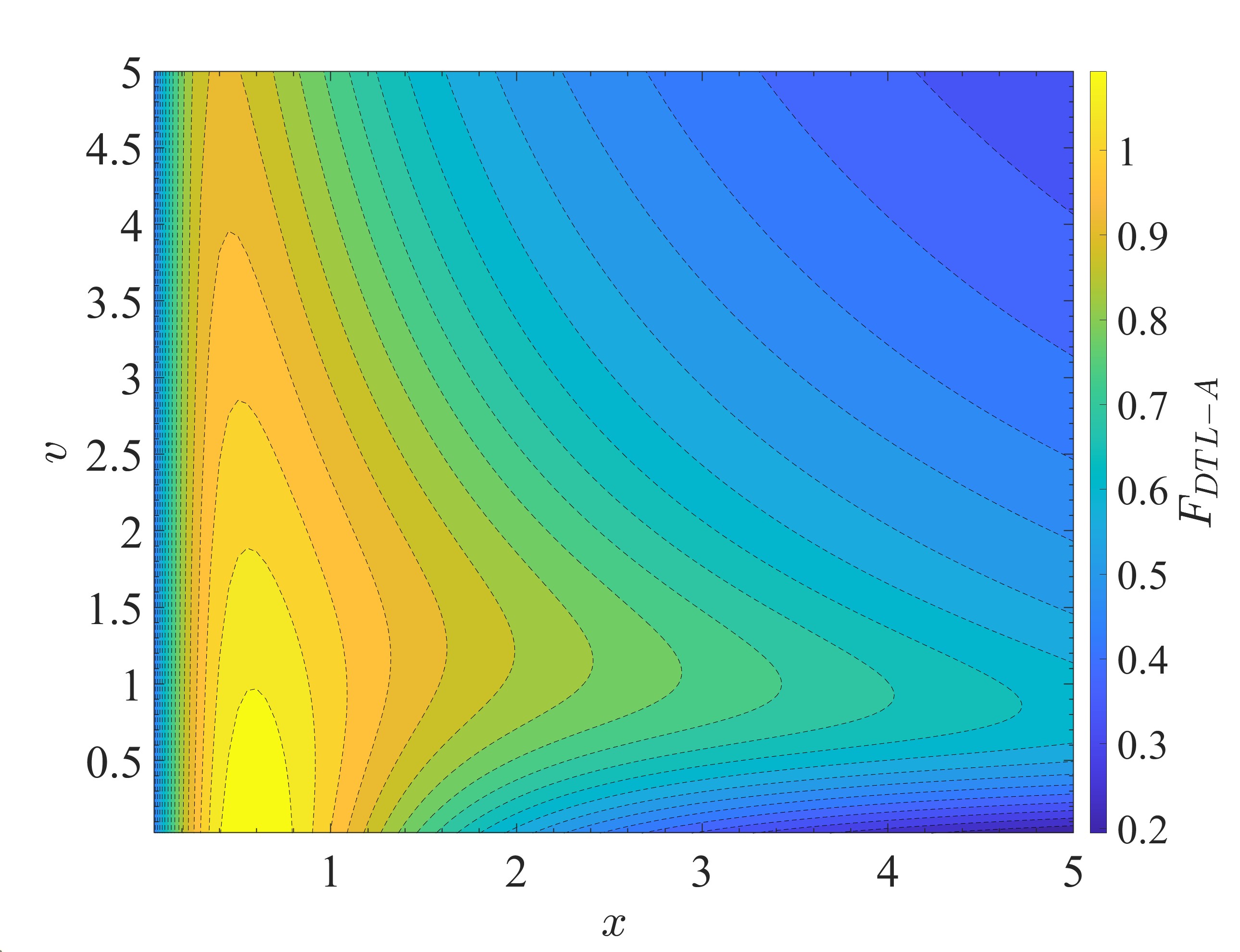}
	\caption{
		\label{fig:FContourChao} 
		Contour plot of $F_{\text{DTL-A}}(x,v)$ given by Eq.~(\ref{eq:DTLF}).  }
\end{figure}

To induce vertical angular modulation, we let the two lasers cross in $y$-$z$ plane and {be} polarized in the horizontal plane. The laser field of a normal indicent TEM00 laser is given by Eq.~(\ref{eq:TEM00}).
Assuming that the two lasers are $\pi$-phase-shifted with respect to each other and have the same amplitude. In addition, we assume that $\theta_{2}=-\theta_{1}=-\theta$, and the two lasers have the same Rayleigh length. Then the superimposed field is
{}
	\begin{equation}
		\begin{aligned}
			\left(
			\begin{matrix}
				E_{x}\\
				E_{y}\\
				E_{z}\\
			\end{matrix}
			\right)_{\text{unrot}}&=E_{x0}e^{ik_{L}\left(-y\sin\theta+s\cos\theta\right)-i\omega_{L} t}\frac{\text{exp}\left[-\frac{k_{L}(x^{2}+\left(y\cos\theta+s\sin\theta\right)^{2})}{2(Z_{R}+i\left(-y\sin\theta+s\cos\theta\right))}\right]}{1+i\frac{\left(-y\sin\theta+s\cos\theta\right)}{Z_{R}}}
			\left(
			\begin{matrix}
				1\\
				-i\frac{x}{Z_{R}+i\left(-y\sin\theta+s\cos\theta\right)}\sin\theta\\
				-i\frac{x}{Z_{R}+i\left(-y\sin\theta+s\cos\theta\right)}\cos\theta\\
			\end{matrix}
			\right)\\
			&\ \ \ \ -E_{x0}e^{ik_{L}\left(y\sin\theta+s\cos\theta\right)-i\omega_{L} t}\frac{\text{exp}\left[-\frac{k_{L}(x^{2}+\left(y\cos\theta-s\sin\theta\right)^{2})}{2(Z_{R}+i\left(y\sin\theta+s\cos\theta\right))}\right]}{1+i\frac{\left(y\sin\theta+s\cos\theta\right)}{Z_{R}}}
			\left(
			\begin{matrix}
				1\\
				i\frac{x}{Z_{R}+i\left(y\sin\theta+s\cos\theta\right)}\sin\theta\\
				-i\frac{x}{Z_{R}+i\left(y\sin\theta+s\cos\theta\right)}\cos\theta\\
			\end{matrix}
			\right).
		\end{aligned}
	\end{equation}
	As before, we will focus on the impact of $E_{x}$ on laser-electron interaction, and ignore the contribution from $E_{z}$. 
	Note that if $y=0$, then $E_{x}=0$. We want to know $\frac{\partial E_{x}}{\partial y}$ when $y$ is close to zero.  
	For this purpose, we do Taylor expansion of the above horizontal electric field expression with respect to $\theta$ when $\theta$ is small,  
	\begin{equation}
		\begin{aligned}
			E_{x}&=E_{x0}\frac{ Z_R \left(2 i s k_L Z_R+2 k_L Z_R^2+x^2 k_L+y^2 k_L-2 Z_R-2 i s\right) \exp \left[-\frac{k_L \left(x^2+y^2\right)}{2 \left(Z_R+i z\right)}+i k_{L}s-i\omega_{L}t \right]}{\left(s-i Z_R\right){}^3} \theta  y.
		\end{aligned}
	\end{equation}
	When $x,y\ll Z_{R}$, $\lambda_{L}\ll Z_{R}$, we have
	\begin{equation}
		\begin{aligned}
			E_{x}&\approx 
			-i2k_{L}E_{x0}\frac{ \exp \left[-\frac{\left(x^2+y^2\right)}{w_{0}^{2} \left(1+i\frac{s}{Z_{R}}\right)}+i k_{L}s-i\omega_{L}t \right]}{\left(1+i\frac{s}{Z_{R}} \right)^2} \theta  y.
		\end{aligned}
	\end{equation}
	But note that when $L_{u}\theta$ is comparable to $w_{0}$, the term $s\sin\theta$ should be kept in the exponential term. Also for the laser phase, we should use the more accurate $s\cos\theta$. These arguments have been explained before also when we analyze the application of DTL for energy modulation.  So the more correct approximate expression of $E_{x}$ is
	\begin{equation}
		\begin{aligned}
			E_{x}&\approx 
			-i2k_{L}E_{x0}\frac{ \exp \left(-\frac{s^{2}\theta^{2}}{w_{0}^{2} \left(1+i\frac{s}{Z_{R}}\right)}+i k_{L}s\cos\theta-i\omega_{L}t \right)}{\left(1+i\frac{s}{Z_{R}} \right)^2} \theta  y.
		\end{aligned}
	\end{equation}
	Again taking the same notation as given in Eq.~(\ref{eq:v}), the integrated modulation voltage induced by a DTL (two lasers $\pi$-phase-shifted with respect to each other) in a planar horizontal undulator can be calculated to be
	\begin{equation}
		\begin{aligned}
			V_{L}&\approx\text{Re}\left[\int_{-\frac{L_{u}}{2}}^{\frac{L_{u}}{2}}v_{x}E_{x}\frac{ds}{c}\right]\\
			&\approx k_{L}E_{x0}\frac{ K[JJ]}{\gamma}\theta  yZ_{R}\\
			&\left[\int_{-\frac{L_{u}}{2Z_{R}}}^{\frac{L_{u}}{2Z_{R}}}\frac{ \exp \left(-v\frac{{u}^{2}}{\left(1+{u}^{2}\right)} \right)}{(1+{u}^{2})^{2}}\left[\left(1-{u}^{2}\right)\cos\left(v\frac{{u}^{3}}{\left(1+{u}^{2}\right)}\right)+2{u}\sin\left(v\frac{{u}^{3}}{\left(1+{u}^{2}\right)}\right)\right] d{u}\right]\sin\phi_{0}.\\
		\end{aligned}
	\end{equation} 
	We choose $\phi_{0}=\frac{\pi}{2}$ to maximize $V_{L}$. 
	Therefore, we get the maximum linear angular chirp strength as
	\begin{equation}
		\begin{aligned}
			{g}&=\frac{\partial{\left(\frac{eV_{L}}{E_{0}}\right)}}{\partial y}=\frac{ek_{L}K[JJ]  }{\gamma^{2}m_{e}c^{2}}\sqrt{\frac{2P_{L}Z_{0}}{\lambda_{L}}}F_{\text{DTL-A}}\left(\frac{L_{u}}{2Z_{R}},\frac{k_{L}Z_{R}\theta^{2}}{2 }\right)\sqrt{L_{u}} \theta,\\
		\end{aligned}
	\end{equation}
	where
	\begin{equation}\label{eq:DTLF}
		\begin{aligned}
			F_{\text{DTL-A}}(x,v)&=\frac{1}{\sqrt{x}}\int_{-x}^{x}\frac{ \exp \left(-v\frac{{u}^{2}}{\left(1+{u}^{2}\right)} \right)}{(1+{u}^{2})^{2}}\left[\left(1-{u}^{2}\right)\cos\left(v\frac{{u}^{3}}{\left(1+{u}^{2}\right)}\right)+2{u}\sin\left(v\frac{{u}^{3}}{\left(1+{u}^{2}\right)}\right)\right]d{u}.
		\end{aligned}
	\end{equation}
{}
A contour plot of $F_{\text{DTL-A}}(x,v)$ is shown in Fig.~\ref{fig:FContourChao}.

\begin{figure}[H]
	\centering 
	\includegraphics[width=0.7\columnwidth]{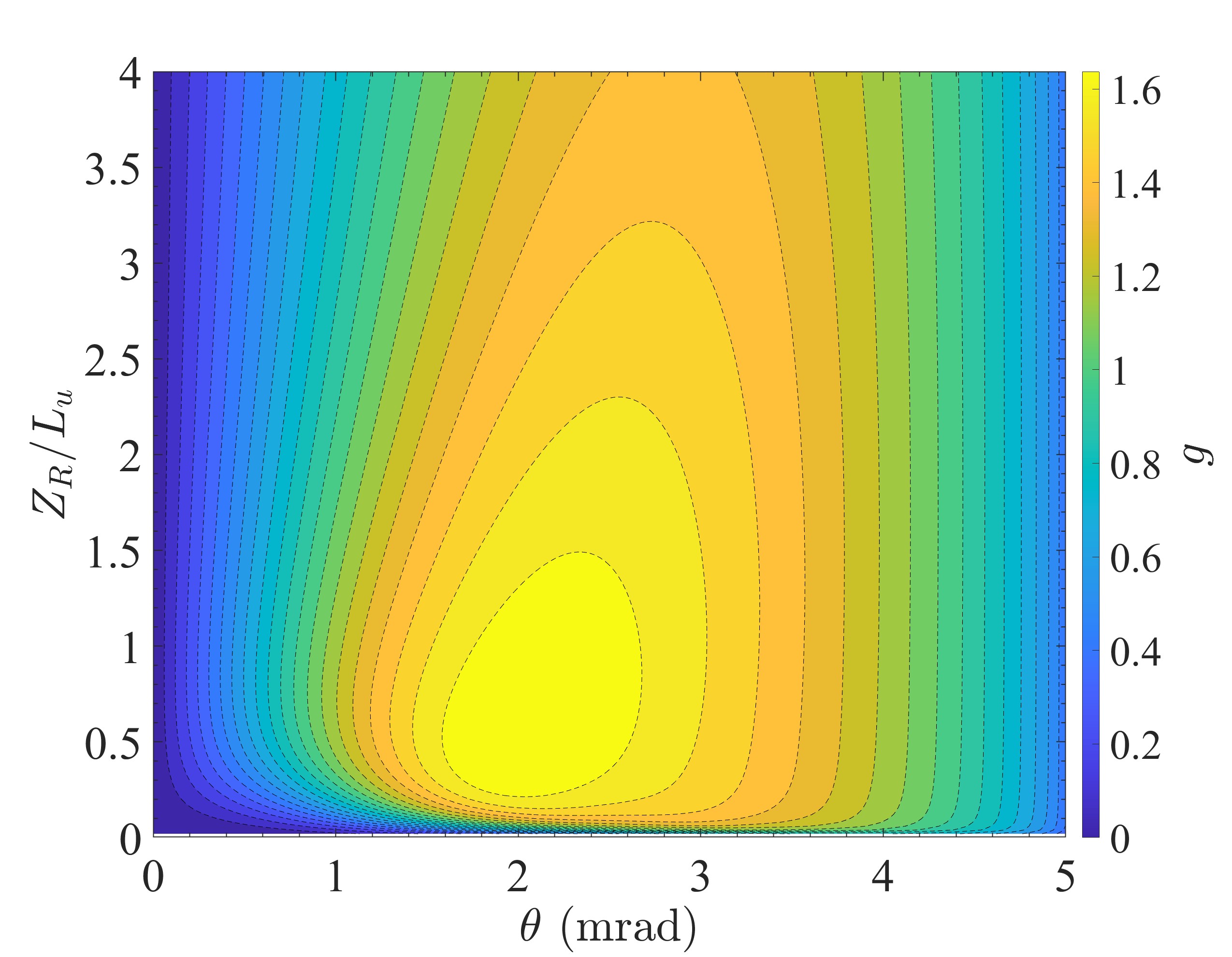}
	\caption{
		\label{fig:tContour800Chao} 
		Angular chirp strength  ${g}$ vs. $\theta$ and $Z_{R}/L_{u}$, keep $\lambda_{u}$ fixed when changing $\theta$. Parameters used: $E_{0}=600$ MeV, $\lambda_{L}=1064$~nm, $P_{L}=1$ MW, $\lambda_{u}=0.08$ m, $N_{u}=10$, $L_{u}=0.8$ m. }
\end{figure}

\begin{figure}[H]
	\centering 
	\includegraphics[width=0.7\columnwidth]{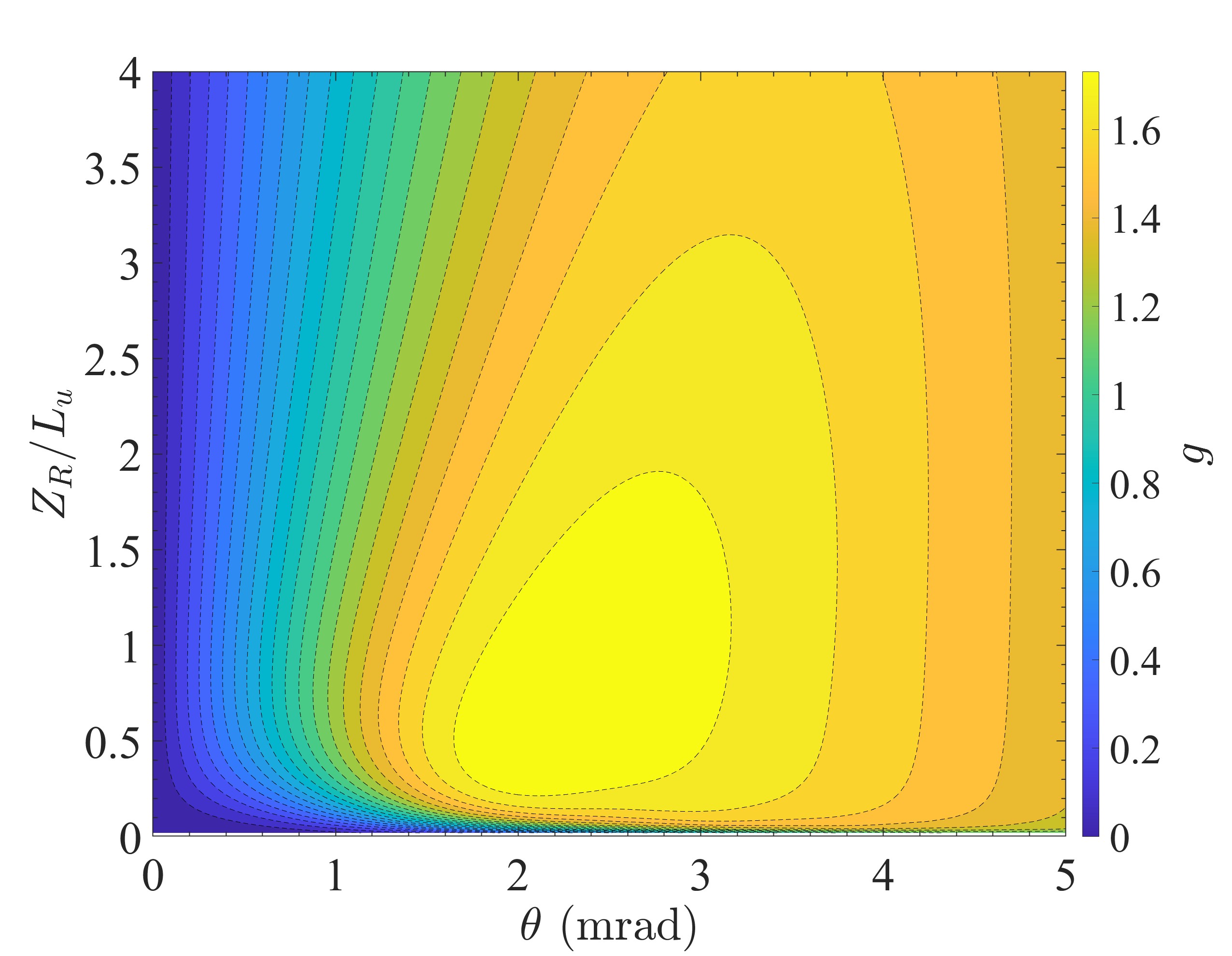}
	\caption{
		\label{fig:tContour800Chao2} 
		Angular chirp strength  ${g}$ vs. $\theta$ and $Z_{R}/L_{u}$, keep $B_{0}$ fixed when changing $\theta$. Parameters used: $E_{0}=600$ MeV, $\lambda_{L}=1064$~nm, $P_{L}=1$ MW, $B_{0}=1.2$~T, $L_{u}=0.8$ m. }
\end{figure}

Then the angular chirp strength introduced by a DTL compared to the energy chirp strength introduced by a single TEM00 laser modulator, i.e., that given in Eq.~(\ref{eq:TEM00h}), with the same laser parameters and undulator length can be expressed as
\begin{equation}
	\frac{{g}}{h}=\xi\theta.
\end{equation}
with
\begin{equation}
	\xi=\frac{F_{\text{DTL-A}} \left(\frac{L_{u}}{2Z_{R}},\frac{k_{L}Z_{R}\theta^{2}}{2 }\right)}{\frac{\tan^{-1}\left(\frac{L_{u}}{2Z_{R}}\right)}{\sqrt{\frac{L_{u}}{2Z_{R}}}}}\frac{K[JJ]}{K[JJ]|_{\theta=0}}.
\end{equation}
Note that $K[JJ]$ in the numerator is a function of $\theta$ {according to the off-axis resonance condition given in Eq.~(\ref{eq:offaxis})}.
Since 
$
\xi\sim1,
$
and $\theta$ is usually in mrad level, given the same laser power, the DTL-induced angular chirp strength will be much smaller than the energy modulation strength induced by a TEM00 mode laser. This observation has been supported by more quantitative calculation of the angular chirp strength induced by DTL as shown in Figs.~\ref{fig:tContour800Chao} and \ref{fig:tContour800Chao2}. As before, we have considered the case of keeping $\lambda_{u}$ or $B_{0}$ unchanged when we change the crossing angle $\theta$, respectively. In both cases, the maximal angular chirp strength induced with $P_{L}=1$ MW is about ${g}\approx1.7$ $\text{m}^{-1}$. While from the evaluation in Sec.~\ref{sec:TEM00}, at the same power level, a TEM00 mode laser modulation can induce an energy chirp strength of $h\sim955\ \text{m}^{-1}$. This as explained is because $\theta$ is only  $1\sim2$~mrad. 

So we can see that the DTL-induced angular chirp strength, although a factor of three larger than that induced by a single normally incident TEM01 mode laser with the same laser power,  is still generally quite small.  There are two reasons why DTL is not effective in imprinting angular modulation:
\begin{itemize}
	\item The crossing angle between the laser and the electron propagating directions results in that they have a rather limited effective interaction region. For example, if the crossing angle is $\theta=5$~mrad, and the undulator length is $L_{u}=0.8$ m. Then the center of electron beam and center of laser beam at the undulator entrance and exit depart from each other with a distance of $\frac{L_{u}}{2}\theta=2$ mm, which is a large value compared to the laser beam waist $w_{0}=\sqrt{\frac{Z_{R}\lambda_{L}}{\pi}}\approx\sqrt{\frac{L_{u}\lambda_{L}}{2\pi}}=368\ \mu$m and results in a very weak laser electric field felt by the electron there. 
	\item The decrease of undulator parameter $K$ with the increase of $\theta$ to meet the off-axis resonance condition, as can be seen in Figs.~\ref{fig:KThetaLmabdaUFixed} and \ref{fig:lambdaUVSTheta2}. 
\end{itemize}

Since the angular chirp strength is small, then according to Theorem Two, the required vertical beta function at the modulator $\beta_{y\text{M}}$ will be large. For example, if $\epsilon_{y}=4$~pm, $\sigma_{z\text{R}}=\sqrt{\epsilon_{y}\mathcal{H}_{y\text{R}}}=2$ nm, and ${g}=2\ \text{m}^{-1}$, then we need $\beta_{y\text{M}}=2.5\times10^{5}$~m, which is too large to be used in practice in a storage ring.   If we want to lower $\beta_{y\text{M}}$, it then means a higher modulation laser power is needed. This is the reason why we tend to use energy modulation-based coupling scheme for bunch compression or microbunching generation in GLSF SSMB. {Our analysis and conclusion here is consistent with what reported in Ref.~\cite{Liu2021} when comparing energy and angular modulation microbunching schemes for laser plasma accelerator based light source.}

Generally, when we compare the energy modulation-based and DTL-induced angular modulation-based bunch compression schemes, from the Theorem One and Two, i.e., Eqs.~(\ref{eq:theorem1}) and~(\ref{eq:theorem2}), for the same modulation laser wavelength $\lambda_{L}$ and power $P_{L}$, vertical emittance $\epsilon_{y}$  and target linear bunch length at the radiator $\sigma_{z\text{R}}=\sqrt{\epsilon_{y}\mathcal{H}_{y\text{R}}}$,  we have
\begin{equation}
\beta_{y\text{M}}({\text{Angular modulation}})\sim\frac{\mathcal{H}_{y\text{M}}({\text{Energy modulation}})}{\theta^{2}}.
\end{equation} 

\subsection{Realization Examples}
Similar to the section on energy modulation-based schemes, here we also introduce some realization examples of angular modulation-based microbunching.
The first proposal of applying the angular modulated beam for harmonic generation to our knowledge is from Ref.~\cite{Xiang2010}. Later an emittance exchange-based harmonic generation scheme is proposed in Ref.~\cite{Jiang2011}. These two schemes apply the TEM01 mode laser to induce angular modulation. Following these development, {there are proposals to realize angular modulation using TEM00 mode laser, with Ref.~\cite{Feng2018} using off-resonance laser, and Ref.~\cite{Wang2019}} using tilted incident laser. And later a dual-tilted-laser (DTL) modulation scheme is applied in emittance exchange at the optical laser wavelength range~\cite{Wang2020}. And most recently, the DTL scheme is proposed to compress the bunch length in SSMB and lower the requirement on laser power by a factor of four compared to a single-tilted laser scheme~\cite{Feng2022}. Note that for these angular modulation-based harmonic or bunch compression schemes, we have the inequality given in Eq.~(\ref{eq:theorem2}), i.e., our Theorem Two.

\section{1 kW GLSF SSMB EUV Source}\label{sec:application}

Our goal in this paper as stated is to find a solution for high-power EUV source based on SSMB, using parameters within the reach of present technology.  According to our analysis, generalized longitudinal strong focusing (GLSF) turns out to be the most promising scenario, compared to longitudinal weak focusing and longitudinal strong focusing. The key of a GLSF SSMB ring is the precision transverse-longitudinal coupling dynamics to utilize the ultrasmall natural vertical emittance in a planar electron storage ring for efficient microbunching formation. For our purpose we find energy modulation-based coupling schemes are preferred than angular modulation-based coupling ones, in lowering the required modulation laser power. So the conclusion is that we will use a TEM00 laser induced energy modulation-based coupling scheme in a GLSF SSMB storage ring. In this section, we first present a solution of 1~kW-average-power EUV source based on GLSF SSMB. More detailed analyses to support our solution are then developed. 



\subsection{A Solution of 1 kW EUV Source}

Based on all what we have studied in previous sections and the various important physical effects to be discussed in this section, here we present an example parameters set of an 1 kW-average-power EUV light source based on GLSF SSMB as shown in Tab.~\ref{tab:Chap6-GLSF-SSMB_para1064_600}.  All the parameters list should be doable from an engineering viewpoint. Such a table summarizes our investigations presented in this paper.

\begin{table}[H]
	\caption{\label{tab:Chap6-GLSF-SSMB_para1064_600}
		An example parameters set of a GLSF SSMB ring for 1~kW-average-power EUV radiation generation.}
	\centering
	\begin{tabular}{lll}  
		\hline
		Parameter & \multicolumn{1}{l}{\textrm{Value}}  & Description \\
		\hline 
		$E_{0}$ & $600$ MeV & Beam energy \\	
		$C_{0}$ & $\sim200$ m & Circumference \\
		${\eta}$ & ${\sim5\times10^{-3}}$ & {Phase slippage factor}\\
		$I_{P}$ & $40$ A & Peak current \\
		$f_{e}$ & 0.5\% & Electron beam filling factor \\
		$I_{A}$ & 200 mA & Average current \\
		$B_{\text{ring}}$ & 1.33 T & Bending magnet field in the ring \\
		$\rho_{\text{ring}}$ & 1.5 m & Bending radius in the ring \\
		$U_{0\text{dipoles}}$ &  7.7 keV & Radiation loss per particle per turn from ring dipoles\\
		$B_{0{w}}$ & 6 T & Bending field of damping wiggler\\
		$L_{{w}}$ & {40 m} & Total length of $N_{wc}$ identical damping wigglers \\
		${N_{wc}}$ & {20} & {Number of identical damping wigglers}\\
		${\lambda_{w}}$ & {$<0.168$ m} & {Wiggler period length}\\
		$U_{0{w}}$ &  {328 keV} & Radiation loss per particle per turn in damping wiggler\\
		$P_{R\text{beam}}$ & {68.3 kW} & Total radiation loss power of the electron beam\\ 
		${P_{RF}}$ & {$100\sim200$ kW} & {Total power consumption of the RF system}\\
		$\sigma_{\delta0}$ &  $4.2\times10^{-4}$ & Natural energy spread (without damping wiggler)\\
		$\sigma_{\delta{\text{IBS}}}$ &  $8.5\times10^{-4}$ & Energy spread with IBS (with damping wiggler)\\
		$\epsilon_{x}$ & 2 nm & Horizontal emittance	 \\ 
		$\epsilon_{y}$ & 40 pm & Vertical emittance	 \\ 
		$\tau_{y\text{RD}}$ & {2.38 ms} & Vertical radiation damping time with damping wiggler\\
		$\tau_{y\text{IBS}}$ & {7.11 ms} & Vertical IBS diffusion time\\
		\hline 
		$\lambda_{L}$ & 1064 nm & Modulation laser wavelength	\\
		$\sigma_{z\text{R}}$ & 2 nm & Linear bunch length at the radiator	\\ 
		$\mathcal{H}_{y\text{R}}$ & 0.1 $\mu$m & $\mathcal{H}_{y}$	at the radiator \\ 
		$\mathcal{H}_{y\text{M}}$ & 0.056 m & $\mathcal{H}_{y}$	at the modulator \\ 
		$h$ &  $1.33\times10^{4}\ \text{m}^{-1}$ & Modulator induced linear energy chirp strength \\ 
		$\lambda_{u\text{M}}$ &  0.1 m &  Modulator undulator period\\  
		$B_{0\text{M}}$ &  0.806 T &  Modulator peak magnetic flux density\\ 
				$K_{u\text{M}}$ &  7.53 & $K$ of modulator undulator\\ 
		$L_{u\text{M}}$ &  {1.5 m}  & Modulator length	\\
		$\Delta\epsilon_{y\text{M}}$ & {13.4 pm} & Modulators' quantum excitation to vertical emittance	 \\  
		$Z_{R}$ &  $\sim\frac{L_{u\text{M}}}{3}$ & Laser Rayleigh length \\ 
		$P_{LP}$ &  130 MW & Peak modulation laser power\\ 
		$f_{L}$ & 0.5\% & Laser beam filling factor \\
		$P_{LA}$ &  651 kW & Average  modulation laser power\\ 
		\hline
		$\lambda_{R}=\frac{\lambda_{L}}{79}$ & $13.5$ nm & Radiation wavelength\\
		{$b_{79}$} & 0.0675 & Bunching factor\\
		$\sigma_{\bot\text{R}}$ & 20 $\mu$m & Effective transverse electron  beam size at the radiator\\
		$\lambda_{u\text{R}}$ &  1.8 cm & Radiator undulator period\\
		$B_{0\text{R}}$ &  0.867 T &  Radiator peak magnetic flux density\\   
		$N_{u\text{R}}$ &  $79\times4$ & Number of undulator periods\\ 
		$L_{u\text{R}}$ &  5.69 m & Radiator length	\\ 
		$P_{RP}$ & 224 kW & Peak radiation power \\
		$P_{RA}$ & 1.12 kW & Average radiation power \\
		\hline 
	\end{tabular}
\end{table}


\subsection{Some Basic Considerations}

Now we present the detailed considerations and calculations to support our solution.
First are some basic considerations on parameters choice. As explained in the beginning of Sec.~\ref{sec:SSMB}, we will use a beam energy of $E_{0}=600$~MeV and a modulation laser wavelength of $\lambda_{L}=1064$~nm. In the GLSF scheme, a small vertical emittance is crucial. We assume that the vertical emittance used to accomplish our goal stated above is $\epsilon_{y}=40$ pm. One may notice that the vertical emittance we use is actually not extremely small. This conservative choice is mainly a refection of consideration for intra-beam scattering (IBS) to be introduced soon. To realize significant coherent EUV generation,  and considering that we may generate microbunching based on a coasting beam or RF bunched beam in the ring, {which is much longer than the modulation laser wavelength}, we may need to compress the linear bunch length $\sigma_{z\text{R}}=\sqrt{\epsilon_{y}\mathcal{H}_{y\text{R}}}$ to be as short as 2 nm. The bunching factor at 13.5 nm according to Eq.~(\ref{eq:BFADM}) is then 0.0675. 

With $\epsilon_{y}=40$~pm, to get the desired linear bunch length 2 nm at the radiator, we need
$
\mathcal{H}_{y\text{R}}=0.1\ \mu\text{m}.
$ {If $\beta_{y}$ at the radiator is around 1 m, then the required control precision of dispersion and dispersion angle at the radiator is at the level of 0.3 mm and 0.3 mrad.} Such a precise control of $\mathcal{H}_{y\text{R}}$ is challenging but realizable using present technology. {We remind the readers that the dispersion function is actually not well defined when the system is transverse-longitudinal coupled, and $\mathcal{H}_{y}$ here should be replaced by the our defined generalized beta function $\beta_{55}^{II}$. But here we still use the classical definition of $\mathcal{H}_{y}$ in getting the number for the above $D_{x}$ and $D_{x}'$ control precision to give the readers a more concrete feeling. Following similar line of thought, we also need a precision control of $\mathcal{H}_{x}$ or $\beta_{55}^{I}$ at the radiator, since the horizontal emittance is even larger than the vertical one. Generally, we require a precision control of both $D_{x,y}$ and $D_{x,y}'$. Besides, we also need to ensure the coupling of horizontal emittance $\epsilon_{x}$ to the vertical plane to be less than 1\% since our applied $\epsilon_{x}=2$~nm is about two orders of magnitude larger than $\epsilon_{y}=40$ pm.} 

{Another important parameter is the beam current, both the average and peak one. First we observe that given the same average beam current, the average radiation power will be higher with the decreasing of the beam filling factor $f_{e}$. This is because the peak power of the coherent radiation is proportional to the peak current squared $P_{RP}\propto I_{P}^{2}\propto I_{A}^{2}f_{e}^{-2}$, where $I_{P}$ and $I_{A}$ are the peak and average current, and $f_{e}$ is the electron beam filling factor. The average radiation power is then $P_{RA}=P_{RP}f_{e}\propto I_{A}^{2}f_{e}^{-1}$.  So we tend to choose as high average current and as small filling factor as we can, as long as collective effects like intra-beam scattering and coherent synchrotron radiation to be analyzed soon do not degrade the beam properties. We have applied a 40 A peak current and a 200 mA average current in our solution. Note that in the calculation of filling factor, for simplicity we have assumed the beam current is like square waves. The real beam distribution is more like a smooth curve, for example a Gaussian profile if there is only a single RF cavity in the ring. But this observation only results in some numerical factor adjustment and does not affect the coral part of our analysis. So in Tab.~\ref{tab:Chap6-GLSF-SSMB_para1064_600} and below, we take this simplification of a square wave current distribution.}

\subsection{Quantum Excitation Contribution to Vertical Emittance}\label{sec:emitContribution}

After the general considerations, let us now take a closer look at the critical parameter $\epsilon_{y}$. It turns out the first contribution of vertical emittance is the quantum excitation in the GLSF section itself, since $\mathcal{H}_{y}$ there is nonzero. More clearly, according to our Theorem One, i.e., Eq.~(\ref{eq:theorem1}), the modulator should be placed at a dispersive location if we use energy modulation-based coupling for bunch compression or harmonic generation. Therefore, the quantum excitation of all bend-like elements, like dipoles, modulators and radiator in the GLSF section, will contribute to the vertical emittance. 

Like the calculation in Eq.~(\ref{eq:EmittanceZ}), the quantum excitation contribution of a radiator to $\epsilon_{y}$ is
\begin{equation}\label{eq:verticalEmittanceRadiator}
	\begin{aligned}
		\Delta\epsilon_{y\text{R}}
		&=C_{q}\frac{\gamma^{2}}{J_{y}}\frac{1}{I_{2}}\times\frac{\mathcal{H}_{y\text{R}}}{\rho_{0\text{R}}^{3}}\frac{4}{3\pi}L_{u\text{R}},
	\end{aligned}
\end{equation}
where $L_{u\text{R}}$ is the radiator undulator length, $\rho_{0\text{R}}$ is the bending radius corresponds to the peak magnetic field of the radiator $B_{0\text{R}}$.
Assuming that the radiation energy loss in the GLSF section is much less than that induced by the bending magnets in the ring, {then we have $I_{2}\approx\frac{2\pi}{\rho_{\text{ring}}}$}. Taking the the approximation $J_{y}\approx1$, {in an easy-to-use form} we have
\begin{equation}\label{eq:verticalEmittanceRadiator2}
	\begin{aligned}
		\Delta\epsilon_{y\text{R}}[\text{nm}]
		&=8.9B_{\text{ring}}^{-1}[\text{T}]B_{0\text{R}}^{3}[\text{T}]\mathcal{H}_{y\text{R}}[\text{m}]L_{u\text{R}}[\text{m}].
	\end{aligned}
\end{equation}
For the parameters given in Tab.~\ref{tab:Chap6-GLSF-SSMB_para1064_600},  $B_{\text{ring}}=1.33$ T, $\lambda_{u\text{R}}=1.8$ cm ($B_{0\text{R}}=0.867$~T), $\mathcal{H}_{y\text{R}}=0.1\ \mu$m, $L_{u\text{R}}=5.69$~m ($N_{u\text{R}}=4\times79$) , then $\Delta\epsilon_{y\text{R}}=2.5$ fm, which is much less than $\epsilon_{y}$. Generally, since $\mathcal{H}_{y}$ at the radiator is quite small, the contribution of radiator to the vertical emittance is not the dominant one, compared to that from modulators and dipoles to be introduced.

Similar to the analysis for radiator, the contribution of two GLSF modulators to $\epsilon_{y}$ is
\begin{equation}\label{eq:verticalEmittance}
	\begin{aligned}
		\Delta\epsilon_{y\text{M}}[\text{nm}]
		&=17.8B_{\text{ring}}^{-1}[\text{T}]B_{0\text{M}}^{3}[\text{T}]\mathcal{H}_{y\text{M}}[\text{m}]L_{u\text{M}}[\text{m}],
	\end{aligned}
\end{equation}
where $B_{0\text{M}}$ and $L_{u\text{M}}$ are the peak magnetic field and length of the modulator undulators, respectively.
For the parameters given in Tab.~\ref{tab:Chap6-GLSF-SSMB_para1064_600}, $B_{\text{ring}}=1.33$ T, $B_{0\text{M}}=0.806$~T, $\mathcal{H}_{y\text{M}}=0.056$ m, {$L_{u\text{M}}=1.5$ m, then $\Delta\epsilon_{y\text{M}}=592$ pm,} which is a quite large value compared to that contributed from radiator. This is mainly due to the fact that $\mathcal{H}_{y\text{M}}\gg\mathcal{H}_{y\text{R}}$. As we will introduce soon, we can apply a horizontal planar damping wiggler to control this contribution by increasing the radiation damping rate.

In the above analysis, we have assumed $\mathcal{H}_{y}$ is a constant value {throughout} an undulator. This strictly speaking is not true, since there will be intrinsic dispersion generated inside an undulator. We can refer to the transfer matrix of an undulator or laser modulator to study the evolution of $\mathcal{H}_{y}$ in an undulator radiator and modulator, to get a more accurate evaluation of their quantum excitation contribution, {as to be shown in Sec.~\ref{sec:DWEmit}}. Our calculation shows that this will result in a significant difference for radiator's contribution to $\epsilon_{y}$, but not much difference for the modulators. However, since the radiator's contribution to $\epsilon_{y}$ is negligible small compared to that of modulators, here we still use the simplified formula Eq.~(\ref{eq:verticalEmittance}) in our {following} discussion. 

There are also vertical bending magnets in the GLSF section for optics manipulation to fulfill the bunch compression or harmonic generation condition. But in principle we can use weak dipole field to minimize their quantum excitation contribution to vertical emittance, and satisfy the symplectic optics requirement at the same time. Of course, the total length of these dipoles should not be too long. Therefore, in our present evaluation, we will assume that the quantum excitation contribution from the two modulators are the dominant source of $\epsilon_{y}$ if we consider single-particle dynamics alone.

\subsection{Application of Damping Wigglers}

\subsubsection{To Speed Up Damping}
It is desired that $\Delta\epsilon_{y\text{M}}$ is only a small portion of our desired $\epsilon_{y}$, since then it provides room for other contribution of vertical emittance, like IBS and that from $x$-$y$ coupling. In principle, we can also use a weaker modulator field to weaken quantum excitation, but this then means the laser-electron interaction will be less efficient, and a larger laser power is needed if we want to imprint the same energy modulation strength.
Instead, the solution we choose is to increase the radiation damping rate per turn. To speed up damping, which is helpful in controling the vertical emittance growth from both quantum excitation and IBS, we may invoke one or multiple damping wigglers. We can put horizontal planar wigglers at dispersion-free locations. In such a way, the damping wiggler will contribute only damping and no excitation to the vertical emittance. Assuming that
\begin{equation}
	U_{0{w}}=R_{{w}} U_{0\text{dipoles}},
\end{equation}
with $U_{0\text{dipoles}}=C_{\gamma}{E_{0}^{4}}/{\rho_{\text{ring}}}$ the radiation energy loss per particle per turn from the bending magnets in the ring, $U_{0{w}}$ the radiation loss from the damping wiggler, and
\begin{equation}\label{eq:R}
	R_{{w}}=\frac{U_{0{w}}}{U_{0\text{dipoles}}}=\frac{1}{2}\left(\frac{B_{0{w}}}{B_{\text{ring}}}\right)^{2}\frac{L_{{w}}}{2\pi\rho_{\text{ring}}},
\end{equation}
where $B_{0{w}}$  and $L_{{w}}$ are the peak magnetic field and length of the damping wiggler, respectively.
Since the damping rate is proportional to the radiation energy loss per turn, then the damping constant will be a factor of $R_{{w}}$ larger by applying the damping wiggler
\begin{equation}
	\alpha_{V}=(1+R_{{w}})\alpha_{V0},
\end{equation}
with $\alpha_{V0}$ the natural vertical damping rate without damping wiggler.
Then the above evaluated emittance growth from radiaotr, modulators and dipoles will become $\frac{1}{1+R_{{w}}}$ of the original value. For example for the modulators, we have
\begin{equation}\label{eq:verticalEmittanceWiggler}
	\begin{aligned}
		\Delta\epsilon_{y\text{M}}[\text{nm}]
		&=\frac{17.8B_{\text{ring}}^{-1}[\text{T}]B_{0\text{M}}^{3}[\text{T}]\mathcal{H}_{y\text{M}}[\text{m}]L_{u\text{M}}[\text{m}]}{1+R_{{w}}}.
	\end{aligned}
\end{equation}
Put the above relation in another way
\begin{equation}
	L_{u\text{M}}[\text{m}]\approx56.2\left(1+R_{{w}}\right)\frac{B_{\text{ring}}[\text{T}]\Delta\epsilon_{y\text{M}}[\text{pm}]}{\mathcal{H}_{y\text{M}}[\mu\text{m}]B_{0\text{M}}^{3}[\text{T}]}.
\end{equation}

We will use a rather strong superconducting damping wiggler to speed up damping to fight against the IBS diffusion and quantum excitation to maintain a small vertical emittance. For the parameters given in Tab.~\ref{tab:Chap6-GLSF-SSMB_para1064_600}, $B_{0\text{ring}}=1.33$ T, $U_{0\text{dipoles}}=7.7$ keV, $B_{0{w}}=6$ T and {$L_{{w}}=40$~m}, then {$U_{0{w}}=328$~keV and $R_{{w}}=42.9$}. By applying such a strong damping wiggler, we now have {$\Delta\epsilon_{y\text{M}}=13.4$~pm}, which is a factor of {three} smaller than the desired $\epsilon_{y}=40$ pm  and should be acceptable. Assuming the circumference of the ring is $C_{0}=200$ m, the longitudinal and vertical radiation damping time are correspondingly
\begin{equation}\label{eq:RDWiggler}
	{\tau_{\delta\text{RD}}=1.19\ \text{ms},\ \tau_{y\text{RD}}=2.38\ \text{ms}.}
\end{equation} 
We will compare this radiation damping speed with that of IBS diffusion later.

\subsubsection{{Impact of Damping Wigglers to Energy Spread, Horizontal Emittance and Phase Slippage}}\label{sec:DWEmit}

Our primary goal of applying damping wiggler is to speed up radiation damping, but the damping wiggler also contributes to quantum excitation, thus may affect the energy spread and horizontal emittance. Let us investigate the energy spread first. Considering both the ring dipoles and damping wiggler, the new equilibrium energy spread is
\begin{equation}\label{eq:equilibriumEnergySpread}
	\sigma_{\delta{w}}=\sigma_{\delta0}\sqrt{\frac{1+\frac{4}{3\pi}\left(\frac{B_{0{w}}}{B_{\text{ring}}}\right)^{3}\frac{L_{u}}{2\pi\rho_{\text{ring}}}}{1+\frac{1}{2}\left(\frac{B_{0{w}}}{B_{\text{ring}}}\right)^{2}\frac{L_{u}}{2\pi\rho_{\text{ring}}}}},
\end{equation}
where 
\begin{equation}
	\sigma_{\delta0}=\sqrt{\frac{C_{q}}{J_{z}}\frac{\gamma^{2}}{\rho_{\text{ring}}}}
\end{equation}
is the natural energy spread if there is no damping wiggler. 
Nominally, we have the longitudinal damping partition $J_{z}\approx2$. 
So 
\begin{equation}
	\sigma_{\delta0}\approx4.69\times10^{-4}B^{\frac{1}{2}}_{\text{ring}}[\text{T}]E_{0}^{\frac{1}{2}}[\text{GeV}].
\end{equation}
For example, if $B_{0{w}}=1.33$ T, and $E_{0}=600$ MeV, then $\sigma_{\delta0}=4.2\times10^{-4}$. When $\frac{1}{2}\left(\frac{B_{0{w}}}{B_{\text{ring}}}\right)^{2}\frac{L_{u}}{2\pi\rho_{\text{ring}}}\gg1$ and $\frac{4}{3\pi}\left(\frac{B_{0{w}}}{B_{\text{ring}}}\right)^{3}\frac{L_{u}}{2\pi\rho_{\text{ring}}}\gg1$, which means the energy spread is dominant by the damping wiggler, we have
\begin{equation}\label{eq:equilibriumEnergySpread2}
	\begin{aligned}
		\sigma_{\delta{w}}&\approx\sigma_{\delta0}\sqrt{\frac{8}{3\pi}\frac{B_{0{w}}}{B_{\text{ring}}}}\\
		&\approx4.32\times10^{-4}B^{\frac{1}{2}}_{0{w}}[\text{T}]E_{0}^{\frac{1}{2}}[\text{GeV}].
	\end{aligned}
\end{equation}
So given the beam energy, the new equilibrium energy spread will depends solely on the peak magnetic field of the damping wiggler $B_{0{w}}$. For example, if  $E_{0}=600$ MeV and $B_{0{w}}=6$~T, then the asymptotic energy spread is $\sigma_{\delta{w}}=8.2\times10^{-4}$. This energy spread can affect coherent EUV radiation power in a long radiator as will be studied in Sec.~\ref{sec:PR}.

Now let us look at the impact on horizontal emittance. We just said that we can place the damping wigglers at horizontally dispersion-free locations to minimize its quantum excitation on $\epsilon_{x}$. But there will be some intrinsic horizontal dispersion and dispersion angle and thus $\mathcal{H}_{x}$ generated inside the wiggler and the strong field strength raises the concern that the quantum excitation of damping wiggler may result in horizontal emittance growth. Now we present a quantitative evaluation of this. Part of the content in this section can also be found in Ref.~\cite{Deng2024Wiggler} 

Usually the central part of the wiggler has a sinusoidal field strength pattern along the longitudinal axis. We set the origin of the global path length coordinate $s=0$ to be the location of the peak magnetic field closest to the wiggler center. Note that this choice of origin location has a correspondence to the transfer matrix of wiggler to be given soon. Then the vertical magnetic field of a horizontal planar wiggler is
\begin{equation}
	B_{y}=B_{0w}\cosh(k_{x}x)\cosh(k_{y}y)\cos(k_{w}s),
\end{equation}
with $B_{0w}$ the peak magnetic field and $k_{w}={2\pi}/{\lambda_{w}}$ the wavenumber of wiggler, and $k_{x}^{2}+k_{y}^{2}=k_{w}^{2}$.
The linear transfer matrix of ${\bf X}$ from $s=0$ to $s\in[-\frac{L_{w}}{2},\frac{L_{w}}{2}]$ with $L_{w}$ the wiggler length is then~\cite{ZhaoJingyuan2023}
{}
	\begin{equation}
		{\bf W}(s|0)=\left(
		\begin{matrix}
			
			1&s&0&0&0&-\frac{K}{\gamma k_{w}}[1-\cos(k_{w}s)]\\
			0&1&0&0&0&-\frac{K}{\gamma}\sin(k_{w}s)\\
			0&0&\cos(k_{y}s)&\frac{\sin(k_{y}s)}{k_{y}}&0&0\\
			0&0&-k_{y}\sin(k_{y}s)&\cos(k_{y}s)&0&0\\
			W_{51}&W_{52}&0&0&1&W_{56}\\
			0&0&0&0&0&1
		\end{matrix}
		\right),
	\end{equation}
{}
where $W_{51}=-W_{26}$, $W_{52}=-sW_{26}+W_{16}$ and 
\begin{equation}
	W_{56}
	=\frac{2\lambda_{0}}{\lambda_{w}}s+\frac{K^{2}}{\gamma^{2}}\left[\frac{\sin(2k_{w}s)-4\sin(k_{w}s)}{4k_{w}}\right],
\end{equation}
with $\lambda_{0}=\frac{1+K^{2}/2}{2\gamma^{2}}\lambda_{w}$ being the fundamental on-axis resonant wavelength, $\gamma$ is the Lorentz factor, $K$ is the dimensionless undulator parameter of the wiggler, with $e$ the elementary charge, $m_{e}$ the electron mass, $c$ the speed of light in free space.

In a planar uncoupled ring, the normalized eigenvector of the storage ring one-turn map corresponding to the horizontal eigenmode at $s=0$ can be expressed as the first vector in Eq.~(\ref{eq:eigenvector}).
We assume $\alpha_{x0},\beta_{x0}$ are the Courant-Snyder functions and $D_{x0},D_{x0}'$ are the dispersion and dispersion angle corresponding to the horizontal plane at $s=0$. 
Then the horizontal chromatic function $\mathcal{H}_{x0}$ at $s=0$ is
\begin{equation}
	\begin{aligned}
		\mathcal{H}_{x0}&\equiv\beta_{55}^{I}(0)=2|E_{I5}(0)|^{2}\\
		&=\frac{D_{x0}^{2}+\left(\alpha_{x0}D_{x0}+\beta_{x0}D_{x0}'\right)^2}{\beta_{x0}}.
	\end{aligned}
\end{equation}
The evolution of $\mathcal{H}_{x}$ from $s=0$ to $s\in[-\frac{L_{w}}{2},\frac{L_{w}}{2}]$  is
{}
	\begin{equation}
		\begin{aligned}
			\mathcal{H}_{x}(s)&\equiv\beta_{55}^{I}(s)=2|E_{I5}(s)|^{2}=2|\left({\bf W}(s|0){\bf E}_{I}(0)\right)_{5}|^{2}\\
			&=\frac{(D_{x0}+W_{16}-W_{26}s)^{2}+\left[\alpha_{x0}(D_{x0}+W_{16}-W_{26}s)+\beta_{x0}(D_{x0}'+W_{26})\right]^2}{\beta_{x0}},
		\end{aligned}
	\end{equation}
	where $W_{ij}$ means the $i$-th row and $j$-th column  matrix term of ${\bf W}(s|0)$.
	Put in the explicit expression of the wiggler matrix terms, we have
	\begin{equation}\label{eq:Hx2}
		\begin{aligned}
			\mathcal{H}_{x}(s)&=\frac{1}{\rho _w^2 k_w^4 \beta _{x0}}\left\{[D_{x0}\rho_{w}k_{w}^{2}+\sin \left(k_{w}s\right)k_{w}s+\cos(k_{w}s)-1]^2\right.\\
			&\left.+[\beta _{x0}k_w\left(\rho_{w}k_{w}D_{x0}'- \sin \left(k_{w}s\right)\right)+\alpha_{x0}\left(D_{x0}\rho_{w}k_{w}^{2}+\sin \left(k_{w}s\right)k_{w}s+\cos(k_{w}s)-1\right)]^2\right\}.\\
		\end{aligned}
	\end{equation}
{}
We assume that the quantum excitation contribution from the entrance and exit region of the wiggler, where the field strength in reality deviates from the ideal sinusoidal pattern, is much smaller than that of the central sinusoidal field region.  Then the quantum excitation of a wiggler to the horizontal beam emittance can be evaluated by the integral
\begin{equation}
	\begin{aligned}
		I_{5w}&=\int_{-\frac{L_{w}}{2}}^{\frac{L_{w}}{2}}\frac{\mathcal{H}_{x}(s)}{|\rho(s)|^{3}}ds\\
		&=\frac{1}{\rho_{w}^{3}}\int_{-\frac{L_{w}}{2}}^{\frac{L_{w}}{2}}\mathcal{H}_{x}(s)|\cos(k_{w}s)|^{3}ds,
	\end{aligned}
\end{equation}
where $\rho_{w}=\frac{\gamma m_{e}\beta c}{eB_{0w}}$ corresponds to the bending radius at the location of peak magnetic field $B_{0w}$, with $\beta=\sqrt{1-\frac{1}{\gamma^{2}}}$.
Put Eq.~(\ref{eq:Hx2}) in, we have
{}
	\begin{equation}\label{eq:I5wmost}
		\begin{aligned}
			I_{5w}
			&=\frac{4}{15\pi}\frac{ L_{w}}{\rho_{w}^{5}k_{w}^{2}}\left[\beta_{x0}+\gamma_{x0}L_{w}^{2}\mathcal{R}+5\rho_{w}^{2}k_{w}^{2}\mathcal{H}_{x0}-\left(10+\frac{15\pi}{8}\right)\frac{\rho_{w}\left(D_{x0}+\alpha_{x0}D_{x0}+\beta_{x0}D_{x0}'\right)}{\beta_{x0}}\right],
		\end{aligned}
	\end{equation}
	where $\gamma_{x0}=\frac{1+\alpha_{x0}^{2}}{\beta_{x0}}$ and
	\begin{equation}\label{eq:correction0}
		\mathcal{R}=\frac{15}{32\pi^{2} }\frac{1}{N_{w}^{3}}\int_{-N_{w}\pi}^{N_{w}\pi}[\sin \left(x\right)x+\cos(x)-1]^2|\cos(x)|^{3}dx,
	\end{equation}
	with $N_{w}={L_{w}}/{\lambda_{w}}$ the number of wiggler period which is assumed to be an integer. Equation~(\ref{eq:I5wmost}) above is the exact formula for wiggler's contribution to the radiation integral $I_{5x}$ in an electron storage ring.

	Given a specific $N_{u}$, $\mathcal{R}$ can be straightforwardly obtained by integration in Eq.~(\ref{eq:correction0}).  When $N_{w}\gg1$, we have  
	$
	\mathcal{R}\approx\frac{1}{12},
	$
	and
	\begin{equation}\label{eq:I5waccurate}
		\begin{aligned}
			I_{5w}
			&\approx\frac{4}{15\pi}\frac{ L_{w}}{\rho_{w}^{5}k_{w}^{2}}\left[\langle\beta_{x}\rangle_{w}+5\rho_{w}^{2}k_{w}^{2}\mathcal{H}_{x0}-\left(10+\frac{15\pi}{8}\right)\frac{\rho_{w}\left(D_{x0}+\alpha_{x0}D_{x0}+\beta_{x0}D_{x0}'\right)}{\beta_{x0}}\right],
		\end{aligned}
	\end{equation}
{}
where $\langle\beta_{x}\rangle_{w}$ is the average $\beta_{x}$ along the wiggler
\begin{equation}
	\langle\beta_{x}\rangle_{w}=\frac{1}{L_{w}}\int_{-\frac{L_{w}}{2}}^{\frac{L_{w}}{2}}\beta_{x}(s)ds=\beta_{x0}+\frac{\gamma_{x0}L_{w}^{2}}{12}.
\end{equation}
Denote $\chi_{x0}=\text{Arg}\left(\frac{E_{I5}(0)}{E_{I1}(0)}\right)$, where $E_{I5}(0)$ and $E_{I1}(0)$ represent the fifth and first term of the first eigenvector in Eq.~(\ref{eq:eigenvector}) and Arg() means the angle of a complex number, then
Eq.~(\ref{eq:I5waccurate}) can be written as
{}
	\begin{equation}\label{eq:I5wNew2}
		\begin{aligned}
			I_{5w}
			&\approx\frac{4}{15\pi}\frac{ L_{w}}{\rho_{w}^{5}k_{w}^{2}}\left[\langle\beta_{x}\rangle_{w}+5\rho_{w}^{2}k_{w}^{2}\mathcal{H}_{x0}-\left(10+\frac{15\pi}{8}\right)\rho_{w}\sqrt{\frac{2\mathcal{H}_{x0}}{\beta_{x0}}}\sin\left(\chi_{x0}-\frac{\pi}{4}\right)\right].
		\end{aligned}
	\end{equation}
{}
The first term in the above bracket corresponds to the approximate formula found in literature
\begin{equation}
	I_{5w,\text{intrinsic}}\approx\frac{4}{15\pi}\frac{ L_{w}\langle\beta_{x}\rangle_{w}}{\rho_{w}^{5}k_{w}^{2}}.
\end{equation} 
It can be viewed as the intrinsic contribution of a wiggler to the radiation integral $I_{5x}$, since there will be intrinsic dispersion and dispersion angle generated inside the wiggler even if $D_{x0}=0$ and $D_{x0}'=0$ as can be seen from the matrix term $W_{16}$ and $W_{26}$ of the wiggler. The second and third term in the bracket arise from a nonzero $\mathcal{H}_{x0}$. When $D_{x0}'$ or $\frac{D_{x0}}{\beta_{x0}}$ is of the order $\frac{K}{\gamma}=\frac{1}{\rho_{w}k_{w}}$, which can easily be the case in a real lattice, the contribution from this nonzero $\mathcal{H}_{x0}$ in the bracket could be comparable or even larger than the first term and cannot be neglected. The more accurate formula derived here should then be invoked to calculate the wiggler's quantum excitation of beam emittance. When the third term is much smaller than the second term, i.e., roughly when $\mathcal{H}_{x0}\beta_{x0}\gg\left(\frac{K}{\gamma}\lambda_{w}\right)^{2}$, Eq.~(\ref{eq:I5wNew2}) can be further approximated as
\begin{equation}\label{eq:I5wNew3}
	\begin{aligned}
		I_{5w}
		&\approx\frac{4}{15\pi}\frac{ L_{w}\langle\beta_{x}\rangle_{w}}{\rho_{w}^{5}k_{w}^{2}}+\frac{4}{3\pi}\frac{ L_{w}\mathcal{H}_{x0}}{\rho_{w}^{3}},
	\end{aligned}
\end{equation}
where the first term accounts for the intrinsic contribution, and the second term for the nonzero $\mathcal{H}_{x0}$.


From the above analysis, we can see that the minimum $I_{5w}$ is realized when
\begin{equation}\label{eq:optimalDW}
	\alpha_{x0}=0,\ \beta_{x0}=\frac{L_{w}}{2\sqrt{3}},\ D_{x0}=0,\ D_{x0}'=0,\ \left(\mathcal{H}_{x0}=0\right),
\end{equation}
and the minimal value is
\begin{equation}\label{eq:miniI5w}
	I_{5w,\text{min}}\approx\frac{4}{15\sqrt{3}\pi}\frac{L_{w}^{2}}{\rho_{w}^{5}k_{w}^{2}}.
\end{equation}

Now we can evaluate the impact of damping wiggler on the equilibrium horizontal emittance to make sure the desired $\epsilon_{x}=2$ nm can be realized.
From Eq.~(\ref{eq:Sands}), the equilibrium emittance with damping wiggler is given by 
\begin{equation}
	\epsilon_{xw}=C_{q}\frac{\gamma^{2}}{J_{x}}\frac{I_{50}+I_{5w}}{I_{20}+I_{2w}},
\end{equation}
where $I_{20}$ and $I_{50}$ are the radiation integrals of the ring without the damping wigglers.
Note that the natural emittance without damping wiggler is given by (re-examine the analysis here)
\begin{equation}
	\epsilon_{x0}=C_{q}\frac{\gamma^{2}}{J_{x}}\frac{I_{50}}{I_{20}}
\end{equation}
If we want $\epsilon_{xw}\leq\epsilon_{x0}$, then we need $C_{q}\frac{\gamma^{2}}{J_{x}}\frac{I_{5w}}{I_{2w}}\leq\epsilon_{x0}$. Using $I_{2w}\approx\frac{L_{w}}{2\rho_{w}^{2}}$ and Eq.~(\ref{eq:miniI5w}), we then have
\begin{equation}
	\lambda_{w}\leq2\pi\sqrt{\frac{15\sqrt{3}\pi J_{x}\epsilon_{x0} \rho_{w}^{3}}{8C_{q}\gamma^{2}L_{w}}}.
\end{equation}
According to the above scaling, a longer wiggler length requires a shorter wiggler period to control the quantum excitation contribution to horizontal emittance. This is because the average $\beta_{x}$ and $\mathcal{H}_{x}$ in the wiggler is linearly proportional to $L_{w}$. But note that the above analysis  assumes there is only a single wiggler. If we split the wiggler into $N_{wc}$ identical cells with the total length fixed, we can make the contribution of wiggler to $I_{5w}$ and thus quantum excitation to horizontal emittance becomes a factor of $N_{wc}$ smaller, while the contribution to $I_{2w}$ and thus the effect on radiation damping is unchanged. Then the tolerance of $\lambda_{w}$ can be a factor of $\sqrt{N_{wc}}$ larger.
Put in the numerical numbers, we have a more useful scaling
\begin{equation}
	\lambda_{w}[\text{m}]\leq 3.19\sqrt{\frac{N_{wc}J_{x}E_{0}[\text{GeV}]\epsilon_{x0}[\text{nm}]}{B_{0w}^{3}[\text{T}]L_{w}[\text{m}]}}.
\end{equation}
Nominally $J_{x}\approx1$. For our example parameters given in Tab.~\ref{tab:Chap6-GLSF-SSMB_para1064_600}, $E_{0}=600$ MeV,  $\epsilon_{x0}=2$ nm,  $B_{0w}=6$ T, $L_{w}=40$~m, $N_{wc}=20$ which means each small wiggler has a length of 2~m, we then to realize $\epsilon_{xw}\leq\epsilon_{x0}$ we need
\begin{equation}
	\lambda_{w}\leq0.168\ \text{m}.
\end{equation}
A wiggler with a period length of 10 cm and peak field of 6 T is doable using superconducting magnet technology. We recognize the impact of such a long (total length) damping wiggler on beam dynamics, single-particle nonlinear dynamics and collective instabilities need further in-depth study, especially considering the fact the gap of such a wiggler is small to realize such a strong field strength.  Actually another practical reason to split the wigglers into shorter sections is to avoid the synchrotron radiation generated by itself hitting on the magnet poles. We also recognize it may take some efforts in the lattice design to make the optimal conditions given by Eq.~(\ref{eq:optimalDW}) fulfilled in practice.

Apart from the quantum excitation, the damping wiggler also contributes to the phase slippage. If $B_{w}=6$ T, $\lambda_{w}=0.1$~m, $L_{w}=40$ which means the total wiggler period number $N_{w}=400$, then the wiggle undulator parameter is $K_{w}=0.934\times6\times10=56.04$, then the fundamental resonance wavelength of the wiggler is $\lambda_{rw}=\frac{1+K_{w}^{2}/2}{2\gamma^{2}}\lambda_{w}=57\ \mu$m. The $R_{56}$ of the whole damping wiggler is twice the fundamental frequency radiation slippage length $R_{w,56}=2N_{w}\lambda_{rw}=45.6$ mm~\cite{deng2024theoretical}.  Since the total $R_{56}$ of the ring used in Tab.~\ref{tab:Chap6-GLSF-SSMB_para1064_600} is about 1 m, then the contribution of damping wiggler to $R_{56}$ is acceptable.



\subsection{Intra-beam Scattering}

We mentioned that our conservative choice of $\epsilon_{y}=40$~pm is mainly out of the consideration for intra-beam scattering (IBS).  Now this can be understood with more quantitative calculations. We will see that IBS turns out to the most fundamental obstacle in obtaining the ultrasmall vertical emittance in GLSF SSMB. This is partially because our choice of beam energy is not too high. In addition, to realize high EUV power, we need a high peak current which means a high charge density in phase space. 

We use Bane's high-energy approximation~\cite{Bane2002} to calculate the IBS diffusion rate

\begin{equation}
	\begin{aligned}
		&\frac{1}{T_{\delta}}\approx \frac{r_{e}^2 cN L_{c}}{16\gamma^{3}\epsilon_{x}^{\frac{3}{4}}\epsilon_{y}^{\frac{3}{4}}\sigma_{z}\sigma_{\delta}^{3}}\left\langle\sigma_{H} g_{\text{Bane}}\left(\frac{a}{b}\right)\left(\beta_{x}\beta_{y}\right)^{-\frac{1}{4}}\right\rangle,\\
		&\frac{1}{T_{x,y}}=\frac{\sigma_{\delta}^{2}\langle \mathcal{H}_{x,y}\rangle}{\epsilon_{x,y}}\frac{1}{T_{\delta}},\\
		&\frac{1}{\sigma_{H}^{2}}=\frac{1}{\sigma_{\delta}^{2}}+\frac{\mathcal{H}_{x}}{\epsilon_{x}}+\frac{\mathcal{H}_{y}}{\epsilon_{y}},\\
		&g_{\text{Bane}}(\alpha)=\frac{2\sqrt{\alpha}}{\pi}\int_{0}^{\infty}\frac{du}{\sqrt{1+u^2}\sqrt{\alpha^{2}+u^2}},\\
		&a=\frac{\sigma_{H}}{\gamma}\sqrt{\frac{\beta_{x}}{\epsilon_{x}}},\  b=\frac{\sigma_{H}}{\gamma}\sqrt{\frac{\beta_{y}}{\epsilon_{y}}}.
	\end{aligned}
\end{equation}
where $r_{e}=2.818\times10^{-15}$ m is the classical electron radius and $L_{c}=\ln\left(\frac{b_{\text{max}}}{b_{\text{min}}}\right)$ is the Coulomb Log factor, {where $b_{\text{max}}$ and $b_{\text{min}}$ are the maximum and minimal impact factor for the scattering process, respectively. Typically $L_{c}$ is in the range of 10 to 20.} $N$ is the number of electrons in the bunch. For a coasting beam, we need to replace $\sigma_{z}\rightarrow L/(2\sqrt{\pi})$ where $L$ is the bunch length. Note that $\frac{eN}{L/c}=I_{P}$ according to our definition, where $I_{P}$ is the peak current. 


Now we put in some example numbers to do an estimation for the IBS diffusition rate in a GLSF EUV SSMB ring:
\begin{equation}
	\begin{aligned}
		&E_{0}=600\ \text{MeV},\ I_{P}=40\ \text{A},\ \sigma_{\delta}=8.5\times10^{-4},\\
		&\epsilon_{x}=2\ \text{nm},\ \epsilon_{y}=40\ \text{pm}, \langle \sigma_{H}\rangle=4\times10^{-5} ,\\
		&\frac{a}{b}=\frac{1}{10},\ g_{\text{Bane}}\left(\frac{1}{10}\right)=0.744,\
		\langle\left(\beta_{x}\beta_{y}\right)^{-\frac{1}{4}}\rangle=0.32,\\ 
		&\langle \mathcal{H}_{y}\rangle=\frac{2\times1.6\times0.056\ \text{m}}{200}=0.9\ \text{mm},\ L_{c}=10.
	\end{aligned}
\end{equation}
Note that in Tab.~\ref{tab:Chap6-GLSF-SSMB_para1064_600} we have $\mathcal{H}_{y}=0.056$ m at the two modulators, whose length are both 1.6 m. In evaluating $\langle \mathcal{H}_{y}\rangle$ we have only considered the contribution from two modulators, where $\mathcal{H}_{y}$ reach the its maximum value. This is a simplification, but should give a correct order of magnitude estimation. Putting in the example numbers, we then have
\begin{equation}
	\begin{aligned}
		\tau_{\delta\text{IBS}}&=113\ \text{ms},\
		\tau_{y\text{IBS}}={7.11\ \text{ms}}.
	\end{aligned}
\end{equation}
Compared with the radiation damping times given in Eq.~(\ref{eq:RDWiggler}), we can see that even for the vertical dimension, the IBS diffusion is more than {three times} slower than radiation damping. Therefore, the IBS diffusion can now be controlled by the strong damping induced by the damping wiggler. This calculation also justifies the necessity or benefit of applying damping wigglers.

\subsection{Microwave Instability}

Now we want to evaluate if the peak current of 40 A we apply in the above example is doable. One of the main limitation of peak current is the microwave instability induced by coherent synchrotron radiation (CSR).  According to Ref.~\cite{Bane2010}, the CSR-induced microwave instability threshold is 
\begin{equation}
	(S_{\text{CSR}})_{\text{th}}=0.5+0.12\Pi,
\end{equation}
with 
\begin{equation}
	S_{\text{CSR}}=\frac{I\rho^{1/3}}{\sigma_{z0}^{4/3}},\ I=\frac{r_{e}N_{b}}{2\pi\nu_{s0}\gamma\sigma_{\delta0}},\ \Pi=\frac{\sigma_{z0}\rho^{1/2}}{g^{3/2}},
\end{equation}
and $2g$ is the separation between the two plates. So the threshold peak current is
{}
	\begin{equation}
		\begin{aligned}
			I_{\text{th,peak}}&=\frac{eN_{b}}{\sqrt{2\pi}\sigma_{z0}/c}
			=\frac{1}{2\sqrt{2\pi}}I_{\text{Alf}}\gamma\left(1+0.24\frac{\sigma_{\delta0}|R_{56}|^{1/2}\rho^{1/2}}{|h_{RF}|^{1/2}g^{3/2}}\right)\frac{\sigma_{\delta0}^{\frac{4}{3}}|R_{56}|^{\frac{2}{3}}|h_{RF}|^{\frac{1}{3}}}{\rho^{1/3}},
		\end{aligned} 
	\end{equation} 
{}
with $I_{\text{Alf}}=\frac{ec}{r_{e}}=17$ kA being the Alfven current. The $R_{56}=-\eta C_{0}$ is that of the whole ring. $h_{RF}=\frac{eV_{RF}\cos\phi_{s}}{E_{0}}k_{RF}$ is the linear energy chirp strength around the synchronous RF phase. Putting in some typical parameters for the EUV SSMB: $E_{0}=600$ MeV, $B_{0}=1.33$~T, $\rho=1.5$ m, $|R_{56}|=1$~m,  $\sigma_{\delta0}=8.5\times10^{-4}$, $g=4$ cm, $h_{RF}=0.01\ \text{m}^{-1}$, then
\begin{equation}
	I_{\text{th,peak}}=60.3\times\left(1+0.31\right)\ \text{A}=79\ \text{A}.
\end{equation}
So our application of a peak current of 40 A should be safe from microwave instability. 

{The astute readers may notice that one of the main reasons we have a large threshold current here is the large phase slippage or $R_{56}$ that we applied to the ring. To avoid confusion, first we need to make clear that in this example GLSF SSMB EUV source, the electron bunch in the ring can be a coasting beam or an RF-bunched beam, and microbunching appears only at the radiator, due to the phase space manipulation of the GLSF section. Actually in our setup, we have used an RF-bunched beam in the ring. Therefore, the phase slippage factor of the ring does not need to be small, like that in a LWF SSMB ring. Then the question becomes whether the required large phase slippage is doable and what's the beam dynamical effect it may have. As a reference, the Metrology Light Source storage ring~\cite{Feikes2011} in standard user mode has an $|R_{56}|\approx1.6$~m, which means  a phase slippage factor of $3.3\times10^{-2}$ given a circumference of 48 m. Therefore we believe our application of $|R_{56}|=1$ m is realizable. But we recognize that such a large $R_{56}$ requires large horizontal dispersion $D_{x}$ at the dipoles, since $-R_{56}=\eta C_{0}\approx2\pi\langle D_{x}\rangle_{\rho}$ where $\langle\rangle_{\rho}$ means average around the dipoles in the ring. A large $D_{x}$ may result in a large $\mathcal{H}_{x}$, and then  the quantum excitation of dipoles to horizontal emittance $\epsilon_{x}$ should be carefully evaluated. A too large $\epsilon_{x}$ will degrade the coherent EUV radiation~\cite{deng2024theoretical,deng2023average}. What we need is $|R_{56}|\sim1$ m, and at the same time $\epsilon_{x}\lesssim2$~nm, and some optimization may be required realize such a goal.}

{Now we check if the required RF system in the above example calculation is feasible. The longitudinal beta function at the RF cavity is $\beta_{z}\approx\sqrt{\eta C_{0}/h_{RF}}=10$ m, then the RMS bunch length is $\sigma_{z}=\sigma_{\delta0}\beta_{z}=8.5$ mm. To get a bean filling factor of 0.5\%, roughly we need an RF wavelength of 1.7~m, which means an RF frequency of 176.5 MHz. Then the required energy chirp strength of $h_{RF}=0.01\ \text{m}^{-1}$, which means the required RF voltage is 1.62 MV. Such an RF voltage should be doable at this frequency range. Multiple cavities can be invoked if it is too demanding for a single cavity to reach the desired voltage.}


We remind the readers again that in this evaluation, we actually assumed that the beam is RF bunched. While in our evaluation of IBS, we have assumed that the beam is a coasting beam. These evaluations here mainly serve as an order of magnitude estimation that support the general feasibility of our parameter choice. A more detailed analysis of collective effects will be necessary in the future development of such a GLSF SSMB light source.

%
%

\subsection{{Energy Compensation System}}

The large radiation loss of the electron beam in the ring, especially that induced by the strong damping wigglers, need to be compensated. Here we present some preliminary analysis on the requirement of the energy compensation system of such an GLSF SSMB EUV light source.  From Tab.~\ref{tab:Chap6-GLSF-SSMB_para1064_600}, the total radiation loss per particle per turn is $U_{0}=U_{0\text{dipoles}}+U_{0w}+U_{0\text{R}}=341.3$~keV, where the three terms on the right hand side represent the radiation loss in dipoles, damping wigglers and radiator, respectively. So the synchronous acceleration phase corresponds to an acceleration voltage of $V_{\text{acc}}=341.3$ kV. Such a high voltage is not easy to be realized using induction linac, considering the required repetition rate is in the MHz level. So we use conventional RF cavities to supply the radiation loss. Following the discussion of last section, and assuming we have used an RF frequency of 166.6 MHz (RF wavelength 1.8 m), then we need an RF voltage of 1.72 MV to realize an energy chirp strength of $h_{RF}=0.01\ \text{m}^{-1}$. Under this parameters set, and consider $V_{RF}\gg V_{\text{acc}}$, the RF bucket half-height is~\cite{deng2024theoretical}
\begin{equation}
	\hat\delta_{\frac{1}{2}}=\frac{2}{\beta_{z}k_{RF}}=\frac{\lambda_{RF}}{\pi}\sqrt{\frac{h}{\eta C_{0}}}=5.73\times10^{-2},
\end{equation} 
which is $67.4\sigma_{\delta 0}$ with the energy spread $\sigma_{\delta 0}=8.5\times10^{-4}$. So the bucket half-height should be large enough to ensure the beam quantum lifetime and Touschek lifetime. 

To relieve the burden on the RF cavities, we may use three RF cavities to achieve the total RF voltage, with each cavity having a voltage of 573 kV. To minimize the power disspated on the cavity, we need a large shunt impedance of each cavity which here we assume to be $R_{s}=20\ \text{M}\Omega$, then the total power dissipated on the three cavity walls is
\begin{equation}
	P_{\text{diss}}=\frac{1}{3}\frac{V_{RF}^{2}}{R_{s}}=49.2\ \text{kW}.
\end{equation}
The power delivered to the 200 mA-average current beam is 
\begin{equation}
	P_{b}=I_{A}\frac{U_{0}}{e}=68.3\ \text{kW}.
\end{equation}
So the power dissipated on the cavity walls is at the same level as that delivered to the beam. We recognize the large shunt impedance may take efforts to realize in practice. If the shunt impedance is lower than the assumed value, then the power consumption on the wall will be correspondingly larger. Superconducting RF cavities can be used to lower the power dissipation on the wall.

Generally, the total power consumption of the RF system of an SSMB ring is at 100 kW to 200 kW level. Together with the power consumption of the other systems like electromagnets, superconducting damping wigglers, vacuum and water cooling system, the overall power consumption of an SSMB storage ring is at the level of several hundred kWs. The output EUV power per radiator is about 1~kW. In principle, an SSMB ring can accommodate multiple GLSF insertions and therefore multiple radiators, but here we consider the case of only one radiator in the ring. So for such an GLSF SSMB storage ring, it takes a couple of 100 kWs electricity power to generate 1 kW EUV light. Such a large power consumption may raise the question on the advantage of such an SSMB-EUV source compared to the superconducting RF-based high-repetition rate FEL-EUV source, in particular, the energy recovery linac-based FEL-EUV source. As a comparison, according to Ref.~\cite{Nakamura2023}, it takes 7 MW overall power consumption to generate 10 kW EUV light in an ERL-FEL EUV source, which means 700~kW electricity power per 1~kW EUV light. So the overall power efficiency from electricity to EUV light for an SSMB-EUV source and an ERL-FEL EUV source is comparable. But we remind the readers that these numbers are only rough estimation, and it requires more indepth study to reach a concrete conclusion.  Another side comment is that the radiation emitted by the damping wigglers may also be useful.

\subsection{Modulation Laser Power}

Now we evaluate the modulation laser power required. Given the laser wavelength, modulator undulator parameters and the required energy chirp strength, we can use Eq.~(\ref{eq:TEM00h}) to calculate the required laser power
\begin{equation}
	P_{L}=\frac{\lambda_{L}}{4Z_{0}Z_{R}}\left(h\frac{1}{\frac{eK[JJ]}{\gamma^{2}m_{e}c^{2}}\tan^{-1}\left(\frac{L_{u}}{2Z_{R}}\right)k_{L}}\right)^{2}.
\end{equation}
Using Theorem one, i.e., Eq.~(\ref{eq:theorem1}), and let the equality in the relation holds, we have
{}
	\begin{equation}\label{eq:PLAccurate}
		\begin{aligned}
			P_{L}
			&\approx\frac{1}{1+R_{{w}}}\frac{\epsilon_{y}}{\Delta\epsilon_{y\text{M}}}\frac{1}{\left(K[JJ]\right)^{2}}\frac{\lambda_{L}^{3}}{3\pi^{3}Z_{0}}\frac{55}{48\sqrt{3}}\frac{\alpha_{F}c^{2}{\lambdabar}_{e}^{2}\gamma^{7}B^{3}_{0\text{M}}}{C_{\gamma}E^{3}_{0}B_{\text{ring}}}\frac{1}{\sigma_{z\text{R}}^{2}}\frac{\frac{L_{u}}{2Z_{R}}}{\left[\tan^{-1}\left(\frac{L_{u}}{2Z_{R}}\right)\right]^{2}},\\
		\end{aligned}
	\end{equation}
{}
where $\alpha_{F}=\frac{1}{137.036}$ is the fine structure constant, and $R_{{w}}$ is given by Eq.~(\ref{eq:R}). In the above derivation we have used the electron momentum $P_{0}=\gamma m_{e}c$ and  approximation $E_{0}\approx P_{0}c$.
Now we try to derive more useful scaling laws to offer guidance in our parameters choice for a GLSF SSMB storage ring. {As shown in Fig.~\ref{fig:PlanarUndulatorEnergyModulation},} to maximize the energy modulation, we need $\frac{Z_{R}}{L_{u}}=0.359\approx\frac{1}{3}$. When $K>\sqrt{2}$, we approximate the resonance condition as
\begin{equation}
	\lambda_{L}=\frac{1+\frac{K^{2}}{2}}{2\gamma^{2}}\lambda_{u}\approx\frac{K^{2}}{4\gamma^{2}}\lambda_{u},
\end{equation}
and $[JJ]\approx0.7$. Then we have
\begin{equation}
	\begin{aligned}
		P_{L}&\propto\frac{1}{1+R_{{w}}}\frac{\epsilon_{y}}{\Delta\epsilon_{y\text{M}}}\frac{\lambda_{L}^{3}}{ K^{2}}\frac{\gamma^{4}B^{3}_{0\text{M}}}{B_{\text{ring}}}\frac{1}{\sigma_{z\text{R}}^{2}}\\
		&\propto\frac{1}{1+R_{{w}}}\frac{\epsilon_{y}}{\Delta\epsilon_{y\text{M}}}\frac{\lambda_{L}^{\frac{7}{3}}\gamma^{\frac{8}{3}}B^{\frac{7}{3}}_{0\text{M}}}{B_{\text{ring}}}\frac{1}{\sigma_{z\text{R}}^{2}}.
	\end{aligned}
\end{equation}
Putting in the numbers for the constants, we obtain the quantitative expressions of the above scalings for practical use
\begin{equation}\label{eq:scaling4}
	\begin{aligned}
		P_{L}[\text{kW}]&\approx 5.7\frac{1}{1+R_{{w}}}\frac{\epsilon_{y}}{\Delta\epsilon_{y\text{M}}}\frac{\lambda_{L}^{\frac{7}{3}}[\text{nm}]E_{0}^{\frac{8}{3}}[\text{GeV}]B_{0\text{M}}^{\frac{7}{3}}[\text{T}]}{\sigma_{z\text{R}}^{2}[\text{nm}]B_{\text{ring}}[\text{T}]}.
	\end{aligned}
\end{equation}
{The above scaling laws are accurate when $K>\sqrt{2}$, and it should be noted that the calculated power refers to the peak power of the laser.}
For completeness, here we give also the modulator length scaling
\begin{equation}
	L_{u\text{M}}[\text{m}]\approx56.2\left(1+R_{{w}}\right)\frac{B_{\text{ring}}[\text{T}]\Delta\epsilon_{y\text{M}}[\text{pm}]}{\mathcal{H}_{y\text{M}}[\mu\text{m}]B_{0\text{M}}^{3}[\text{T}]}.
\end{equation}
Therefore, to lower the required modulation laser power, we can apply a large $R_{{w}}$, which means a strong damping wiggler. A  low beam energy and short laser wavelength are also preferred. But their choices should take more factors into account, for example IBS and engineering experience of optical enhancement cavity as explained in Sec.~\ref{sec:SSMB}. A strong bending magnet field in the ring is also desired. Concerning the modulator field strength, a weaker one is favored to lower the required laser power. But note that the required modulator length may be longer, if we keep the quantum excitation contribution of modulators to vertical emittance $\Delta\epsilon_{y\text{M}}$ unchanged in this process.


\subsection{Radiation Power}\label{sec:PR}

Having formed the microbunching, now comes the radiation generation. We will use a planar undulator as the radiator. Coherent undulator radiation power at the odd-$H$-th harmonic from a transversely-round electron beam is~\cite{deng2024theoretical,deng2023average}
\begin{equation}\label{eq:PR}
	P_{H,\text{peak}}[\text{kW}]=1.183N_{u}H\chi[JJ]_{H}^{2}FF_{\bot}(S)|b_{z,H}|^{2}I_{P}^{2}[\text{A}],
\end{equation}
where $N_{u}$ is the number of undulator periods, 
\begin{equation}
	[JJ]_{H}^{2}=\left[J_{\frac{H-1}{2}}\left(H\chi\right)-J_{\frac{H+1}{2}}\left(H\chi\right)\right]^{2},
\end{equation}
with $\chi=\frac{K^{2}}{4+2K^{2}}$, and the transverse form factor is
\begin{equation}
	FF_{\bot}(S)=\frac{2}{\pi}\left[\tan^{-1}\left(\frac{1}{2S}\right)+S\ln\left(\frac{(2S)^{2}}{(2S)^{2}+1}\right)\right],
\end{equation}
with $S=\frac{\sigma^{2}_{\bot}\frac{\omega}{c}}{L_{u}}$ and $\sigma_{\bot}$ the RMS transverse electron beam size, $b_{z,H}$ is the bunching factor at the $H$-th harmonic decided by the longitudinal current distribution, and $I_{P}$ is the peak current.

The above formula is derived by assuming that the longitdudinal and transverse distribution of the electron beam do not change much in the radiator. Actually the energy spread of electron beam can lead to current distribution change inside the undulator, considering that the undulator has an $R_{56}=2N_{u}\lambda_{0}$, where $\lambda_{0}$ is the fundamental on-axis resonant wavelength of the undulator. If we consider the impact of energy spread on coherent radiation, and assuming that the microbunching length reach its minimum at the radiator center, there will be a correction or reduction factor multiplied to the radiation power given by Eq.~(\ref{eq:PR})
\begin{equation}
	\mathcal{C}=\frac{\sqrt{\pi}}{2}\frac{\text{erf}\left(\frac{\omega}{c}\sigma_{\delta}N_{u}\lambda_{0}\right)}{\frac{\omega}{c}\sigma_{\delta}N_{u}\lambda_{0}}.
\end{equation}
Here we remind the readers that in our case of GLSF SSMB, the beam is actually not transversely round at the radiator. Therefore, we have used an effective round beam size in evaluating the radiation power using the above formula. This effective transverse beam size $\sigma_{\bot}$ is in between $\sigma_{x}$ and $\sigma_{y}$. For a more accurate calculation of radiation power, we can use the real beam distribution and invoke numerical method~\cite{deng2024theoretical}.

With all the beam physics issues properly handled, we have finally obtained a solution of a 1 kW EUV source based on GLSF SSMB as shown in Tab.~\ref{tab:Chap6-GLSF-SSMB_para1064_600}.


\section{Summary}\label{sec:summary}

This paper is about our efforts in obtaining a solution for 1~kW EUV light source based on SSMB. Here we give a short summary of this endeavor. {We start by presenting the generalized Courant-Snyder formalism to build the theoretical framework for the following investigations.} Based on the formalism we conducted theoretical minimum emittance analysis in an electron storage ring, from which we know that to get small longitudinal emittance, we need to decrease the bending angle of each bending magnet which means increasing the number of bending magnets in the ring. In principle we can get as small longitudinal emittance and as short bunch length as we want along this line. But there is actually practical limitation. To get short bunch length, we need not only  to increase the bending magnet number, but also to lower the phase slippage factor of the ring. Using present realizable phase slippage, which at minimum is in the order of $1\times10^{-6}$, a bunch length {of} a couple of 10 nm is the lower limit if we apply the longitudinal weak focusing regime. To compress the bunch length further, longitudinal strong focusing regime can be invoked, not unlike its transverse counterpart in the final focus of a collider, to compress the longitudinal beta function thus the bunch length at the radiator significantly. This scheme can realize a bunch length of nm level, thus allowing coherent EUV radiation generation. However, since the compression of longitudinal beta function requires a strong energy chirp of the electron beam, which is similar to a strong quadrupole focusing strength in the transverse dimension, the modulation laser power required is in GW level, making the optical enhancement cavity of SSMB can only work in a low duty cycle pulsed mode, and thus limits the filling factor of microbunched beam in the ring, and thus the average output EUV power. This then leads us to the generalized longitudinal strong focusing (GLSF) regime, which is the focus of this paper. The basic idea of GLSF is to exploit the ultrasmall natural vertical emittance in a planar electron storage ring and apply partial transverse-longitudinal emittance exchange to compress the bunch length or generate high-harmonic bunching with a shallow energy modulation strength, thus lowering the requirement on modulation laser power. The backbone of such a scheme is the transverse-longitudinal phase space coupling. To find a solution based on the GLSF scheme, we first conduct some formal mathematical analysis of transverse-longitudinal coupling (TLC)-based bunch compression and harmonic generation schemes, and prove three related theorems which are useful in later choice of parameters and evaluation of laser power. We then go into the details of different specific coupling schemes, grouping them into two categories, i.e., energy modulation-based coupling schemes and angular modulation-based coupling schemes. We derive the formulas of bunching factor and laser-induced modulation strength in each case, and use them for quantitative calculations and comparisons. Our conclusion from these {analyses} is that the commonly used TEM00 mode laser-induced energy modulation-based schemes are favored for our application in SSMB, as its requirement on modulation laser power is lower than that in angular modulation-based schemes. There are also other various important physical issues to be taken into account in finding a solution, like the quantum excitation contribution of GLSF section to the vertical emittance, the application of damping wiggler to speed up damping and its impact on transverse emittance, the intra-beam scattering and coherent synchrotron radiation-induced microwave instability, and the energy compensation system. The motivation of these studies is to make our choice of parameters as {self-consistent} as possible from a beam physics perspective. Finally, based on all the {analyses} and calculations we present an example parameters set of a GLSF SSMB light source which can deliver 1~kW-average-power EUV radiation. This 1 kW EUV solution given in Tab.~\ref{tab:Chap6-GLSF-SSMB_para1064_600} can be viewed as a summary of our investigations presented in this paper. Our work provides the basis for the future development of SSMB.

\section*{Acknowledgements}
	We thank Chao Feng, Ji Li and Yujie Lu for helpful discussions on storage ring-based coherent light source, tilted laser modulation schemes, and the issue of quantum excitation contribution of damping wiggler to the horizontal beam emittance. Many helpful discussions with our colleagues in Tsinghua and other institutes are also much appreciated. This work is supported by the National Key Research and Development Program of China (Grant No.~2022YFA1603401), the National Natural Science Foundation of China (NSFC Grant No. 12035010 and No. 12342501), the Beijing Outstanding Young Scientist Program (No. JWZQ20240101006) and the Tsinghua University Dushi Program.

\end{document}